# Lifetimes of vibro-rotational levels for excited electronic states of the diatomic hydrogen isotopologues

**S. A. Astashkevich[1], B. P. Lavrov[2]**

*Faculty of Physics, St.-Petersburg State University, 198504, Russian Federation*

## Abstract

Current situation in studies of lifetimes of excited rovibronic levels for the $H_2$, $D_2$, $T_2$, HD, HT, and DT molecules is analyzed. All measured values of the lifetimes, reported before November 2014, are compiled and listed in the tabular format together with annotated bibliography. Experimental data are now available for the $H_2$, HD and $D_2$ molecules only. The data collected in the present work show fragmentariness of experimental data. For vast majority of the levels the lifetime value was reported in one paper only, and has no independent experimental verification. Complete bibliography of publications concerning semi-empirical determination and non-empirical calculations of the lifetimes is presented. Numerical results obtained by these two approaches are listed in an explicit form only in the cases when experimental data are available. For more than half of the levels the differences between measured and calculated values are 3 times higher than experimental errors. This direct contradiction shows necessity of more precise experimental and non-empirical studies. For finite number of rovibronic levels our analysis made it possible to propose certain set of recommended data.

---

[1] astashkevich@mail.ru.

[2] b.lavrov@spbu.ru (author to whom correspondence should be addressed).

**CONTENTS**

## List of Symbols and Abbreviations

| Symbols | Meanings |
|---|---|
| $\tau$ | The lifetime. |
| $\sigma$ | An estimate of experimental errors reported by authors of original papers. |
| $\delta = \sigma/\tau$ | Relative error estimates of measured lifetime values presented in original papers. |
| $P$ | The molecular gas pressure. |
| $v$ | Vibrational quantum number. |
| $J$ | Quantum number of total angular momentum of the molecule excluding nuclear spin. |
| $N$ | Quantum number of total angular momentum of the molecule excluding electron and nuclear spins. |
| $\Lambda$ | Quantum number of the component of the resultant orbital angular momentum of the electrons about the internuclear axis. Designations $\Sigma$, $\Pi$, and $\Delta$ correspond to $\Lambda = 0$, 1, and 2. |
| $nl\lambda$ | The state of a single electron in a molecule (an orbital) corresponding to the united atom limiting case |
| $n$ | Principle quantum number. |
| $l$ | Quantum number of the orbital angular momentum of an electron in the molecule. Designations $s, p,$ and $d$ correspond to $l = 0$, 1, and 2. |
| $\lambda$ | Quantum number of the component of the orbital angular momentum of an electron about the internuclear axis. Designations $\sigma, \pi,$ and $\delta$ correspond to $\lambda = 0$, 1, and 2. |
| DC | Method of Delayed Coincidences |
| TF | Time-of-Flight spectroscopy |
| Hanle | Method Hanle |
| MOMRIE | Molecular Optical Magnetic Resonance Induced by Electrons |
| PS | Phase Shift method |
| PLASMA | The plasma experiments |
| ei | Electron impact excitation |
| sr | Absorption of synchrotron radiation |
| ch.tr. | Charge transfer in ion-molecule collisions |
| laser | Absorption of the tunable laser radiation |
| SE | Semi-empirical methods |
| NE | Non-empirical calculations |
| AA | Adiabatic approximation |
| NA | Non-adiabatic models |

# List of Tables

# List of Figures

# 1.     Introduction

Experimental, theoretical and computational studies of optical spectra and structure of the hydrogen molecule[1] have rather long history [34RIC, 58DIE, 72CRO, 79HUB/HER, 85FRE/SCI]. In spite of great many efforts and victories along this way many unknowns are remaining so far. Permanent interest to studies of spectra and structure of the diatomic hydrogen isotopologues is stimulated mainly by two kinds of motivations.

On the one hand, diatomic hydrogen is one of the fundamental objects of molecular spectroscopy and quantum mechanics (the simplest neutral molecule – the four-particle system). It is very useful for studies of various non-adiabatic effects of intramolecular dynamics and for testing new experimental and computational techniques (see e.g. [11ROS/TSU, 11GLA/JUN, 13SPR/JUN, 14SAL/BAI]).

---

[1] Here and below the term *hydrogen molecule* is used in a broad sense to denote a system consisting of two nuclei and two electrons. In cases when some statement concerns to a particular isotopologue only, we use corresponding chemical symbols: $H_2$, $D_2$, $T_2$, HD, HT and DT.

On the other hand (for various applications of quantitative spectroscopy), atomic and molecular hydrogen is widely spread in nature, objects of basic research, and various technical applications (astrophysics [00MOL], [13TIE], controlled nuclear fusion [09AAS], nonthermal plasma chemistry and physics [84LAV, 01RÖP/DAV, 13NPCP], ultraviolet sources of radiation [70ZAI, 99LAV/MEL], plasma and surface modern technologies [08LOW, 13FUK] and others). Therefore, researchers and engineers engaged in various fields of science and technology need information concerning spectroscopic characteristics of hydrogen molecules. Experimental determining cross sections and rate coefficients of various collision processes by detecting an emission of molecular hydrogen also needs reliable data on radiative transition probabilities and mean lifetimes of the excited states.[2]

There is a certain contradiction between those two aspects. Indeed, the most important in investigations inspired by the first kind of motivations is to achieve maximum possible precision of experimental data and/or results of calculations. This is usually made for rather limited number of rovibronic levels because obtaining systematic data for a large number of vibro-rotational levels for the same electronic state is not of much interest and significance. On the contrary, our ability to use spectroscopic data in various applications assumes an existence of certain complete sets of data for a wide range of vibro-rotational levels within electronic states of interest [96AST/KÄN, 84LAV, 99LAV/MEL, 00ABG/ROU, 01RÖP/DAV, 06FAN/WÜN, 13NPCP]. Therefore, nowadays there exists some kind of a contradiction between goals of molecular spectroscopy and needs of the quantitative spectroscopy and computer modeling of ionized gases and plasmas.

In spite of integral nature, experimental lifetime values of excited electronic-vibro-rotational states are of particular interest being the observable characteristics of the states. They may be used as reference data for a comparison with results of non-empirical calculations obtained in the framework of various theoretical models (so called *the direct spectroscopic problem*). Within *the inverse spectroscopic problem*, they represent a channel of information about dipole moments and wave functions. It is necessary to notice, that one have to distinct between two different terms *a mean radiative lifetime* (or simply *a radiative lifetime*) and *a mean lifetime* (*a lifetime*), both relating to an excited state of an isolated molecule. The first one is used to characterize the lifetime of an excited state caused by the spontaneous emission only. It is equal to the reciprocal value of the sum of probabilities for all allowed spontaneous transitions to the underlying states. In the presence of significant predissociative effects, this value is extremely difficult for direct experimental determining, but can be calculated by non-empirical and some semi-empirical methods. The second term represents a total mean lifetime of an excited state of an isolated molecule due to all possible decay processes. It takes into account, in addition to the spontaneous emission, some other de-excitation intramolecular processes caused by perturbations, predissociation and/or preionization. This magnitude is measured in experiments when all secondary collisional-radiative processes are excluded or taken into account. The difference between the two terms is usually insignificant in studying lifetimes of atomic states, since de-excitation via autoionization is a rare exclusion for double excited states. Finally, "*an effective mean lifetime*" (*an effective lifetime)* of an excited state of a molecule in ionized gases and plasmas is smaller than the lifetime of the same excited state of an isolated molecule due to an additional de-excitation (stepwise processes and/or quenching) caused by collisions with electrons, other particles and walls.

---

[2] See e.g. the qualitative difference in final results between the cross sections for the electron impact excitation of vibro-rotational levels of the $d^3\Pi_u^-$ electronic state of $H_2$ obtained in [81LAV/OST] and in [99LAV/MEL]. The results reported in [81LAV/OST] were based on rovibronic transition probabilities from [82KIR/LAV] and lifetimes from [78DAY/AND]. Similar data reported in [99LAV/MEL] were obtained with transition probabilities from [89LAV/POZ] and the lifetimes from [90BUR/LAV]. In the first case, the cross sections have shown strong non-Franck-Condon behavior, while in the second one the data were in excellent accordance with the Franck-Condon approximation.

The aim of the present work was:

1) to collect a complete set of bibliography devoted to experimental, semi-empirical and non-empirical studies of lifetimes for vibro-rotational levels of excited electronic states of the diatomic hydrogen isotopologues;

2) to present in the tabular format all available experimental values of the lifetimes for rovibronic levels of all isotopologues of molecular hydrogen and to compare, when it is possible, experimental and semi-empirical data with results of non-empirical calculations;

3) to study an opportunity to develop a certain set of reference data suitable for spectroscopic diagnostics and modeling of non-equilibrium hydrogen-containing plasmas in various fields of science and technology.

Present compilation is based on the results reported in all papers (known to authors) published before November 2014 in the following journals: Anales de Fisica (Spain), Astronomy and Astrophysics Supplement Series, Astrophysics Journal, Atomic Data & Nuclear Data Tables, Bulletin of American Physics Society, Chemical Physics, Chemical Physics Letters, Journal of Chemical Physics, Journale de Phisique, Journal of Molecular Spectroscopy; Journal of Physics B: Atomic, Molecular and Optical Physics; Journal of the Physical Chemistry, Journal of Physics Society of Japan, Journal of Quantitative Spectroscopy & Radiative Transfer, Journal of Optics Society of America, New Journal of Physics, Optics and Spectroscopy, Physics Abstracts, Physics Letters A., Physical Review, Physical Review A., Physical Review E., Physical Review Letters, Soviet Physics JETP, Spectrochemical Acta A., Vestn.Leningr.Univ. Fiz.&Khim.(USSR), Zeitschrift fur Physik D.

## 2. Specificity of optical spectra of the hydrogen molecule isotopologues

The peculiarity of molecular hydrogen and its isotopic species – abnormally small nuclear masses – leads to large separations of vibrational and rotational levels within electronic states. Therefore, spectral lines belonging to different branches, bands and band systems are located in the same spectral regions and often overlap each other. Under sufficiently high resolution, the rovibronic spectrum of diatomic hydrogen has no easily recognizable band structure, but looks like multiline atomic spectrum. For dependable measurements of the separate line intensities one needs to provide high spectral resolution together with high sensitivity of experimental setup. On the other hand, small nuclear masses stimulate a breakdown of the Born–Oppenheimer approximation due to adiabatic and non-adiabatic effects of perturbations which may have regular [69KOV, 70HOU, 75MIZ, 83KOV/LAV, 86LEF/FIE] or irregular [35aDIE, 91AST/LAV, 93KIY/SAT] character. This seriously complicates an interpretation of optical spectra of hydrogen isotopologues and unambiguous identification of rovibronic spectral lines. Symmetry rules for the permutation of identical nuclei in homonuclear isotopologues ($H_2$, $D_2$ and $T_2$) cause the known effect of an alternation of line intensities within the rotational structure of bands due to the alternation in degeneracy of successive rotational levels with odd and even values of the rotational quantum number $N$. For example in the $(1s\sigma)$ $X$ $^1\Sigma_g^+$ ground electronic states the ratios of the degeneracy order for even and odd $N$ values are equal to 1:3 in the case of $H_2$ and $T_2$, and to 6:3 for $D_2$. This effect brings serious additional difficulties for an identification of spectral lines. Thus, most of the lines (about two thirds) observed in the optical spectra of $H_2$ and $D_2$ have not yet been assigned in spite of tremendous efforts of researchers over the last century [34RIC, 58DIE, 72CRO, 79HUB/HER, 85FRE/SCI] (see also [08LAV/UMR]). Results of non-empirical calculations are often used in the process of the line assignment. But currently for exited electronic states the precision of calculations of the rovibronic term values and wavenumbers of spectral lines (about $0.1 \div 10$ sm$^{-1}$) is still significantly worse than that of experimental data (about $0.0001 \div 0.05$ sm$^{-1}$) [88DAV/GUE, 90bQUA/DRE, 94RON/LAU, 98REI/LAN, 03SAT/YOS, 05AIT/OGI, 06ROU/LAU, 07ROU/TCH, 10BAI/SAL, 11DIC/IVA, 12DIC/SAL, 12LAV/UMR].

Because of absence of well-separated bands and band systems of hydrogen molecule an experimentalist has to work with intensities of individual electronic-vibro-rotational lines and levels. This makes rather strong limitations for the lifetime measurements of singular rovibronic states of hydrogen molecule. On the one hand, such measurements need high resolving power together with high sensitivity of spectroscopic setup.[3] On the other hand, experiments have to be carried out with low enough gas pressure for excluding or taking into account collisional quenching of excited molecules. This inevitably leads to corresponding decrease in excitation rates and line intensities. Therefore, it is not surprising that experimental data are fragmentary and the data obtained by different methods and/or by different authors are in discordance quite often (see below).

For the low-lying excited electronic states of diatomic hydrogen ($n$=2÷4) the angular momentum coupling is not too far from the Hund's limiting case "$b$". Fine and hyperfine structures of levels and lines are less than ≈0.05 cm$^{-1}$ except ≈0.2 cm$^{-1}$ in the (1σ2pπ) $c^3\Pi_u^-$ state. They are usually unresolved by optical spectrometers used in the lifetime measurements. Therefore in the present work the rotational levels are characterized by the quantum number $N$ of the total angular momentum of a molecule excluding spins of electrons and nuclei. In rare cases when rather thin triplet structure was resolved, we used the quantum number $J$ of total angular momentum excluding spins of nuclei as well.

Non-adiabatic interactions between electronic and nuclear degrees of freedom are rather significant for almost all excited electronic states of the hydrogen molecule. In experiments, they become apparent as the so-called perturbations – deviations of observable molecular characteristics from those predicted by an adiabatic approximation. The examples are significant perturbations of the following observable magnitudes:

(i)       The rovibronic terms and spectral line wavenumbers, including especially noticeable effects: abnormally large values of the Λ-doubling of the vibro-rotational levels in the Π and Δ electronic states (up to 15 cm$^{-1}$ [35bDIE, 72CRO, 85FRE/SCI, 90POZ, 12LAV]), and a disappearance of the part of emission lines coming from vibro-rotational levels lying above the dissociation limit of a perturbing electronic state due to the predissociation;

(ii)       The g-factors up to two orders of magnitude [04bAST, 06aAST/LAV, 06AST, 07AST];

(iii)       Relative transition probabilities (the branching ratios) for spontaneous emission (by a factor of 2÷10 [83KOV/LAV, 84LAV/TYU, 88LAV/TYU, 93AST/LAV, 94bAST/LAV, 12LAV] or even above 30 [99AST/KAL, 07cAST/LAV] (see also our review paper [99LAV/AST]);

(iv)       The mean lifetimes including significant differences between the lifetimes of vibro-rotational levels of the Λ$^+$ and Λ$^-$ electronic states (see e.g. Fig. 1) and the sharp decrease of the experimental lifetimes for the $d^3\Pi_u^-$,$v$,N rovibronic levels lying above dissociation limit of the c$^3\Pi_u^-$ state of H$_2$ [88LAV, 90BUR/LAV, 99KIO/SAT]).

---

[3] In the case of heavy diatomic molecules, it is possible to measure total intensities of separate branches or even of separate bands. That is realizable by spectrometers with much smaller resolving power and sensitivity.

Therefore, an adequate description of the energy, radiative and magnetic characteristics of the rovibronic levels in excited electronic states of hydrogen isotopologues often requires the non-adiabatic models [63KOL/WOL, 83KOV/LAV, 90LAV/UST, 90bQUA/DRE, 94aAST/LAV, 00ADA/PAZ, 04aAST]. This peculiarity of intramolecular dynamics complicates non-empirical calculations and semi-empirical determination of the lifetimes for singular rovibronic levels.

For marking the electronic states of the hydrogen molecule, various authors used mainly a notation system introduced by G.H. Dieke [58DIE, 72CRO] and/or that being employed in famous reference book [79HUB/HER]. Nowadays Herzberg's notation system [79HUB/HER] is generally accepted and therefore used in the present paper. At the same time, experimentalists who study the lifetimes by means of measuring the emission intensities of separate rovibronic lines have to use Dieke's notation employed in the wavelength tables [72CRO] and [85FRE/SCI] for an identification of the lines in the spectra under the study. Therefore, a correspondence between two systems is presented in Table 1 for the states with the electron configuration $(1s\sigma nl\lambda)$ ($n = 1 \div 4$).

The $EF^1\Sigma_g^+$, $GK^1\Sigma_g^+$ and $H\bar{H}\,^1\Sigma_g^+$ electronic states we denoted according to [90bQUA/DRE]. They represent the superpositions of the $1s\sigma nl\lambda$ and $(2p\sigma)^2$ adiabatic electron configurations. For the $EF^1\Sigma_g^+$, $GK^1\Sigma_g^+$ and и $H\bar{H}\,^1\Sigma_g^+$ states the orbitals of an outer electron are $2s\sigma$, $3d\sigma$ and $3s\sigma$ respectively [06ROS/YOS]. Frequently used short notations $(E^1\Sigma_g^+$ and $F^1\Sigma_g^+)$ for the state $EF^1\Sigma_g^+$, $(G^1\Sigma_g^+$ and $K^1\Sigma_g^+)$ for $GK^1\Sigma_g^+$, and $(H^1\Sigma_g^+$ and $\bar{H}\,^1\Sigma_g^+)$ for $H\bar{H}\,^1\Sigma_g^+$ refer to the parts of these states characterized by the potential wells located at smaller $(E^1\Sigma_g^+$, $G^1\Sigma_g^+$ and $H^1\Sigma_g^+)$ (inner well) and at greater $(F^1\Sigma_g^+$, $K^1\Sigma_g^+$ and $\bar{H}\,^1\Sigma_g^+)$ (outer well) internuclear distances. The $B''\bar{B}\,^1\Sigma_g^+$ state is one more such double-well state which is the superposition of the $1s\sigma3p\sigma$ and $(2p\sigma)^2$ adiabatic states having inner well $(B''^1\Sigma_g^+$ state) and outer well $(\bar{B}\,^1\Sigma_g^+$ state) [01LAN/HOG]. Sometimes they were considered as distinct electronic states (see Table 2).

## 3. Current situation in studies of the lifetimes

The studies of mean lifetimes of excited electronic-vibro-rotational states of hydrogen molecules were previously reviewed in [80KUZ/KUZ, 96AST/KÄN, 02aAST/LAV]. Published more than thirty years ago, the book [80KUZ/KUZ] contains only experimental lifetimes and for five electronic states of $H_2$ only. Our compilation [96AST/KÄN] included results of semi-empirical determination, but it was also very selective being focused on seven electronic states of $H_2$ only. Our first attempt to overview studies of the lifetimes of excited rovibronic states of hydrogen isotopologues appeared relatively recently [02aAST/LAV], but we were not able to present all numerical data even for experimental results because of the journal limitations for the allowed volume of publications.

The main problems of the lifetime studies for separate rovibronic levels of the diatomic hydrogen isotopologues are connected with the peculiarities of this molecule described above. They are:

a)        in experimental studies it is necessary to provide selective recording of the decay curves for individual electronic-vibro-rotational levels under low enough particle density of molecules, what requires high resolving power and high sensitivity of an experimental setup;

b)        in semi-empiric determining and non-empirical calculating the lifetime values one have to take into account both adiabatic and non-adiabatic effects.

Current progress in experimental studies is connected with serious improvements in experimental technique caused by an accessibility of tunable lasers, new photo detectors and fast electronics. Significant improvements in non-empirical methods have been achieved by more precise calculation of the adiabatic electronic wave functions [95WOL], [99WOL/STA], [03WOL/STA] and by taking into account non-adiabatic effects [84aGLA/QUA], [87ROS/JUN], [90bQUA/DRE], [94ROS/JUN], [98MAT/JUN], [00ABG/ROU], [00MAT/JUN], [01 ROS/JUN], [03KIY/SAT]. As a result, in the last two decades there was a considerable increase in the number of publications devoted to studies of the lifetimes for separate rovibronic levels of the molecular hydrogen isotopologues as well as in the volume of reported data. This is illustrated by corresponding numerical data presented in Fig. 1. One may see that the volume of the data is increasing even at higher rate, especially for HD and $D_2$ isotopologues.

To date experimental data on the lifetimes were reported only for the $H_2$, HD and $D_2$ molecules. Chronological list of publications containing results of experimental determining the lifetimes of vibro-rotational levels for excited electronic states of the $H_2$, HD and $D_2$ molecules is presented in Table 2. The lifetimes reported in 61 papers were obtained by following experimental techniques: delayed coincidence method (35 papers), time-of-flight spectroscopy (8 papers), Hanle effect (8 papers), MOMRIE (4 papers), phase shift measurements (3 papers) and afterglow in pulsed beam plasma (3 papers). The rovibronic levels were excited by various methods, namely: the electron impact excitation (61% of the total number of papers), laser (29%), synchrotron radiation (7%) and charge transfer (by $H_2^+$ ions) (3%). Various combinations of these methods (e.g., electron impact and photo excitation) were employed as well. In the part of works the decay curves were measured under various gas pressures, and the lifetime values (together with the rate coefficients for collisional quenching) were obtained by an extrapolating measured effective lifetimes to the zero pressure. For most of the rovibronic states the relative experimental errors $\delta$ are within 5÷10%. Some data are reported as being amazingly precise $\delta$=0.2÷0.6% [04YOS/OGI, 05AIT/YOS, 06ROS/YOS, 11ROS/TSU]. The values as high as $\delta$≈30÷50% may be found in earlier works [62LIC, 66HES/DRE, 68HES, 80BRY/KOT, 84BRU/NEU, 94BER/OTT]. It should be stressed, that we calculated the $\delta$ values shown in Table 2 by using lifetime and error values reported by authors of original experimental works. Further comparative analysis of the lifetime values obtained for the same rovibronic levels in different experimental works have shown that those error bars are underestimated quite often.

Chronological list of publications containing results of semi-empirical determining the lifetimes of vibro-rotational levels for excited electronic states of the $H_2$, HD and $D_2$ molecules is presented in Table 3. One may see that semi-empirical values of the lifetimes were obtained for wide range of rotational (up to N=6) and vibrational (up to $v$=7) levels for 12 electronic states of $H_2$ and only for the $d^3\Pi_u^-$ state of the HD and $D_2$ molecules. A specific advantage of the semi-empirical approach [83LAV/TYU] is that the data for a large number of rovibronic levels and transitions may be obtained from restricted amount of experimental data on the branching ratios and the lifetimes of separate rovibronic levels. Completeness of the obtained data sets is very important for their application in spectroscopic diagnostics and modeling of non-equilibrium plasmas. Moreover, an opportunity of transferring errors in the *inverse* and *direct* problems makes it possible to estimate error bars of semi-empirical values and to study limits of an applicability of various models. An appearance of the error bars should be considered as quite natural because in principle semi-empirical methods may be considered as a part of an experiment – data processing in the framework of certain theoretical model.

Chronological list of a bibliography concerning non-empirical calculating lifetimes of excited electronic-vibro-rotational levels for various isotopologues of hydrogen molecule is given in Table 4. To date the non-empirical lifetime values were obtained in the framework of adiabatic (23 papers) and non-adiabatic (20 papers) models. In both cases the electronic wave functions are calculated by variational *ab initio* approach [39JAM/COO, 60KOL/ROO, 63KOL/WOL, 95WOL, 06SIM/HAG] or by means of more specific methods, such as the quantum defect theory (QDT) [99KIO/SAT, 00ADA/PAZ]) and the multichannel QDT method [94ROS/JUN, 01ROS/JUN, 02PAZ/PUP]. Last two are based on the assumption of a quasi-Coulomb interaction between an excited electron and the positively charged molecular core [83SEA]. Therefore, the QDT methods are applicable primarily to highly excited (Rydberg) states.

Comprehensive and complete list of bibliography concerning experimental, semi-empiric and non-empirical data on the lifetimes for all six isotopologues of hydrogen molecule is presented in Table 5. For convenience, the table is organized according to the growth of excitation energy of singlet and triplet electronic states. One may see that lifetimes for 20 singlet and 13 triplet electronic states of various isotopologues were studied by different methods. An attention of authors to various electronic states is not homogeneous. For example, the lifetime studies of the $GK^1\Sigma_g^+$ electronic state were reported in 23 papers (52% of publications devoted to the studies of the singlet states). Such an interest could be motivated by sophisticated molecular dynamics responsible for formation of this electronic state. Among the triplet electronic states the most attractive for researchers was the $d^3\Pi_u^-$ state (16 papers representing the 27% of publications devoted to the studies of the triplet states). This is an upper state of the Fulcher-$\alpha$ band system (the $d^3\Pi_u^- \rightarrow a^3\Sigma_g^+$ electronic transition), which is widely used in spectroscopic diagnostics of hydrogen-containing non-equilibrium plasma (see bibl. in [84LAV, 96AST/KÄN, 01RÖP/DAV, 13NPCP]).

Recently, the results of adiabatic calculations of the lifetimes were reported in [06FAN/WÜN]. The data were obtained for variety of the excited electronic states (with the united atom principal quantum numbers n = 2÷4) for all known isotopologues of the hydrogen molecule ($H_2$, $D_2$, $T_2$, HD, HT, DT). The authors used latest data for adiabatic potential curves and the electronic transition dipole moments obtained in *ab initio* calculations [99WOL/STA, 03WOL/STA]. Vibrational wave functions were calculated by numerical solution of the radial Schrödinger equation. Radiative lifetimes of the vibrational levels in the electronically excited states are obtained by the summation of the transition probabilities over all optically allowed transitions to lower-lying states. An existence of the rotational degree of freedom and non-adiabatic effects were not taken into account, in spite of the well-known importance of both things in the case of isotopic species of diatomic hydrogen. The authors motivated the goal of such very simplified calculations by a necessity to get complete database for spectroscopic studies and computer modeling of non-equilibrium plasma used in the control fusion experiments. From our point of view, the data reported in [06FAN/WÜN] may be used only for very rough estimations in kinetics of collisional and radiative processes in plasma. Their application in optical spectroscopy may be dangerous because:

(i) the radiative lifetimes were calculated for electronic-vibrational states of a rotationless molecule, but as it was already mentioned above, in the case of hydrogen molecules experimentalists have to work with individual electronic-vibro-rotational lines and levels only;

(ii) many vibronic levels appeared in the tables of [06FAN/WÜN] are not observable in emission being unstable due to predissociation (for example, Ref. [06FAN/WÜN] reports lifetime values for the $v$=0÷19 levels of the $c^3\Pi_u$ and $d^3\Pi_u$ states of $H_2$, whereas only the $c^3\Pi_u^-$,$v$=0÷14 and $d^3\Pi_u^-$,$v$=0÷9 levels are the bond states [01ROS/JUN]);

(iii) the lifetime values reported in [06FAN/WÜN] are often in dramatic contradiction with experimental data and results of non-adiabatic calculations for even relatively low lying electronic states of the hydrogen molecule;

(iv) Kronig's "+/-" symmetry of electronic wave functions was ignored in [06FAN/WÜN], and the $\Pi$ and $\Delta$ states were considered as doubly degenerated singular electronic states. In the light molecules (hydrogen and hydrides), two different sets of vibro-rotational levels appeared due to the $\Lambda$-doubling effect (when $\Lambda \neq 0$) have to be considered as belonging to two different electronic states − $\Lambda^+$ and $\Lambda^-$, because they have different characteristics and different channels of an excitation and a deactivation (see Fig. 1 and comments in [80LAV]). This is valid not only for molecules in ionized gases and plasmas but for isolated molecules as well. In particular, they may have essentially different lifetime values. As an example, experimental and calculated lifetime values for various rotational levels of the $c^3\Pi_u^+$,$v$=0 and $c^3\Pi_u^-$,$v$=0 electronic-vibrational states of the $H_2$ molecule are shown in Fig. 2. The calculations reported in [79CHO/BHA, 85COM/BRU, 86FLE/CHO, 93BHA/BHA] were performed in the framework of non-adiabatic models. One may see that they are in good agreement with experimental data obtained in [84BRU/NEU, 94BER/OTT]. In contrast the authors of [06FAN/WÜN] ignoring existence of two different real vibrational states ($c^3\Pi_u^+$,$v$=0 and $c^3\Pi_u^-$,$v$=0) reported one common lifetime value $\tau$=∞ for the $c^3\Pi_u$,$v$=0 state. This "value" is obviously in direct contradiction with experimental data and results of non-adiabatic calculations.

Another contradiction may be seen from Fig. 3, which represents measured and calculated values of lowest rotational levels ($N$=0) for various $B^1\Sigma_u^+$,$v$ electronic-vibrational states of the $H_2$ molecule. The experimental data were obtained in [66HES/DRE], [72SMI/CHE], [84SCH/IMS]. The calculations were performed in the framework of adiabatic [06FAN/WÜN] and non-adiabatic [00ABG/ROU] models. One may see that the adiabatic data of [06FAN/WÜN] agree with both experimental data and results of the non-adiabatic calculation for $v \leq 5$ only. The discrepancy between adiabatic and experimental data is increasing with growth of vibrational quantum number reaching 4 times for $v$=17. It should be noted that the (1s$\sigma$2p$\sigma$) $B^1\Sigma_u^+$ state is the lowest excited bond electronic state of hydrogen molecule. For higher lying electronic states non-adiabatic effects are often more strong. For example, the discrepancy between results of adiabatic [06FAN/WÜN] and non-adiabatic [06ROS/YOS] calculations reaches from 15 up to $2\times10^4$ times for the rovibronic levels with $v$=12÷16.

Despite a rather large number of experimental studies devoted to determination of the lifetimes, the data obtained so far exhibit a rather fragmentary character. The lifetimes were measured only for a very restricted number of rotational levels (in many cases, only for one lowest rotational level) and one (or very few) vibrational levels of the electronic state under the study. The results of various experimentalists often refer to different rovibronic levels, which hinders a direct comparison of the data obtained by different methods and/or by different authors.

One more remark should be made at the end of our brief review of current situation in studies of lifetimes for rovibronic levels of hydrogen molecules. This data are sometimes less sensitive to the perturbations than the differences between the rovibronic term values and the radiative transition probabilities, because they are mainly determined by the $r$-dependencies of the electronic transition moment and relative position of adiabatic potential curves [06aAST/LAV]. But quite often non-adiabatic effects are easily observable in the lifetimes of separate rovibronic levels (see e.g. Fig. 2 and 3). Thus, the experimental values of the lifetimes should be considered as an important channel of information about intramolecular dynamics including both adiabatic and non-adiabatic characteristics of the molecule under the study.

## 4. Collecting and analysis of the data

All reported in 61 publications experimental lifetime values for $H_2$, HD and $D_2$ are given in Tables 6÷8 correspondingly. The tables contain 792 data for 618 different vibro-rotational levels belonging to the 33 different electronic states corresponding to the united atom principle quantum numbers $n$=2÷4. The data were obtained by means of the UV, visible and near IR spectroscopy. An attention of authors to various isotopologues is not homogeneous. The results concerning to the $H_2$, HD and $D_2$ molecules are reported in 54, 5 and 12 publications. The lifetime values were obtained for 317 vibro-rotational levels of 32 electronic states of $H_2$, 96 levels of 9 electronic states of HD and for 205 levels of 16 electronic states of $D_2$. The distribution of studied vibrational and rotational levels over various electronic states is too nonhomogeneous for understanding dependencies of the lifetimes from vibrational and rotational quantum numbers. Most often only lowest rotational levels were studied. Non-empirical data are often reported for significantly greater range of the rotational quantum numbers $N$ (e.g. up to $N$=30 for vibrational levels of the $B^1\Sigma_u^+$ state of $H_2$ [98PAR]). Systematic experimental and non-empirical studies of lifetimes for the fine structure sublevels of triplet electronic states were not reported so far. The exclusions are experimental data for the fine structure sublevels of the $c^3\Pi_u^-$, $v$=0÷3, $N$=1÷4 [94BER/OTT] and the $i^3\Pi_g^-$, $v$=0, $N$=3 rovibronic states [81EYL/PIP] of the $H_2$ molecule.

One of the goals of the present work was to examine an opportunity of selecting the most reliable lifetime data, which could be recommended for use in various spectroscopic applications. For that purpose, we performed a comparative analysis of all available data, obtained by different methods and authors. The available data are too fragmentary to make such analyzing easy and dependable. The fragmentariness is characterized by lack of experimental data for: 1) highly-excited electronic states corresponding to $n$>4; 2) even for relatively low vibrational and rotational levels of the vast majority of electronic states; 3) the tritium-containing isotopologues. Moreover, experimental data reported in at least two different publications are available only for 110 from 317 studied vibro-rotational levels of $H_2$ and for 25 (from 205) of the $D_2$ molecule. For the HD isotopologue, the experimental data were obtained only in one work and a comparison is not possible at all. Thus, an application of statistical methods for extracting most reliable data is now impossible for the vast majority (almost 80%) of rovibronic states of the hydrogen isotopologues. When we were able to find or to select only one dependable datum for one rovibronic level we decided to refrain from making any comments or recommendations. To our mind, such cases should be verified somehow. In the cases of multiple data for the same rovibronic levels, for recognizing more dependable data we used our own expert opinion based on taking into account the information contained in original papers about the experimental technique, the range of experimental conditions, the spectral resolution, a way of primary data processing and others. When multiple dependable lifetime values are available for the same rovibronic level, these data are italicized in Tables 6 and 8 and averaged. The average values thus obtained are printed in bold face, and marked as PW (present work) in the "ref." column and as "Recom" in the "Method" column. From our point of view they may be recommended for use in quantitative spectroscopy and modeling of collisional and radiative processes in ionized gases and plasmas.

Another opportunity to test the reliability of the data is the comparison of experimental data with the results of non-empirical calculations. Here, again, we must pay attention that these calculations were also performed for a limited number of electronic states and of vibro-rotational levels within these states. Currently the results of non-empirical calculations are not reported for about 36% of rovibronic levels studied experimentally. Therefore, such a comparison is possible only for 241 from 317 experimentally studied rovibronic levels of $H_2$, for 10 (from 96) levels of HD and for 143 (from 205) experimentally studied rovibronic levels of $D_2$. All those cases may be divided into four classes (A, B, C and D) according to the criteria:

$$\left| \tau_{expt} - \tau_{calc} \right| \leq K \times \sigma,$$

where $\tau_{expt}$ and $\tau_{calc}$ are measured and non-empirical lifetime values, and $K$=1, 2 and 3. Cases of agreement between experimental and non-empirical data within $1\sigma$ were included into the A-class, within $1\sigma \div 2\sigma$ range – into the B-class, and within the $2\sigma \div 3\sigma$ range – in C-class. The cases in which the deviations $|\tau_{expt} - \tau_{calc}|$ exceeded $3\sigma$ were included into the D-class. According to Shovene's criterion, those cases have to be considered as the direct contradiction between experiments and non-empirical calculations. The belonging to one of those classes is shown in the column O-C (Observed minus Calculated) of the Tables 6÷8. The distributions over the classes A, B, C and D for various isotopologues ($H_2$, HD, $D_2$) are presented in Fig. 4 together with the sum of them. One may see that a coincidence of measured and calculated data within experimental uncertainty (class A) takes place for only 22% of levels studied by both methods. The direct contradiction (class D) is observed in more than half (55%) of the cases when such a comparison can be conducted. The contradictions between the experimental data and results of the most reliable calculations take place for 22 from 31 investigated electronic states of the $H_2$ molecule, for 12 from 17 electronic states of $D_2$, and for 5 from 9 electronic states of the HD molecules (see Tables 6÷8). The reasons of so poor agreement between theory and experiment are certainly unclear. It could be due to unaccounted experimental errors and/or due to insufficient accuracy of non-empirical calculations. In any case, it is hard to say that such a situation in studies of the simplest molecule is normal. Therefore, further experimental studies and non-empirical calculations should be focused on the cancellation of the contradiction observed in the present work.

# 5. Conclusion

Complete bibliography and numerical data collected, analyzed and listed in the present work are mainly destined to experimenters and theorists working in great variety of basic and applied researches of hydrogen-containing media by means of quantitative spectroscopy and modeling of collision-radiative processes, especially in astrophysics, non-equilibrium gas dynamics and plasma physics. The results of our compilation may be also used by experimentalists engaged in studies of the lifetimes and the collision-induced transitions for their own analysis of the situation concerning various excited electronic states of hydrogen isotopologues. For theoreticians dealing with non-empirical calculations of diatomic molecules it should be convenient and useful to see all available experimental data together. Moreover, the authors hope that the data collected in tables of the present paper will stimulate further evolution of the lifetime studies for electronic-vibro-rotational levels of isotopologues of diatomic hydrogen - the simplest and most distributed molecule in the Universe, Science and Technology.

# Acknowledgement

One of the authors (B.P Lavrov) sincerely appreciates financial support from Russian Foundation for Basic Research (project number 13-03-00786a).

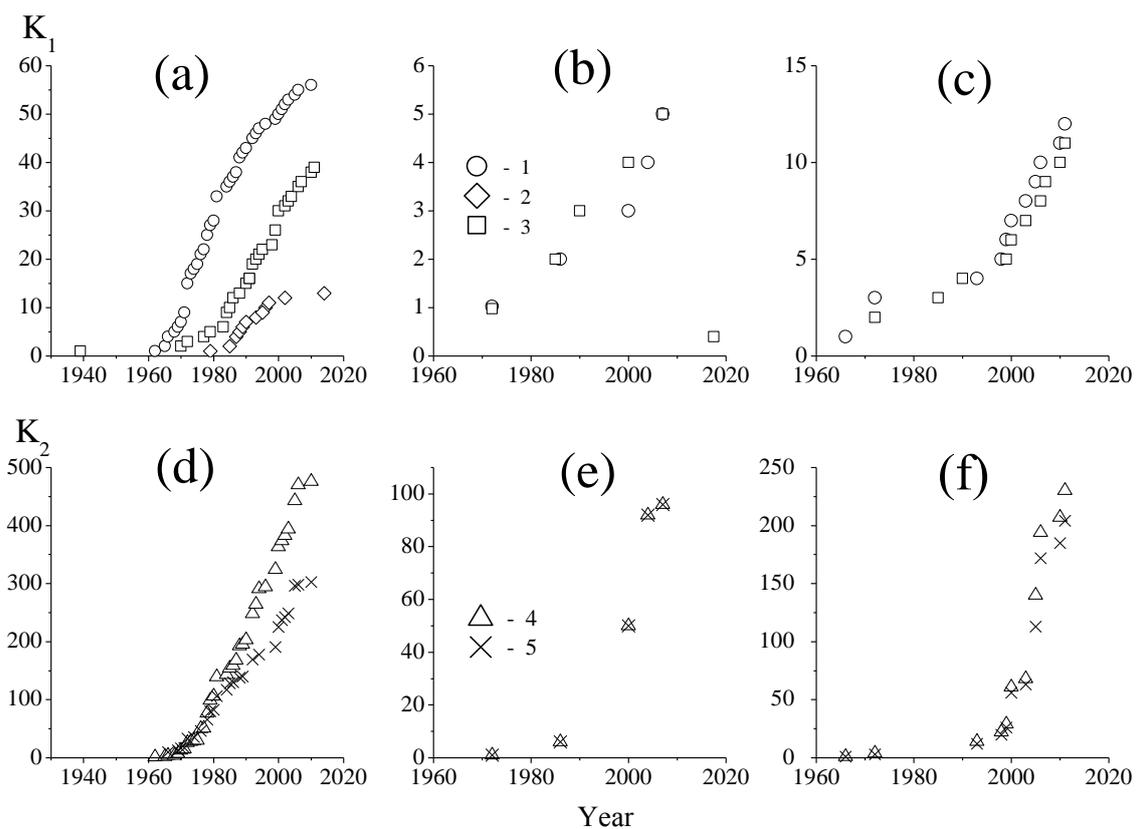

Fig. 1. Temporal dynamics for the growth of interest and results obtained by experimental (1), semi-empiric (2), and non-empirical (3) determination of the lifetimes for rovibronic levels of the $H_2$ (a, d), HD (b, e), and $D_2$ (c, f) molecules from 1939 up to November 2014. $K_1$ is the total number of publications; $K_2$ − total number of reported experimental data (4) and of investigated rovibronic levels (5).

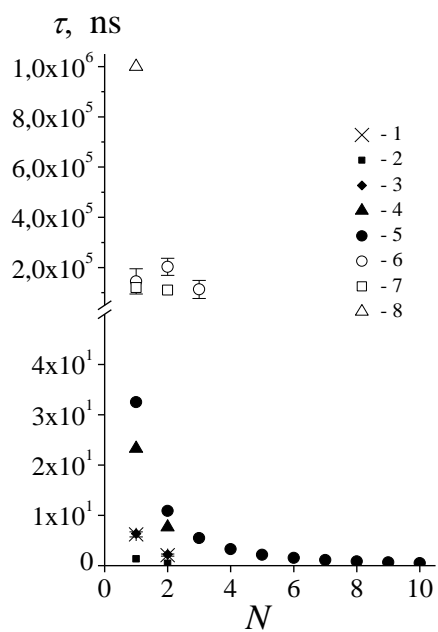

Fig. 2. Experimental (points 1,6) and calculated (points 2-5, 7-8) values of the lifetimes for various rotational levels (with quantum numbers $N$) for the lowest vibrational levels ($v$=0) of the $c^3\Pi_u^+$ (1-5) and $c^3\Pi_u^-$ (6-8) electronic states of the $H_2$ molecule. The data 1, (2, 7), 3, (4, 8), 5 and 6 are reported in [84BRU/NEU], [79CHO/BHA], [85COM/BRU], [86FLE/CHO], [93BHA/BHA] and [94BER/OTT] correspondingly.

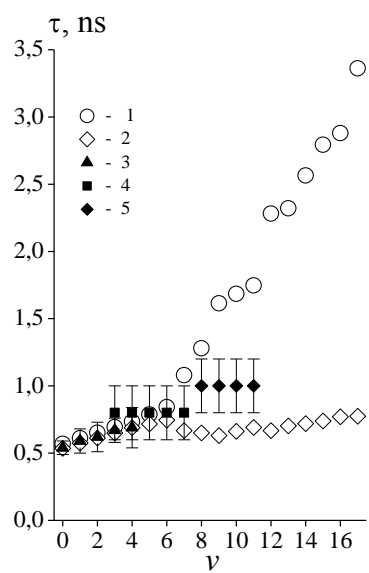

Fig. 3. Experimental and calculated lifetime values for lowest rotational levels ($N$=0) of various $B^1\Sigma_u^+,v$ electronic-vibrational states of the $H_2$ molecule. The calculations 1 and 2 were performed in the framework of adiabatic [06FAN/WÜN] and non-adiabatic [00ABG/ROU] models. The experimental data 3, 4 and 5 were reported in [84SCH/IMS], [66HES/DRE] and [72SMI/CHE].

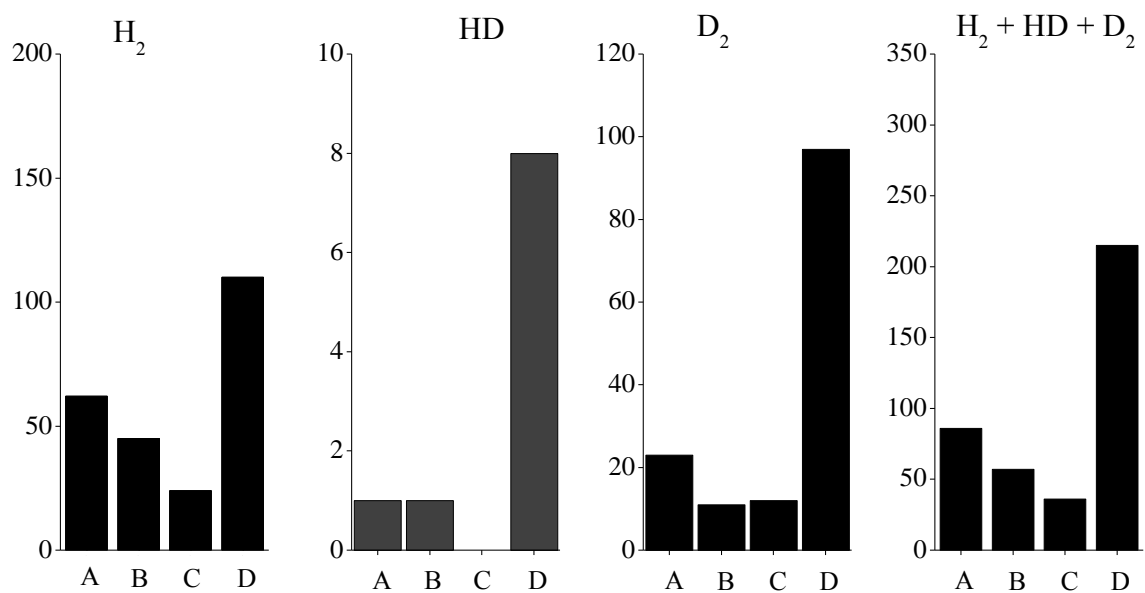

Fig. 4. Distributions of the quantity of rovibronic levels, studied by both experimental and non-empirical methods, over the classes A, B, C and D according to an agreement of experimental and most reliable non-empirical values of the lifetimes for various isotopologues ($H_2$, HD, $D_2$) and the sum of this distributions.

Table 1. The list of various notations used for designation of electronic states of hydrogen molecule corresponding to the $1s\sigma nl\lambda$ electron configurations for $n=1\div4$ in the united atom limiting case.

| State of outer electron $nl\lambda$ | Singlet electronic states | | Triplet electronic states | |
|---|---|---|---|---|
| | G.H.Dieke [58DIE], [72CRO] | G.Herzberg [79HUB/HER] | G.H.Dieke [58DIE], [72CRO] | G.Herzberg [79HUB/HER] |
| $1s\sigma$ | 1A | $X^1\Sigma_g^+$ | — | — |
| $2s\sigma$ | 2A | $E^1\Sigma_g^+$ | 2a | $a^3\Sigma_g^+$ |
| $2p\sigma$ | 2B | $B^1\Sigma_u^+$ | 2b | $b^3\Sigma_u^+$ |
| $2p\pi$ | 2C | $C^1\Pi_u$ | 2c | $c^3\Pi_u$ |
| $3s\sigma$ | 3A | $H^1\Sigma_g^+$ | 3a | $h^3\Sigma_g^+$ |
| $3p\sigma$ | 3B | $B'^1\Sigma_u^+$ | 3b | $e^3\Sigma_u^+$ |
| $3p\pi$ | 3C | $D^1\Pi_u$ | 3c | $d^3\Pi_u$ |
| $3d\sigma$ | 3D | $G^1\Sigma_g^+$ | 3d | $g^3\Sigma_g^+$ |
| $3d\pi$ | 3E | $I^1\Pi_g$ | 3e | $i^3\Pi_g$ |
| $3d\delta$ | 3F | $J^1\Delta_g$ | 3f | $j^3\Delta_g$ |
| $4s\sigma$ | 4A | $O^1\Sigma_g^+$ | 4a | $o^3\Sigma_g^+$ |
| $4p\sigma$ | 4B | $B''^1\Sigma_u^+$ | 4b | $f^3\Sigma_u^+$ |
| $4p\pi$ | 4C | $D'^1\Pi_u$ | 4c | $k^3\Pi_u$ |
| $4d\sigma$ | 4D | $P^1\Sigma_g^+$ | 4d | $p^3\Sigma_g^+$ |
| $4d\pi$ | 4E | $R^1\Pi_g$ | 4e | $r^3\Pi_g$ |
| $4d\delta$ | 4F | $S^1\Delta_g$ | 4f | $s^3\Delta_g$ |

Table 2. Chronological list of publications reported experimental data for the lifetimes of excited electronic-vibro-rotational levels of the $H_2$, HD and $D_2$ molecules.

| | | | | | | | |
|---|---|---|---|---|---|---|---|
| | | | | The $H_2$ molecule | | | |
| Ref. | Electronic state under the study | $v$ | $N$ | Method | Exci-tation | $P$, $10^{-3}$ Torr | $\delta$ (%) |
| 62LIC | $c^3\Pi_u^-$ | 0 | 2 | TF | ei | | 50 |
| 65FOW/HOL, 66FOW/HOL | $a^3\Sigma_g^+$ | a | | PLASMA | ei | 20÷40000[d] | 23 |
| 66HES/DRE | $B^1\Sigma_u^+$; $C^1\Pi_u^+$ | 3÷7[b]; 0÷3[b] | c | PS | sr | | 25÷33 |
| 68HES | $B^1\Sigma_u^+$; $C^1\Pi_u^+$ | 3÷7[b]; 0÷3[b] | c | PS | sr | | 25÷33 |
| 69CAH | $d^3\Pi_u^-$ | 0÷2 | 1 | DC | ei | | 7÷9 |
| 70LIN | $G^1\Sigma_g^+$ | 0 | 1÷3 | Hanle | ei | | 5 |
| 71LIN | $G^1\Sigma_g^+$ | 0 | 1÷3 | Hanle | ei | | 5÷6 |
| 71IMH/REA | $a^3\Sigma_g^+$ | 0,1 | c | DC | ei | 2÷7 | 4÷6 |
| 72LIN/DAL | $G^1\Sigma_g^+$, $I^1\Pi_g^-$, $I^1\Pi_g^+$, $K^1\Sigma_g^+$ | 0÷2 | 1÷3 | Hanle | ei | 15÷90[d] | 4÷7 |
| 72THO/FOW | $a^3\Sigma_g^+$ | 0 | c | DC | ei | 34÷800[d] | 8 |
| 72SMI/CHE | $a^3\Sigma_g^+$; $B^1\Sigma_u^+$ | 0÷3; 8÷11 | c | PS | ei | 0.75÷1.5 | 10÷20 |
| 72MAR/JOS | $d^3\Pi_u^-$, $d^3\Pi_u^+$ | 0÷3[b] | 1,2 | Hanle | ei | 10÷1000[d] | 10 |
| 72JOH | $c^3\Pi_u^-$ | 0 | 1÷2 | TF | ei | | 5 |
| 72FRE/MIL | $G^1\Sigma_g^+$ | 0 | 1 | MOMRIE | ei | 0.2÷1 | 19 |
| 73aFRE/MIL | $d^3\Pi_u^-$ | 0÷3[b] | 1 | MOMRIE | ei | 10 | 16 |
| 73bFRE/MIL | $d^3\Pi_u^+$ | 0÷3[b] | 1 | MOMRIE | ei | 10 | 11 |
| 74MIL/FRE | $k^3\Pi_u^-$ | 1÷3[b] | 1 | MOMRIE | ei | 4 | 16 |
| 75KIN/REA | $a^3\Sigma_g^+$ | a | | DC | ei | 0.5 | 2 |
| 76MEL/LOM | $G^1\Sigma_g^+$ | 1 | 1 | Hanle | ei | 20 | 4 |
| 76GOM/CAM | $G^1\Sigma_g^+$, $I^1\Pi_g^-$, $I^1\Pi_g^+$ | 0÷3 | 1÷4 | DC | ei | 25÷50 | 7÷10 |
| 77VOG/MEI[a] | $c^3\Pi_u^+$ | 0 | c | TF | ch.tr. | | |
| 78KLI/RHO[a] | $EF^1\Sigma_g^+$ | 6 | 0÷2 | DC | laser | 10÷300[d] | 20 |
| 78CHI/DAL | $GK^1\Sigma_g^+$, $I^1\Pi_g^-$, $I^1\Pi_g^+$, $EF^1\Sigma_g^+$ | 0÷3; 26 | 1÷3 | Hanle | ei | 10÷100 | 6 |
| 78DAY/AND | $d^3\Pi_u^-$; $k^3\Pi_u^-$ | 0÷3; 0÷3 | 1÷3[c]; 1÷2 | DC | ei | 3÷50 | 6 |

| | | | | The $H_2$ molecule | | | |
|---|---|---|---|---|---|---|---|
| Ref. | Electronic states under the study | $v$ | $N$ | Method | Exci-tation | $P$, $10^{-3}$ Torr | $\delta$, % |
| 79DAY/AND | $GK^1\Sigma_g^+$; | $0\div6$; | $1\div5$; | DC | ei | $10\div400^d$ | 8 |
| | $I^1\Pi_g^-$, $I^1\Pi_g^+$ | 0 | $1\div2$ | | | | |
| 79MOH/KIN | $a^3\Sigma_g^+$ | $0\div4$ | c | DC | ei | 0.3 | $2\div11$ |
| 80BRY/KOT | $I^1\Pi_g^-$, $I^1\Pi_g^+$ | 0 | $1\div6$ | Hanle | ei | $80\div600$ | $18\div38$ |
| 81BRE/GOD | $a^3\Sigma_g^+$, $d^3\Pi_u^-$ | 0 | c | DC | ei | $100\div1.6\times10^{4\,d}$ | 7 |
| 81BÖS/LIN[a] | $H^1\Sigma_g^+$, $G^1\Sigma_g^+$, $I^1\Pi_g$, $J^1\Delta_g$ | 0 | c | DC | ei | $1.5\div2.3$ | 6 |
| 81EYL/PIP | $h^3\Sigma_g^+$, $g^3\Sigma_g^+$, $i^3\Pi_g$, $j^3\Delta_g^+$ | $0\div1$ | 2 | DC | laser | | $2\div16$ |
| | $i^3\Pi_g^-$, $j^3\Delta_g^-$ | $0\div1$ | 1,3 | | | | |
| 81BOG/EFR | $d^3\Pi_u^-$ | 0 | $1,3^b$ | DC | ei | 20 | 14 |
| 81MON | $j^3\Delta_g^+$, $j^3\Delta_g^-$ | $1\div3$ | 2,3 | Hanle | ei | | $2\div11$ |
| 84BRU/NEU | $c^3\Pi_u^+$ | $0\div15$ | 1,2 | TF | ch.tr. | | $10\div50$ |
| 84SCH/IMS | $B^1\Sigma_g^+$ | $0\div4$ | $1,2^b$ | DC | sr | $8\div1.5\times10^5$ | $9\div22$ |
| 85GLA/BRE | $D^1\Pi_u^-$ | $3\div15$ | 1 | DC | sr | $<30$ | $7\div22$ |
| 86CHA/THO | $EF^1\Sigma_g^+$ | 0, 3, 6 | 0,1 | TF | laser | | $1.5\div8$ |
| 87KIY | $d^3\Pi_u^-$, $d^3\Pi_u^+$ | $0\div3$ | 1,2 | DC | ei | 1.5 | $4\div8$ |
| 88LAV | $d^3\Pi_u^-$ | $4\div6$ | 1,3 | PLASMA | ei | | 13 |
| 88SAN/CAM | $EF^1\Sigma_g^+$, | 22,23,26; | $1\div5$; | DC | ei | $5\div100$ | $5\div15$ |
| | $GK^1\Sigma_g^+$ | 1,3,4; | $1\div5$; | | | | |
| | $I^1\Pi_g^+$; | $0\div3$; | $1\div4$; | | | | |
| | $I^1\Pi_g^-$ | 1; | 4; | | | | |
| | $J^1\Delta_g^-$, $P^1\Sigma_g^+$, $j^3\Delta_g^-$, | 0; | $1\div5$; | | | | |
| | $d^3\Pi_u^-$, | $0\div3$; | 1; | | | | |
| | $u$ | 1 | $3,5^b$ | | | | |
| 88WED/PHE | $a^3\Sigma_g^+$ | $0\div1$ | 1 | DC | laser | $300\div1000^d$ | 3 |
| 89KOO/ZAN | $i^3\Pi_g^-$, $j^3\Delta_g^+$, $j^3\Delta_g^-$ | $0\div2$ | $1\div3$ | TF | laser | | $7\div10$ |
| 90BUR/LAV | $d^3\Pi_u^-$ | $0\div6$ | 1, 3 | DC | ei | $1\div200^d$ | $2\div19$ |
| 92TSU/ISH | $EF^1\Sigma_g^+$; | $19\div21$; | $1\div4$; | DC | laser | 0.03 | $1.1\div10$ |
| | $GK^1\Sigma_g^+$; | 0,2; | $1\div4$; | | | | |
| | $H^1\Sigma_g^+$; $I^1\Pi_g^+$, $I^1\Pi_g^-$, | 0; | $1\div4$; | | | | |
| | $J^1\Delta_g^-$, $J^1\Delta_g^+$ | 0 | $1\div4$; | | | | |
| 92TSU/SHI | $EF^1\Sigma_g^+$; | 32; | $0\div3$; | DC | laser | 0.03 | $6\div8$ |
| | $GK^1\Sigma_g^+$ | 8 | $0\div1$ | | | | |
| 93KIY/SAT | $d^3\Pi_u^-$, | $0\div3$; | 1; | DC | ei | $1.5\div30$ | $2\div30$ |
| | $d^3\Pi_u^+$ | $0\div3$ | $1\div4$ | | | | |
| 94BER/OTT | $c^3\Pi_u^-$ | $0\div3$ | $1\div4$ | TF | laser | | $11\div35$ |
| | | | | | The $H_2$ molecule | | |
| Ref. | Electronic state under the study | $v$ | $N$ | Method | Exci-tation | $P$ ($10^{-3}$ Torr) | $\delta$ (%) |
| 96RAY/LAF | $g^3\Sigma_g^+$; | 1; | 2; | DC | laser | | 2 |
| | $i^3\Pi_g^-$; | 0; | 1; | | | | |
| | $j^3\Delta_g^-$ | 1 | 3 | | | | |
| 99KIY/SAT | $d^3\Pi_u^-$; | 4÷6; | 1÷3; | DC | ei | 9 | 4÷33; |
| | $e^3\Sigma_u^+$ | 0÷5 | 0÷3 | | | | 7÷47 |
| 00REI/HOG | $\bar{H}\,^1\Sigma_g^+$ | 6÷15 | 0÷5 | TF | laser | | 10÷20 |
| 01KIY/SAT | $B'^1\Sigma_u^+$; | 2; | 2; | DC | ei | 5÷12 | 2÷10 |
| | $D^1\Pi_u^-$; | 0; | 1; | | | | |
| | $D^1\Pi_u^+$; | 2; | 1; | | | | |
| | $EF^1\Sigma_g^+$ | 6,7 | 0÷4 | | | | |
| 02PAZ/PUP | $h^3\Sigma_g^+$; $g^3\Sigma_g^+$; $i^3\Pi_g^+$; | 0,1; | 1,2 | DC | ei | 7÷12 | 4÷18 |
| | $j^3\Delta_g^-$ | 0 | 2 | | | | |
| 03KIY/SAT | $k^3\Pi_u^-$ | 0÷6 | 1÷5 | DC | ei | 2÷9 | 1÷3 |
| 05AIT/OGI | $EF^1\Sigma_g^+$; | 29÷33; | 0÷5; | DC | laser | | 0.4÷7 |
| | $GK\,^1\Sigma_g^+$; | 6÷7; | 0÷5; | | | | |
| | $H^1\Sigma_g^+$; | 2; | 0÷5; | | | | |
| | $O^1\Sigma_g^+$; | 0; | 0,1,3; | | | | |
| | $P^1\Sigma_g^+$; | 0; | 0÷5; | | | | |
| | $I^1\Pi_g^+$; | 3; | 1÷4; | | | | |
| | $I^1\Pi_g^-$; | 3; | 1÷4; | | | | |
| | $R^1\Pi_g^+$; | 0; | 1÷3; | | | | |
| | $R^1\Pi_g^-$; | 0; | 1; | | | | |
| | $J^1\Delta_g^+$; | 2; | 5; | | | | |
| | $S^1\Delta_g^+$; | 0; | 4; | | | | |
| | $S^1\Delta_g^-$ | 0 | 3÷4 | | | | |
| 06ROS/YOS | $\bar{H}\,^1\Sigma_g^+$ | 6÷11 | 1÷3 | DC | laser | | 1÷3 |
| 10ROS/AND | $I^1\Pi_g^+$; | 4[e] | 1÷3 | DC | laser | | 9÷11 |
| | $I^1\Pi_g^-$ | 4[e] | 1÷2 | | | | 3÷6 |

| | | | The HD molecule | | | | |
|---|---|---|---|---|---|---|---|
| Ref. | Electronic state under the study | $v$ | $N$ | Method | Exci- tation | $P$ ($10^{-3}$ Torr) | $\delta$ (%) |
| 72JOH | $c^3\Pi_u^-$ | 0 | 1÷2 | TF | ei | | 5 |
| 86CHA/THO | $EF^1\Sigma_g^+$ | 0,3,6 | 0,1 | TF | laser | | 2÷6 |
| 00REI/HOG | $\bar{H}\,^1\Sigma_g^+$ | 16÷19; | 0÷3; | TF | laser | | 20 |
| | $\bar{B}\,^1\Sigma_g^+$ | 16÷22 | 0÷3 | | | | |
| 04YOS/OGI | $EF^1\Sigma_g^+$; | 31,33,35÷37; | 1÷4; | DC | laser | | 0.3÷17 |
| | $GK^1\Sigma_g^+$; | 6÷9; | 1÷4; | | | | |
| | $H^1\Sigma_g^+$; | 0÷2; | 1÷3; | | | | |
| | $I^1\Pi_g^+$, | 3,5; | 1,3; | | | | |
| | $I^1\Pi_g^-$; | 3; | 1÷3; | | | | |
| | $J^1\Delta_g^+$; | 3; | 2÷4; | | | | |
| | $J^1\Delta_g^-$ | 3 | 2 | | | | |
| 07ROS/FUJ | $I^1\Pi_g^-$ | 4 | 1÷4 | DC | laser | | 2÷3 |

Table 2 – Continued.

| | | | | The D$_2$ molecule | | | |
|---|---|---|---|---|---|---|---|
| Ref. | Electronic state under the study | $v$ | $N$ | Method | Excitation | $P$ (10$^{-3}$ Torr) | $\delta$ (%) |
| 66FOW/HOL | $a^3\Sigma_g^+$ | 0÷1[b] | [c] | PLASMA | ei | 40÷40000[d] | 23 |
| 72SMI/CHE | $a^3\Sigma_g^+$ | 0÷1 | [c] | PS | ei | 0.75÷1.5 | 10÷12 |
| 72JOH | $c^3\Pi_u^-$ | 0 | 1÷2 | TF | ei | | 5 |
| 93KIY/SAT | $d^3\Pi_u^-$; | 1÷2; | 1÷3; | DC | ei | 3 | 4÷5 |
| | $d^3\Pi_u^+$ | 1÷2 | 1÷5 | | | | |
| 98SUZ/NAK | $EF^1\Sigma_g^+$; | 28,32; | 2; | DC | laser | | 1÷10 |
| | $GK^1\Sigma_g^+$; | 2; | 0,2; | | | | |
| | $H^1\Sigma_g^+$ | 0,1 | 0÷2 | | | | |
| 99KIY/SAT | $d^3\Pi_u^-$; | 4÷6; | ≤5 | DC | ei | 9 | 4÷12 |
| | $e^3\Sigma_u^+$ | 3÷5 | | | | | |
| 00REI/HOG | $\bar{H}^1\Sigma_g^+$ | 15÷23 | 0÷5 | TF | laser | | 10÷20 |
| 03KIY/SAT | $k^3\Pi_u^-$ | 0÷6 | 2 | DC | ei | 2÷9 | 2÷4 |
| 05AIT/YOS | $EF^1\Sigma_g^+$; | 41; | 0÷5; | DC | laser | | 0.2÷14 |
| | $GK^1\Sigma_g^+$; | 8÷11; | 0÷5; | | | | |
| | $H^1\Sigma_g^+$; | 3; | 0÷4; | | | | |
| | $I^1\Pi_g^+$; | 4; | 1÷4; | | | | |
| | $I^1\Pi_g^-$; | 4÷5; | 1÷4; | | | | |
| | $P^1\Sigma_g^+$; | 0; | 1÷5; | | | | |
| | $R^1\Pi_g^+$; | 0; | 1÷2 | | | | |
| | $R^1\Pi_g^-$; | 0,1,4÷7; | 1÷4; | | | | |
| | $S^1\Delta_g^+$; | 0; | 2÷5; | | | | |
| | $S^1\Delta_g^-$ | 0 | 3 | | | | |
| 06ROS/YOS | $\bar{H}^1\Sigma_g^+$ | 9÷19 | 0÷4 | DC | laser | | 0.3÷9 |
| 10ROS/AND | $I^1\Pi_g^+$; | 5÷6[e] | 1÷4 | DC | laser | | 6÷14 |
| | $I^1\Pi_g^-$ | 5÷6[e] | 1÷3 | | | | 5÷12 |
| 11ROS/TSU | $EF^1\Sigma_g^+$; | 44÷46 | 0÷5 | DC | laser | | 0.6÷6 |
| | $GK^1\Sigma_g^+$ | 11 | 1÷5 | | | | 1÷5 |

[a] – the datum for the electronic state was obtained without taking into account an existence of vibrational and rotational structure of the levels;

[b] – in these works the spectrometer resolution was insufficient for correct separating rovibronic spectral lines;

[c] - the data were calculated for electronic-vibrational states without taking into account an existence of rotational structure of the levels;

[d] – these papers contain also the data about the rate coefficients for collisional quenching of the rovibronic states under the study;

[e] - the levels of outer potential well of the electronic state.

Table 3. Chronological list of publications containing results of semi-empirical determining the lifetimes of vibro-rotational levels for various excited electronic states of the $H_2$, HD, $D_2$ and $T_2$ molecules.

| Reference | Isotopic species | Electronic states | $v$ | $N$ | Theoretical model |
|---|---|---|---|---|---|
| 79GLA | $H_2$ | $D'\,^1\Pi_u^+$ | 1÷7 | 1÷2 | NA |
| 85DRA/LAV | $H_2$, HD, $D_2$ | $d^3\Pi_u^-$ | 0÷4 | 1 | AA |
| 87DRA/LAV | $H_2$, HD, $D_2$ | $d^3\Pi_u^-$ | 0÷4 | 1÷3 | AA |
| 87KIY | $H_2$ | $d^3\Pi_u^+$ | 0÷3 | 1÷3 | NA |
| | $H_2$ | $e^3\Sigma_u^+$ | 3÷6 | 1÷3 | NA |
| 88LAV/TYU | $H_2$ | $d^3\Pi_u^+$ | 0,2,3 | 1÷6 | NA |
| | $H_2$ | $e^3\Sigma_u^+$ | 1÷3 | 1÷6 | NA |
| 89KOO/ZAN | $H_2$ | $i^3\Pi_g^+$, $j^3\Delta_g^+$, $j^3\Delta_g^-$ | 0÷2 | 1÷3 | NA |
| 90POZ | $H_2$, HD, $D_2$ | $d^3\Pi_u^-$ | 0÷4 | 1 | AA |
| | $T_2$ | $d^3\Pi_u^-$ | 0÷6 | 1 | AA |
| | $H_2$, $D_2$ | $d^3\Pi_u^+$ | 0÷3 | 1÷8 | NA |
| | $H_2$, $D_2$ | $e^3\Sigma_u^+$ | 0÷4 | 1÷8 | NA |
| 93KIY/SAT | $H_2$ | $d^3\Pi_u^+$ | 1 | 1÷3 | NA |
| | $H_2$ | $e^3\Sigma_u^+$ | 4 | 1÷3 | NA |
| | $D_2$ | $d^3\Pi_u^+$ | 1 | 1÷5 | NA |
| | $D_2$ | $e^3\Sigma_u^+$ | 5 | 1÷5 | NA |
| 95bAST/LAV | $H_2$ | $h^3\Sigma_g^+$, $g^3\Sigma_g^+$, $i^3\Pi_g^+$ | 0÷3 | 1÷6 | NA |
| | $H_2$ | $i^3\Pi_g^-$, $j^3\Delta_g^+$, $j^3\Delta_g^-$ | 0÷3 | 1÷6 | NA |
| 96AST/KÄN | $H_2$ | $d^3\Pi_u^-$ | 0÷6 | 1÷3 | AA |
| | $H_2$ | $e^3\Sigma_u^+$ | 0÷1; | 1÷6 | NA |
| | $H_2$ | $d^3\Pi_u^+$; $i^3\Pi_g^-$; $I^1\Pi_g^-$ | 0÷3; | 1÷6 | NA |
| | $H_2$ | $J^1\Delta_g^+$, $j^1\Delta_g^-$ | 0÷3 | 1÷6 | NA |
| 97AST/KOK | $H_2$ | $I^1\Pi_g^-$; $J^1\Delta_g^-$ | 0÷3 | 1÷6 | NA |
| 02bAST/LAV | $H_2$ | $h^3\Sigma_g^+$, $g^3\Sigma_g^+$, $i^3\Pi_g^+$, $j^3\Delta_g^+$ | 0÷3 | 1÷6 | NA |
| PW | $H_2$ | $h^3\Sigma_g^+$, $g^3\Sigma_g^+$, $i^3\Pi_g^+$, $j^3\Delta_g^+$ | 0÷3 | 1÷6 | NA |

Table 4. Chronological list of publications containing results of non-empirical calculating the lifetimes of excited electronic-vibro-rotational levels for various isotopologues of hydrogen molecule. The rotational quantum numbers N are listed when transition probabilities were averaged over initial and summarized over final $n,v,N,J$ sublevels of the multiplet structure. Quantum numbers J are indicated when calculations of the lifetime values were performed for certain $n,v,N,J$ sublevels.

| Ref. | Isotopic species | Electronic states under the study | $v$ | $N$ | Theoretical models |
|---|---|---|---|---|---|
| 39JAM/COO | $H_2$; | $a^3\Sigma_g^+$; | 0÷3; | [a] | AA |
| | $D_2$ | $a^3\Sigma_g^+$ | 0÷4 | | |
| 70FRE/HIS | $H_2$ | $c^3\Pi_u^-$ | 0÷14 | [a] | AA |
| 72STE/DAL | $H_2$; | $B^1\Sigma_u^+$; | 0÷36; | [a] | AA |
| | HD; | $B^1\Sigma_u^+$; | 0÷41; | [a] | |
| | $D_2$; | $B^1\Sigma_u^+$; | 0÷50; | 0, 1 | |
| | $H_2$ | $C^1\Pi_u$ | 0÷13 | | |
| 77BHA/CHO | $H_2$ | $c^3\Pi_u^-$ | 0 | [a] | NA |
| 79CHO/BHA | $H_2$ | $c^3\Pi_u^-$ | 0 | $N=1, 2$ $(J=N,N\pm1)$ | NA |
| 83GLA/QUA | $H_2$ | $EF^1\Sigma_g^+$; | 0÷32; | [a] | AA, NA |
| | | $GK^1\Sigma_g^+$; | 0÷7; | | |
| | | $H^1\Sigma_g^+$ | 0÷2 | | |
| 84aGLA/QUA | $H_2$ | $EF^1\Sigma_g^+$; | 0÷32; | 1 | NA |
| | | $GK^1\Sigma_g^+$; | 0÷7; | | |
| | | $H^1\Sigma_g^+$; | 0÷2; | | |
| | | $I^1\Pi_g^+$; $I^1\Pi_g^-$ | 0÷3 | | |
| 84bGLA/QUA | $H_2$ | $EF^1\Sigma_g^+$; | 0÷32; | [a] | NA |
| | | $GK^1\Sigma_g^+$; | 0÷7; | | |
| | | $H^1\Sigma_g^+$ | 0÷2 | | |
| 84GLA | $H_2$ | $B'^1\Sigma_u^+$; | 0÷8; | [a] | AA |
| | | $D^1\Pi_u^-$ | 0-15 | | |
| 85COM/BRU | $H_2$; | $c^3\Pi_u^+$; | 0÷17; | 1, 2; | NA |
| | HD, $D_2$ | $c^3\Pi_u^+$ | 0÷20 | 1 | |
| 86KWO/GUB | $H_2$ | $a^3\Sigma_g^+$ | 0÷20 | [a] | AA |
| 86FLE/CHO | $H_2$ | $c^3\Pi_u^-$; | 0 | $N=1,2$ $(J=N,N\pm1)$ | AA |
| | | $c^3\Pi_u^+$ | 0 | | |
| 88CHO/FLE | $H_2$ | $c^3\Pi_u^-$ | 0 | [a] | AA |
| 90aQUA/DRE | $H_2$; HD, $D_2$ | $EF^1\Sigma_g^+$; | 0÷34; | 0, 1 | NA |
| | | $GK^1\Sigma_g^+$; | 0÷9; | | |
| | | $H^1\Sigma_g^+$; | 0÷3; | | |
| | | $I^1\Pi_g^+$ | 0÷4 | | |

Table 4 − Continued.

| Ref. | Isotopic species | Electronic states under the study | $v$ | $N$ | Theoretical models |
|---|---|---|---|---|---|
| 90bQUA/DRE | $H_2$ | $EF^1\Sigma_g^+$; | $0\div32$; | 0, 1 | NA |
|  |  | $GK^1\Sigma_g^+$; | $0\div7$; |  |  |
|  |  | $H^1\Sigma_g^+$ | $0\div2$ |  |  |
| 91SHI/SIE | $H_2$ | $h^3\Sigma_g^+, g^3\Sigma_g^+, i^3\Pi_g^+,$ | 0,1 | $1\div3$ | NA |
|  |  | $i^3\Pi_g^-, j^3\Delta_g^+, j^3\Delta_g^-$ |  |  |  |
| 92TSU/ISH | $H_2$ | $EF^1\Sigma_g^+$; | $19\div21$; | $\leq4$ | NA |
|  |  | $GK^1\Sigma_g^+$; | 0, 2; |  |  |
|  |  | $H^1\Sigma_g^+$; | 0; |  |  |
|  |  | $I^1\Pi_g^+, I^1\Pi_g^-,$ | 0; |  |  |
|  |  | $J^1\Delta_g^+, J^1\Delta_g^-$ | 0 |  |  |
| 92TSU/SHI | $H_2$ | $GK^1\Sigma_g^+$ | 8 | [a] | AA, NA |
| 92GUB/DAL | $H_2$ | $c^3\Pi_u^-,$ | $1\div20$; | [a] | AA |
|  |  | $i^3\Pi_g^-$ | $0\div3$ | 1 |  |
| 93BHA/BHA | $H_2$ | $c^3\Pi_u^+$ | 0 | $1\div10$ | NA |
| 94BER/OTT | $H_2$ | $c^3\Pi_u^-$ | $0\div3$ | $N=1\div4$ $(J=N,N\pm1)$ | NA |
| 95SNO/SIE | $H_2$ | $i^3\Pi_g^+$ | 4, 5 | 1 | AA, NA |
| 98PAR | $H_2$ | $B^1\Sigma_u^+$ | $0\div38$ | $\leq30$ | AA |
| 99PAR | $H_2$ | $C^1\Pi_u^-, C^1\Pi_u^+$ | $0\div13$ | $\leq28$ | AA, NA |
| 99KIY/SAT | $H_2$; | $d^3\Pi_u^-,$ | $0\div9$; | $\leq3$ | NA |
|  | $D_2$ | $e^3\Sigma_u^+$; | $0\div7$; |  |  |
|  |  | $d^3\Pi_u^-$ | $0\div10$ |  |  |
| 99LAV/MEL | $H_2$ | $a^3\Sigma_g^+$ | $0\div6$ | [a] | AA |
| 00ABG/ROU | $H_2$ | $B^1\Sigma_u^+$; | $0\div37$; | $\leq10$ | NA |
|  |  | $C^1\Pi_u^-, C^1\Pi_u^+$; | $0\div13$; |  |  |
|  |  | $B'^1\Sigma_u^+$; | $0\div9$; |  |  |
|  |  | $D^1\Pi_u^-, D^1\Pi_u^+$ | $0\div2$ |  |  |
| 00ADA/PAZ | $H_2$ | $i^3\Pi_g^-$; | $0\div3$; | $1\div8$; | NA |
|  |  | $j^3\Delta_g^-$ | $0\div3$ | $2\div8$ |  |
| 00bAST/LAV | $H_2$ | $I^1\Pi_g^-, J^1\Delta_g^-$ | $0\div3$ | $\leq6$ | NA |
| 00FAN/SCH | $H_2$, HD, $D_2$, DT, $T_2$ | $a^3\Sigma_g^+$ | $0\div14$ | [a] | AA |
| 02PAZ/PUP | $H_2$ | $h^3\Sigma_g^+; g^3\Sigma_g^+; i^3\Pi_g^+$ | $0\div1$; | $1\div2$; | NA |
|  |  | $j^3\Delta_g^+$ | $0\div3$ | $2\div3$ |  |

Table 4 – Continued.

| Ref. | Isotopic species | Electronic states under the study | $v$ | $N$ | Theoretical models |
|---|---|---|---|---|---|
| 03KIY/SAT | H$_2$ | $k^3\Pi_u^-$ | 0÷7; | 1÷7; | NA |
| | D$_2$ | $k^3\Pi_u^-$ | 0÷6 | 1÷10 | |
| 04KIL/LEH | H$_2$ | $e^3\Sigma_u^+$ | 0÷4 | [a] | AA |
| 06ROS/YOS | H$_2$ | $\bar{H}\,^1\Sigma_g^+$ | 0÷16; | 0÷5; | NA |
| | D$_2$ | $\bar{H}\,^1\Sigma_g^+$ | 0÷24 | 0÷5 | |
| 06FAN/WÜN | H$_2$, D$_2$, T$_2$, HD, HT, DT | $B^1\Sigma_u^+$ | 0÷36 (H$_2$); 0÷51 (D$_2$); 0÷63 (T$_2$);  0÷42 (HD); 0÷44  (HT); 0÷56 (DT) | [a] | AA |
| | H$_2$, D$_2$, T$_2$, HD, HT, DT | $B'^1\Sigma_u^+$ | 0÷8 (H$_2$); 0÷12 (D$_2$); 0÷15 (T$_2$);  0÷10 (HD); 0÷11  (HT); 0÷14 (DT) | | |
| | H$_2$, D$_2$, T$_2$, HD, HT, DT | $B''\bar{B}\,^1\Sigma_u^+$ | 0÷68 (H$_2$); 0÷97 (D$_2$); 0÷119 (T$_2$);  0÷79 (HD); 0÷84  (HT); 0÷106 (DT) | | |
| | H$_2$, D$_2$, T$_2$, HD, HT, DT | $C^1\Pi_u$ | 0÷13 (H$_2$); 0÷19 (D$_2$); 0÷23 (T$_2$);  0÷15 (HD); 0÷16  (HT); 0÷20 (DT) | | |
| | H$_2$, D$_2$, T$_2$, HD, HT, DT | $D^1\Pi_u$ | 0÷17 (H$_2$); 0÷24 (D$_2$); 0÷30 (T$_2$);  0÷20 (HD); 0÷21  (HT); 0÷27 (DT) | | |
| | H$_2$, D$_2$, T$_2$, HD, HT, DT | $D'^1\Pi_u$ | 0÷17 (H$_2$); 0÷24 (D$_2$); 0÷30 (T$_2$);  0÷20 (HD); 0÷21  (HT); 0÷27 (DT) | | |
| | H$_2$, D$_2$, T$_2$, HD, HT, DT | $EF^1\Sigma_g^+$ | 0÷32 (H$_2$); 0÷46 (D$_2$); 0÷56 (T$_2$);  0÷37 (HD); 0÷39  (HT); 0÷50 (DT) | | |
| | H$_2$, D$_2$, T$_2$, HD, HT, DT | $GK^1\Sigma_g^+$ | 0÷8 (H$_2$); 0÷11 (D$_2$); 0÷14 (T$_2$);  0÷9 (HD); 0÷10  (HT); 0÷13 (DT) | | |
| | H$_2$, D$_2$, T$_2$, HD, HT, DT | $H\bar{H}\,^1\Sigma_g^+$ | 0÷71 (H$_2$); 0÷101 (D$_2$); 0÷124 (T$_2$);  0÷82 (HD); 0÷87  (HT); 0÷111 (DT) | | |
| | H$_2$, D$_2$, T$_2$, HD, HT, DT | $I^1\Pi_g$ | 0÷3 (H$_2$); 0÷4 (D$_2$); 0÷6 (T$_2$);  0÷3 (HD); 0÷4  (HT); 0÷5 (DT) | | |
| | H$_2$, D$_2$, T$_2$, HD, HT, DT | $J^1\Delta_g$ | 0÷17 (H$_2$); 0÷25 (D$_2$); 0÷30 (T$_2$);  0÷19 (HD); 021  (HT); 0÷27 (DT) | | |
| | H$_2$, D$_2$, T$_2$, HD, HT, DT | $O^1\Sigma_g^+$ | 0÷9 (H$_2$); 0÷14 (D$_2$); 0÷17 (T$_2$);  0÷11 (HD); 0÷12  (HT); 0÷15 (DT) | | |
| | H$_2$, D$_2$, T$_2$, HD, HT, DT | $P^1\Sigma_g^+$ | 0÷15 (H$_2$); 0÷22 (D$_2$); 0÷27 (T$_2$);  0÷18 (HD); 0÷19  (HT); 0÷24 (DT) | | |
| | H$_2$, D$_2$, T$_2$, HD, HT, DT | $R^1\Pi_g$ | 0÷13 (H$_2$); 0÷19 (D$_2$); 0÷24 (T$_2$);  0÷16 (HD); 0÷17  (HT); 0÷21 (DT) | | |
| | H$_2$, D$_2$, T$_2$, HD, HT, DT | $S^1\Delta_g$ | 0÷17 (H$_2$); 0÷24 (D$_2$); 0÷30 (T$_2$);  0÷20 (HD); 0÷21  (HT); 0÷27 (DT) | | |
| | H$_2$, D$_2$, T$_2$, HD, HT, DT | $a^3\Sigma_g^+$ | 0÷18 (H$_2$); 0÷26 (D$_2$); 0÷33 (T$_2$);  0÷21 (HD); 0÷22  (HT); 0÷29 (DT) | | |
| | H$_2$, D$_2$, T$_2$, HD, HT, DT | $c^3\Pi_u$ | 0÷19 (H$_2$); 0÷28 (D$_2$); 0÷34 (T$_2$);  0÷22 (HD); 0÷24  (HT); 0÷30 (DT) | | |
| | H$_2$, D$_2$, T$_2$, HD, HT, DT | $d^3\Pi_u$ | 0÷19 (H$_2$); 0÷28 (D$_2$); 0÷34 (T$_2$);  0÷22 (HD); 0÷24 (HT); 0÷31 (DT) | | |
| | H$_2$, D$_2$, T$_2$, HD, HT, DT | $e^3\Sigma_g^+$ | 0÷7 (H$_2$); 0÷10 (D$_2$); 0÷13 (T$_2$);  0÷8 (HD); 0÷9 (HT); 0÷11 (DT) | | |
| | H$_2$, D$_2$, T$_2$, HD, HT, DT | $f^3\Sigma_g^+$ | 0÷1 (H$_2$); 0÷2 (D$_2$); 0÷2 (T$_2$);  0÷1 (HD); 0÷1 (HT); 0÷2 (DT) | | |
| | H$_2$, D$_2$, T$_2$, HD, HT, DT | $g^3\Sigma_g^+$ | 0÷18 (H$_2$); 0÷26 (D$_2$); 0÷32 (T$_2$);  0÷21 (HD); 0÷22 (HT); 0÷29 (DT) | | |
| | H$_2$, D$_2$, T$_2$, HD, HT, DT | $h^3\Sigma_g^+$ | 0÷3 (H$_2$); 0÷5 (D$_2$); 0÷6 (T$_2$);  0÷4 (HD); 0÷4 (HT); 0÷5 (DT) | | |
| | H$_2$, D$_2$, T$_2$, HD, HT, DT | $i^3\Sigma_g^+$ | 0÷3 (H$_2$); 0÷4 (D$_2$); 0÷6 (T$_2$);  0÷3 (HD); 0÷4 (HT); 0÷5 (DT) | | |
| | H$_2$, D$_2$, T$_2$, HD, HT, DT | $j^3\Pi_g$ | 0÷17 (H$_2$); 0÷25 (D$_2$); 0÷31 (T$_2$);  0÷20 (HD); 0÷21 (HT); 0÷27 (DT) | | |
| | H$_2$, D$_2$, T$_2$, HD, HT, DT | $k^3\Pi_u$ | 0÷12 (H$_2$); 0÷18 (D$_2$); 0÷22 (T$_2$);  0÷14 (HD); 0÷15  (HT); 0÷20 (DT) | | |
| | H$_2$, D$_2$, T$_2$, HD, HT, DT | $r^3\Pi_g$ | 0÷16 (H$_2$); 0÷24 (D$_2$); 0÷29 (T$_2$);  0÷19 (HD); 0÷20  (HT); 0÷26 (DT) | | |
| | H$_2$, D$_2$, T$_2$, HD, HT, DT | $s^3\Delta_g$ | 0÷17 (H$_2$); 0÷26 (D$_2$); 0÷32 (T$_2$);  0÷21 (HD); 0÷22  (HT); 0÷28 (DT) | | |

Table 4 – Continued.

| Ref. | Isotopic species | Electronic states under the study | $v$ | $N$ | Theoretical models |
|---|---|---|---|---|---|
| 07ROS/FUJ | $H_2$ | $I^1\Pi_g^-$ | 3÷4 | 1÷5 | AA |
| | HD | $I^1\Pi_g^-$ | 4÷5 | 1÷5 | |
| | $D_2$ | $I^1\Pi_g^-$ | 5÷6 | 1÷5 | |
| 10ROS/AND | $H_2$ | $I^1\Pi_g^-$; $I^1\Pi_g^+$ | 4÷5[b] | 1÷3 | AA |
| | $D_2$ | $I^1\Pi_g^-$; $I^1\Pi_g^+$ | 5÷6[b] | 1÷4 | |
| 10LIU/JOH | $H_2$ | $a^3\Sigma_g^+$ | 0÷20 | 0÷14 | AA |
| 11ROS/TSU | $H_2$ | $EF^1\Sigma_g^+$ | 33 | 0÷1 | NA |
| | $D_2$ | $EF^1\Sigma_g^+$; | 44÷47 | 0÷5 | NA |
| | | $GK^1\Sigma_g^+$; | 11 | 0÷5 | |
| | | $I^1\Pi_g^+$; | 8[c] | 1÷5 | |
| | | $J^1\Delta_g^+$ | 4 | 2÷5 | |

[a] – the rotational structure of electronic-rotational states was not taken into account;

[b] – in private communication;

[c] – the levels of outer potential well of the electronic state.

Table 5. The list of excited electronic states of hydrogen molecule with complete bibliography concerning experimental, semi-empiric and non-empirical data on the lifetimes for various isotopologues: $H_2$ (a), $D_2$ (b), $T_2$ (c), HD (d), DT (e) and HT (f).

| Singlet electronic state | Experiment | Semi-empiric determination | Non-empirical calculations |
|---|---|---|---|
| $B\,^1\Sigma_u^+$ | [66HES/DRE][a], [68HES][a], [72SMI/CHE][a], [84SCH/IMS][a] | | [72STE/DAL][a,b,d], [98PAR][a], [00ABG/ROU][a], [06FAN/WÜN][a,b,c,d,e,f] |
| $C\,^1\Pi_u^{\pm}$ | [66HES/DRE][a], [68HES][a] | | [72STE/DAL][a], [99PAR][a], [00ABG/ROU][a], [06FAN/WÜN][a,b,c,d,e,f] |
| $B'\,^1\Sigma_u^+$ | [01KIY/SAT][a] | | [00ABG/ROU][a], [ 06FAN/WÜN][a,b,c,d,e,f] |
| $D\,^1\Pi_u^-$ | [85GLA/BRE][a], [01KIY/SAT][a] | | [84GLA][a], [00ABG/ROU][a], [06FAN/WÜN][a,b,c,d,e,f] |
| $D\,^1\Pi_u^+$ | [01KIY/SAT][a] | | [00ABG/ROU][a], [06FAN/WÜN][a,b,c,d,e,f] |
| $D'\,^1\Pi_u^+$ | | [79GLA][a] | [06FAN/WÜN][a,b,c,d,e,f] |
| $EF\,^1\Sigma_g^+$ | [78KLI/RHO][a], [78CHI/DAL][a], [79DAY/AND][a], [86CHA/THO][a,d], [88SAN/CAM][a], [92TSU/ISH][a], [92TSU/SHI][a], [98SUZ/NAK][b], [01KIY/SAT][a], [04YOS/OGI][d], [05AIT/OGI][a], [05AIT/YOS][b], [11ROS/TSU][b] | | [83GLA/QUA][a], [84aGLA/QUA][a], [84bGLA/QUA][a], [90aQUA/DRE][a,b,d], [90bQUA/DRE][a], [92TSU/ISH][a], [06FAN/WÜN][a,b,c,d,e,f], [11ROS/TSU][a,b] |
| $GK\,^1\Sigma_g^+$ | [70LIN][a], [71LIN][a], [72LIN/DAL][a], [72FRE/MIL][a], [76MEL/LOM][a], [76GOM/CAM][a], [78CHI/DAL][a], [79DAY/AND][a], [81BÖS/LIN][a], [88SAN/CAM][a], [92TSU/ISH][a], [92TSU/SHI][a], [98SUZ/NAK][b], [04YOS/OGI][d], [05AIT/OGI][a], [05AIT/YOS][b], [11ROS/TSU][b] | | [83GLA/QUA][a], [84aGLA/QUA][a], [84bGLA/QUA][a], [90aQUA/DRE][a,b,d], [90bQUA/DRE][a], [92TSU/ISH][a], [92TSU/SHI][a], [06FAN/WÜN][a,b,c,d,e,f], [11ROS/TSU][b] |
| $H\,^1\Sigma_g^+$ | [81BÖS/LIN][a], [92TSU/ISH][a], [98SUZ/NAK][b], [04YOS/OGI][d], [05AIT/OGI][a], [05AIT/YOS][b] | | [83GLA/QUA][a], [84aGLA/QUA][a], [84bGLA/QUA][a], [90aQUA/DRE][a,b,d], [90bQUA/DRE][a], [92TSU/ISH][a], [06FAN/WÜN][a,b,c,d,e,f] |
| $\overline{H}\,^1\Sigma_g^+$ | [00REI/HOG][a,b,d], [06ROS/YOS][a,b] | | [06ROS/YOS][a,b], [06FAN/WÜN][a,b,c,d,e,f] |
| $I\,^1\Pi_g^-$ | [72LIN/DAL][a], [76GOM/CAM][a] [78CHI/DAL][a], [79DAY/AND][a], [80BRY/KOT][a], [81BÖS/LIN][a], [88SAN/CAM][a], [92TSU/ISH][a], [04YOS/OGI][d], [05AIT/OGI][a], [05AIT/YOS][b], [07ROS/FUJ][d], [10ROS/AND][a,b], | [96AST/KÄN][a], [97AST/KOK][a] | [84aGLA/QUA][a], [00bAST/LAV][a], [06FAN/WÜN][a,b,c,d,e,f] [07ROS/FUJ][a,b,d], [10ROS/AND][a,b] |
| $I\,^1\Pi_g^+$ | [76GOM/CAM][a], [78CHI/DAL][a], [79DAY/AND][a], [80BRY/KOT][a], [81BÖS/LIN][a], [88SAN/CAM][a], [92TSU/ISH][a], [04YOS/OGI][d], [05AIT/OGI][a], [05AIT/YOS][b], [10ROS/AND][a,b] | | [84aGLA/QUA][a], [90aQUA/DRE][a,b,d], [92TSU/ISH][a], [06FAN/WÜN][a,b,c,d,e,f] [10ROS/AND][a,b], [11ROS/TSU][b] |
| $J\,^1\Delta_g^-$ | [81BÖS/LIN][a], [88SAN/CAM][a], [92TSU/ISH][a], [04YOS/OGI][d] | [96AST/KÄN][a], [97AST/KOK][a] | [00bAST/LAV][a], [06FAN/WÜN][a,b,c,d,e,f] |
| $J\,^1\Delta_g^+$ | [81BÖS/LIN][a], [92TSU/ISH][a], [04YOS/OGI][d], [05AIT/OGI][a] | | [92TSU/ISH][a], [06FAN/WÜN][a,b,c,d,e,f] [11ROS/TSU][b] |
| $\overline{B}\,^1\Sigma_g^+$ | [00REI/HOG][d] | | [06FAN/WÜN][a,b,c,d,e,f] |
| $O\,^1\Sigma_g^+$ | [05AIT/OGI][a] | | [06FAN/WÜN][a,b,c,d,e,f] |
| $P\,^1\Sigma_g^+$ | [88SAN/CAM][a], [05AIT/OGI][a], [05AIT/YOS][b] | | [06FAN/WÜN][a,b,c,d,e,f] |
| $R\,^1\Pi_g^-$ | [05AIT/OGI][a], [05AIT/YOS][b] | | [06FAN/WÜN][a,b,c,d,e,f] |
| $R\,^1\Pi_g^+$ | [05AIT/OGI][a], [05AIT/YOS][b] | | [06FAN/WÜN][a,b,c,d,e,f] |
| $S\,^1\Delta_g^-$ | [05AIT/OGI][a], [05AIT/YOS][b] | | [06FAN/WÜN][a,b,c,d,e,f] |
| $S\,^1\Delta_g^+$ | [05AIT/OGI][a], [05AIT/YOS][b] | | [06FAN/WÜN][a,b,c,d,e,f] |

| Triplet electronic state | Experiment | Semi-empirical determination | Non-empirical calculations |
|---|---|---|---|
| $a^3\Sigma_g^+$ | [65FOW/HOL][a], [66FOW/HOL][a,b], [71IMH/REA][a], [72THO/FOW][a], [72SMI/CHE][a,b], [75KIN/REA][a], [79MOH/KIN][a], [81BRE/GOD][a], [88WED/PHE][a] | | [39JAM/COO][a,b], [86KWO/GUB][a], [99LAV/MEL][a], [00FAN/SCH][a,b,c,d,e], [06FAN/WÜN][a,b,c,d,e,f], [10LIU/JOH][a] |
| $c^3\Pi_u^-$ | [62LIC][a], [72JOH][a,b,d], [94BER/OTT][a], | | [70FRE/HIS][a], [77BHA/CHO][a], [79CHO/BHA][a], [86FLE/CHO][a], [88CHO/FLE][a], [92GUB/DAL][a], [94BER/OTT][a], [06FAN/WÜN][a,b,c,d,e,f] |
| $c^3\Pi_u^+$ | [77VOG/MEI][a], [84BRU/NEU][a] | | [86FLE/CHO][a], [85COM/BRU][a,b,d], [06FAN/WÜN][a,b,c,d,e,f] |
| $e^3\Sigma_u^+$ | [99KIY/SAT][a,b] | [87KIY][a], [88LAV/TYU][a], [89LAV/POZ][a,b], [90POZ][a,b], [93KIY/SAT][a,b], [96AST/KÄN][a] | [99KIY/SAT][a,b], [04KIL/LEH][a], [06FAN/WÜN][a,b,c,d,e,f] |
| $d^3\Pi_u^-$ | [69CAH][a], [72MAR/JOS][a], [73aFRE/MIL][a], [78DAY/AND][a], [81BRE/GOD][a], [81BOG/EFR][a], [87KIY][a], [88SAN/CAM][a], [90BUR/LAV][a], [93KIY/SAT][a,b], [99KIY/SAT][a,b] | [85DRA/LAV][a,b,d], [87DRA/LAV][a,b,d], [90POZ][a,b,c,d], [96AST/KÄN][a] | [99KIY/SAT][a,b], [06FAN/WÜN][a,b,c,d,e,f] |
| $d^3\Pi_u^+$ | [72MAR/JOS][a], [73bFRE/MIL][a], [87KIY][a], [93KIY/SAT][a,b] | [87KIY][a], [90POZ][a,b], [96AST/KÄN][a], [93KIY/SAT][a,b] | [06FAN/WÜN][a,b,c,d,e,f] |
| $h^3\Sigma_g^+$ | [81EYL/PIP][a], [02PAZ/PUP][a] | [95bAST/LAV][a], [PW][a] | [91SHI/SIE][a], [02PAZ/PUP][a], [06FAN/WÜN][a,b,c,d,e,f] |
| $g^3\Sigma_g^+$ | [81EYL/PIP][a], [96RAY/LAF][a], [02PAZ/PUP][a] | [95bAST/LAV][a], [PW][a] | [91SHI/SIE][a], [02PAZ/PUP][a], [06FAN/WÜN][a,b,c,d,e,f] |
| $i^3\Pi_g^-$ | [81EYL/PIP][a], [96RAY/LAF][a] | [95bAST/LAV][a], [96AST/KÄN][a] | [91SHI/SIE][a], [92GUB/DAL][a], [00ADA/PAZ][a], [06FAN/WÜN][a,b,c,d,e,f] |
| $i^3\Pi_g^+$ | [81EYL/PIP][a], [89KOO/ZAN][a], [02PAZ/PUP][a] | [89KOO/ZAN][a], [95bAST/LAV][a], [PW][a] | [89KOO/ZAN][a], [91SHI/SIE][a], [95SNO/SIE][a], [02PAZ/PUP][a], [06FAN/WÜN][a,b,c,d,e,f] |
| $j^3\Delta_g^-$ | [81EYL/PIP][a], [81MON][a], [88SAN/CAM][a], [89KOO/ZAN][a], [96RAY/LAF][a] | [89KOO/ZAN][a], [95bAST/LAV][a], [96AST/KÄN][a] | [89KOO/ZAN][a], [91SHI/SIE][a], [00ADA/PAZ][a], [06FAN/WÜN][a,b,c,d,e,f] |
| $j^3\Delta_g^+$ | [81EYL/PIP][a], [81MON][a], [89KOO/ZAN][a], [02PAZ/PUP][a] | [89KOO/ZAN][a], [95bAST/LAV][a], [PW][a] | [89KOO/ZAN][a], [91SHI/SIE][a], [02PAZ/PUP][a], [06FAN/WÜN][a,b,c,d,e,f] |
| $k^3\Pi_u^-$ | [74MIL/FRE][a], [78DAY/AND][a], [03KIY/SAT][a,b] | | [03KIY/SAT][a,b], [06FAN/WÜN][a,b,c,d,e,f] |

Table 6. All experimental and most reliable semi-empiric and non-empirical lifetime values for vibro-rotational levels of various excited electronic states of the $H_2$ molecule. If several experimental lifetime values were reported for the same rovibronic level then data that are more reliable are italicized. Recommended data are marked as PW and printed in bold face. Letters A, B, C and D characterize the differences between observed and calculated values (O-C) when they are less or equal to $1\sigma$, $2\sigma$, $3\sigma$ and greater than $3\sigma$ correspondingly.

| $B^1\Sigma_u^+$,$v$,$N$ rovibronic levels | | | | | | | |
|---|---|---|---|---|---|---|---|
| $v$ | $N$ | Experimental studies | | | Non-empirical calculation | | O-C |
| | | $\tau$, ns | Ref. | Method | $\tau$, ns | Ref. | Model | |
| 0 | 1,2 | 0.54(5) | 84SCH/IMS | DC | 0.538 | 00ABG/ROU | NA | A |
| 1 | 1,2 | 0.59(9) | 84SCH/IMS | DC | 0.575 | 00ABG/ROU | NA | A |
| 2 | 1,2 | 0.62(11) | 84SCH/IMS | DC | 0.613 | 00ABG/ROU | NA | A |
| 3 | 0 | 0.8(2)[a] | 66HES/DRE | PS | 0.649 | 00ABG/ROU | NA | A |
| | 1,2 | *0.67(9)* | 84SCH/IMS | DC | | | | A |
| 4 | 0 | 0.8(2)[a] | 66HES/DRE | PS | 0.689 | 00ABG/ROU | NA | A |
| | 1,2 | 0.69(15) | 84SCH/IMS | DC | | | | A |
| 5 | 0 | 0.8(2)[a] | 66HES/DRE | PS | 0.717 | 00ABG/ROU | NA | A |
| 6 | 0 | 0.8(2)[a] | 66HES/DRE | PS | 0.746 | 00ABG/ROU | NA | A |
| 7 | 0 | 0.8(2)[a] | 66HES/DRE | PS | 0.668 | 00ABG/ROU | NA | A |
| 8 | 0 | 1.0(2)[a] | 72SMI/CHE | PS | 0.651 | 00ABG/ROU | NA | B |
| 9 | 0 | 1.0(2)[a] | 72SMI/CHE | PS | 0.631 | 00ABG/ROU | NA | B |
| 10 | 0 | 1.0(2)[a] | 72SMI/CHE | PS | 0.662 | 00ABG/ROU | NA | B |
| 11 | 0 | 1.0(2)[a] | 72SMI/CHE | PS | 0.691 | 00ABG/ROU | NA | B |

| $C^1\Pi_u^\pm$,$v$,$N$ rovibronic levels | | | | | | | |
|---|---|---|---|---|---|---|---|
| $v$ | $N$ | Experimental studies | | | Non-empirical calculation | | O-C |
| | | $\tau$, ns | Ref. | Method | $\tau$, ns | Ref. | Model | |
| 0 | 1 | 0.6(2) | 66HES/DRE | PS | 0.847 | 00ABG/ROU | NA | B |
| 1 | 1 | 0.6(2) | 66HES/DRE | PS | 0.862 | 00ABG/ROU | NA | B |
| 2 | 1 | 0.6(2) | 66HES/DRE | PS | 0.877 | 00ABG/ROU | NA | B |
| 3 | 1 | 0.6(2) | 66HES/DRE | PS | 0.893 | 00ABG/ROU | NA | B |

| $B'^1\Sigma_u^+$,$v$,$N$ rovibronic level | | | | | | | |
|---|---|---|---|---|---|---|---|
| $v$ | $N$ | Experimental studies | | | Non-empirical calculation | | O-C |
| | | $\tau$, ns | Ref. | Method | $\tau$, ns | Ref. | Model | |
| 2 | 2 | 5.0(3) | 01KIY/SAT | DC | 3.79 | 00ABG/ROU | NA | D |

Table 6 – Continued.

| | | $D^1\Pi_u^+,v,N$ rovibronic level | | | | | |
|---|---|---|---|---|---|---|---|
| | | Experimental studies | | | Non-empirical calculation | | O-C |
| $v$ | $N$ | $\tau$, ns | Ref. | Method | $\tau$, ns | Ref. | Model |
| 2 | 1 | 3.0(3) | 01KIY/SAT | DC | 2.84 | 00ABG/ROU | NA | A |

| | | $D^1\Pi_u^-,v,N$ rovibronic levels | | | | | |
|---|---|---|---|---|---|---|---|
| | | Experimental studies | | | Non-empirical calculation | | O-C |
| $v$ | $N$ | $\tau$, ns | Ref. | Method | $\tau$, ns | Ref. | Model |
| 0 | 1 | 2.9(2) | 01KIY/SAT | DC | 2.82 | 00ABG/ROU | NA | A |
| 3 | 1 | 3.0(4) | 85GLA/BRE | DC | 2.999[a] | 06FAN/WÜN | AA | A |
| 4 | 1 | 2.9(2) | 85GLA/BRE | DC | 3.042[a] | 06FAN/WÜN | AA | A |
| 6 | 1 | 2.7(5) | 85GLA/BRE | DC | 3.111[a] | 06FAN/WÜN | AA | A |
| 7 | 1 | 2.3(5) | 85GLA/BRE | DC | 3.144[a] | 06FAN/WÜN | AA | B |
| 8 | 1 | 2.3(5) | 85GLA/BRE | DC | 3.293[a] | 06FAN/WÜN | AA | B |
| 9 | 1 | 2.3(5) | 85GLA/BRE | DC | 3.516[a] | 06FAN/WÜN | AA | C |
| 10 | 1 | 2.3(5) | 85GLA/BRE | DC | 5.292[a] | 06FAN/WÜN | AA | D |
| 12 | 1 | 2.3(5) | 85GLA/BRE | DC | 7.452[a] | 06FAN/WÜN | AA | D |
| 13 | 1 | 2.3(5) | 85GLA/BRE | DC | 9.289[a] | 06FAN/WÜN | AA | D |
| 14 | 1 | 2.3(5) | 85GLA/BRE | DC | 19.140[a] | 06FAN/WÜN | AA | D |
| 15 | 1 | 2.7(5) | 85GLA/BRE | DC | 22.585[a] | 06FAN/WÜN | AA | D |

Table 6 – Continued.

| $v$ | $N$ | Experimental studies | | | Non-empirical calculation | | | O-C |
|---|---|---|---|---|---|---|---|---|
| | | $\tau$, ns | Ref. | Method | $\tau$, ns | Ref. | Model | |
| | | | | *$EF^1\Sigma_g^+,v,N$ rovibronic levels* | | | | |
| 0 | 0 | 200(3) | 86CHA/THO | TF | 204 | 90bQUA/DRE | NA | B |
| | 1 | 213(13) | 86CHA/THO | TF | 202 | 90bQUA/DRE | NA | A |
| 3 | 0 | 136(6) | 86CHA/THO | TF | 139 | 90bQUA/DRE | NA | A |
| | 1 | 148(12) | 86CHA/THO | TF | 138 | 90bQUA/DRE | NA | A |
| 6 | 0,1,2 | 100(20)[a] | 78KLI/RHO | DC | | | | |
| | 0 | *101(6)* | 86CHA/THO | TF | 100 | 90bQUA/DRE | NA | A |
| | | *101(2)* | 01KIY/SAT | DC | | | | A |
| | | **101(4)** | PW | Recom. | | | | |
| | 1 | *101(2)* | 01KIY/SAT | DC | 101 | 90bQUA/DRE | NA | A |
| | 2 | *110(4)* | 01KIY/SAT | DC | | | | |
| | 3 | 134(6) | 01KIY/SAT | DC | | | | |
| | 4 | 170(9)[b] | 01KIY/SAT | DC | | | | |
| 7 | 0 | 246(13)[b] | 01KIY/SAT | DC | 257 | 90bQUA/DRE | NA | A |
| | 1 | 233(11) | 01KIY/SAT | DC | 249 | 90bQUA/DRE | NA | B |
| | 2 | 209(11) | 01KIY/SAT | DC | | | | |
| | 3 | 157(5) | 01KIY/SAT | DC | | | | |
| | 4 | 129(6) | 01KIY/SAT | DC | | | | |
| 19 | 1 | 239.5(4.8) | 92TSU/ISH | DC | 142 | 92TSU/ISH | NA | A |
| | 2 | 354.1(14.1) | 92TSU/ISH | DC | 368 | 92TSU/ISH | NA | A |
| | 3 | 287.4(11.2) | 92TSU/ISH | DC | 364 | 92TSU/ISH | NA | D |
| | 4 | 345.3(16.2) | 92TSU/ISH | DC | 361 | 92TSU/ISH | NA | A |
| 20 | 1 | 449.0(30.5) | 92TSU/ISH | DC | 464 | 92TSU/ISH | NA | A |
| | 2 | 478.0(45.2) | 92TSU/ISH | DC | 463 | 92TSU/ISH | NA | A |
| | 3 | 240.6(9.5) | 92TSU/ISH | DC | 273 | 92TSU/ISH | NA | D |
| | 4 | 409.4(23.4) | 92TSU/ISH | DC | 454 | 92TSU/ISH | NA | C |
| 21 | 1 | 92.0(7.1) | 92TSU/ISH | DC | 136 | 92TSU/ISH | NA | D |
| | 2 | 106.3(2.1) | 92TSU/ISH | DC | 159 | 92TSU/ISH | NA | D |
| | 3 | 179.0(2.0) | 92TSU/ISH | DC | 178 | 92TSU/ISH | NA | A |
| | 4 | 120.9(1.4) | 92TSU/ISH | DC | 125 | 92TSU/ISH | NA | C |
| 22 | 1 | ***69.1(5.5)*** | 79DAY/AND | DC | | | | |
| | 1,2,3 | ***61.9(4.3)*** | 88SAN/CAM | DC | | | | |
| | 1 | **66(5)** | PW | Recom. | 109 | 90bQUA/DRE | NA | D |
| | 3 | ***68.5(5.5)*** | 79DAY/AND | DC | | | | |
| | 3 | **65(5)** | PW | Recom. | | | | |
| 23 | 3,5 | 110(17) | 88SAN/CAM | DC | | | | |

| | | | | | | | |
|---|---|---|---|---|---|---|---|
| | | | | | | | |

$EF^1\Sigma_g^+, v, N$ rovibronic levels

| $v$ | $N$ | Experimental studies | | | Non-empirical calculation | | | O-C |
|-----|-----|----------------------|------|--------|---------------------------|------|-------|-----|
| | | $\tau$, ns | Ref. | Method | $\tau$, ns | Ref. | Model | |
| 26 | 1 | 48.5(3.0) | 78CHI/DAL | Hanle | 36.8 | 90bQUA/DRE | NA | D |
| | | *38.8(3.1)* | 79DAY/AND | DC | | | | A |
| | | *34.5(2.4)* | 88SAN/CAM | DC | | | | A |
| | | **37(3)** | PW | Recom. | | | | A |
| | 2 | ≅107(6) | 78CHI/DAL | Hanle | | | | |
| | | *39.1(3.1)* | 79DAY/AND | DC | | | | |
| | | **39(3)** | PW | Recom. | | | | |
| | 3 | 59.4(3.6) | 78CHI/DAL | Hanle | | | | |
| | | *44.2(3.5)* | 79DAY/AND | DC | | | | |
| | | *39.4(2.8)* | 88SAN/CAM | DC | | | | |
| | | **42(3)** | PW | Recom. | | | | |
| 29 | 0 | 189.4(1.4) | 05AIT/OGI | DC | 116 | 90bQUA/DRE | NA | D |
| | 1 | 247.2(2.5) | 05AIT/OGI | DC | 218 | 90bQUA/DRE | NA | D |
| | 2 | 149.0(1.4) | 05AIT/OGI | DC | | | | |
| | 5 | 117.9(2.3) | 05AIT/OGI | DC | | | | |
| 30 | 0 | 189.9(4.9) | 05AIT/OGI | DC | 194 | 90bQUA/DRE | NA | A |
| | 3 | 131.3(4.5) | 05AIT/OGI | DC | | | | |
| | 4 | 142.0(0.6) | 05AIT/OGI | DC | | | | |
| | 5 | 171.1(1.2) | 05AIT/OGI | DC | | | | |
| 32 | 0 | 198.2(11.1) | 92TSU/SHI | DC | 245 | 90bQUA/DRE | NA | D |
| | 1 | *233.4(17.1)* | 92TSU/SHI | DC | 260 | 90bQUA/DRE | NA | B |
| | | *245.8(3.9)* | 05AIT/OGI | DC | | | | D |
| | | **240(10)** | PW | Recom. | | | | |
| | 2 | *341.7(24.4)* | 92TSU/SHI | DC | | | | |
| | | *332.9(12.7)* | 05AIT/OGI | DC | | | | |
| | | **337(19)** | PW | Recom. | | | | |
| | 3 | *355.1(29.8)* | 92TSU/SHI | DC | | | | |
| | | *375.0(16.1)* | 05AIT/OGI | DC | | | | |
| | | **365(28)** | PW | Recom. | | | | |
| | 4 | 297.0(7.2) | 05AIT/OGI | DC | | | | |
| | 5 | 289.0(2.1) | 05AIT/OGI | DC | | | | |
| 33 | 0 | 130.3(3.9) | 05AIT/OGI | DC | | | | |
| | 1 | 162.5(3.6) | 05AIT/OGI | DC | | | | |

$GK^1\Sigma_g^+,v,N$ rovibronic levels

| $v$ | $N$ | Experimental studies | | | Non-empirical calculation | | | O-C |
|---|---|---|---|---|---|---|---|---|
| | | τ, ns | Ref. | Method | τ, ns | Ref. | Model | |
| 0 | 1 | ***75(5)*** | 76GOM/CAM | DC | 77.2 | 90bQUA/DRE | NA | A |
| | | 56.3(3.9) | 78CHI/DAL | Hanle | | | | D |
| | | ***75.6(6.0)*** | 79DAY/AND | DC | | | | A |
| | | ***71.0(2.2)*** | 92TSU/ISH | DC | | | | C |
| | | ***74(4)*** | PW | Recom. | | | | C |
| | 2 | ***69(6)*** | 76GOM/CAM | DC | 79.9 | 92TSU/ISH | NA | B |
| | | 56.3(3.9) | 78CHI/DAL | Hanle | | | | D |
| | | ***77.7(6.2)*** | 79DAY/AND | DC | | | | A |
| | | ***60.7(5.2)*** | 92TSU/ISH | DC | | | | D |
| | | ***69(6)*** | PW | Recom. | | | | |
| | 3 | ***56.4(3.9)*** | 78CHI/DAL | Hanle | 82.0 | 92TSU/ISH | NA | D |
| | | ***63.2(3.2)*** | 92TSU/ISH | DC | | | | D |
| | | ***60(4)*** | PW | Recom. | | | | |
| | 4 | 79.9(6.4) | 92TSU/ISH | DC | 107 | 92TSU/ISH | NA | D |
| 1 | 1 | 25.6(1.3) | 70LIN | Hanle | 20.7 | 90bQUA/DRE | NA | D |
| | | 26.6(1.2) | 71LIN | Hanle | | | | D |
| | | 25.5(1.0)[c] | 72LIN/DAL | Hanle | | | | D |
| | | 27.8(1.0)[d] | 72LIN/DAL | Hanle | | | | D |
| | | 21(4) | 72FRE/MIL | MOMRIE | | | | A |
| | | ***24.8(4.0)*** | 79DAY/AND | DC | | | | A |
| | | ***23.7(1.7)*** | 88SAN/CAM | DC | | | | B |
| | | ***24(3)*** | PW | Recom. | | | | |
| | 2 | 38.2(1.9) | 70LIN | Hanle | | | | |
| | | 38.3(2.0) | 71LIN | Hanle | | | | |
| | | 38.3(1.0)[d] | 72LIN/DAL | Hanle | | | | |
| | | 34.0(2.0) | 78CHI/DAL | Hanle | | | | |
| | | ***23.7(1.9)[e]*** | 79DAY/AND | DC | | | | |
| | | ***25.0(1.3)*** | 88SAN/CAM | DC | | | | |
| | | ***24.4(1.5)*** | PW | Recom. | | | | |
| | 3 | 37.8(1.9) | 70LIN | Hanle | | | | |
| | | 39.3(2.5) | 71LIN | Hanle | | | | |
| | | 37.4(1.0)[b] | 72LIN/DAL | Hanle | | | | |
| | | 41.2(3.0)[d] | 72LIN/DAL | Hanle | | | | |
| | | 39.7(2.4) | 78CHI/DAL | Hanle | | | | |
| | | ***25.0(1.3)*** | 88SAN/CAM | DC | | | | |
| | 5 | 23.7(1.9)[e] | 79DAY/AND | DC | | | | |
| 2 | 1 | ***89.8(7.0)*** | 79DAY/AND | DC | 110 | 90bQUA/DRE | NA | C |
| | | ***85.5(3.8)*** | 92TSU/ISH | DC | | | | D |
| | | ***87(5)*** | PW | Recom. | | | | |
| | 2 | ***70.1(5.6)[e]*** | 79DAY/AND | DC | 96.7 | 92TSU/ISH | NA | C |
| | | ***67.8(4.9)*** | 92TSU/ISH | DC | | | | D |
| | | ***69(5)*** | PW | Recom. | | | | |
| | 3 | ***70.1(5.6)[e]*** | 79DAY/AND | DC | 83.5 | 92TSU/ISH | NA | C |
| | | ***62.8(2.4)*** | 92TSU/ISH | DC | | | | D |
| | | ***66(4)*** | PW | Recom. | | | | |
| | 4 | ***70.1(5.6)[e]*** | 79DAY/AND | DC | 106 | 92TSU/ISH | NA | D |
| | | ***57.8(4.9)*** | 92TSU/ISH | DC | | | | D |
| | | ***64(6)*** | PW | Recom. | | | | |

| | | $GK^1\Sigma_g^+,v,N$ rovibronic levels | | | | | | |
|---|---|---|---|---|---|---|---|---|
| | | Experimental studies | | | Non-empirical calculation | | | O-C |
| $v$ | $N$ | $\tau$, ns | Ref. | Method | $\tau$, ns | Ref. | Model | |
| 3 | 1 | 23.8(1.0) | 76MEL/LOM | Hanle | 30.4 | 90bQUA/DRE | NA | D |
| | | *36.1(2.1)* | 78CHI/DAL | Hanle | | | | C |
| | | *39.5(1.9)* | 79DAY/AND | DC | | | | D |
| | | **38(2)** | PW | Recom. | | | | B |
| | 2 | 40.7(2.4) | 78CHI/DAL | Hanle | | | | |
| | 3 | *46.2(2.8)* | 78CHI/DAL | Hanle | | | | |
| | | *40.4(3.2)* | 79DAY/AND | DC | | | | |
| | | **43(3)** | PW | Recom. | | | | |
| | 4,5 | 25(2) | 88SAN/CAM | DC | | | | |
| 4 | 1 | *50.4(4.0)* | 79DAY/AND | DC | 36.1 | 90bQUA/DRE | NA | D |
| | | *40.4(2.8)* | 88SAN/CAM | DC | | | | B |
| | | **45(5)** | PW | Recom. | | | | B |
| | 3 | 55.7(4.4) | 79DAY/AND | DC | | | | |
| 5 | 1 | *47.3(2.8)* | 78CHI/DAL | Hanle | 27.5 | 90bQUA/DRE | NA | D |
| | | 48.9(4.0)[e] | 79DAY/AND | DC | | | | D |
| | | **48(4)** | PW | Recom. | | | | D |
| | 2,3 | 48.9(4.0)[e] | 79DAY/AND | DC | | | | |
| 6 | 0 | 69.4(2.0) | 05AIT/OGI | DC | 52.3 | 90bQUA/DRE | NA | D |
| | 1 | *64.1(5.1)* | 79DAY/AND | DC | 51.3 | 90bQUA/DRE | NA | C |
| | | *67.1(1.0)* | 05AIT/OGI | DC | | | | D |
| | | **67(3)** | PW | Recom. | | | | |
| | 2 | 51.8(4.1)[e] | 79DAY/AND | DC | | | | |
| | | **59.1(1.2)** | 05AIT/OGI | DC | | | | |
| | 3 | 51.8(4.1)[e] | 79DAY/AND | DC | | | | |
| | | **51.8(0.6)** | 05AIT/OGI | DC | | | | |
| | 4 | 51.8(4.1)[e] | 79DAY/AND | DC | | | | |
| | | **43.2(0.2)** | 05AIT/OGI | DC | | | | |
| | 5 | 29.1(0.2) | 05AIT/OGI | DC | | | | |
| 7 | 0 | 94.4(2.8) | 05AIT/OGI | DC | 63.8 | 90bQUA/DRE | NA | D |
| | 1 | *29.2(1.8)* | 78CHI/DAL | Hanle | 21.7 | 90bQUA/DRE | NA | D |
| | | *32.1(2.6)* | 79DAY/AND | DC | | | | D |
| | | *30.9(1.8)* | 05AIT/OGI | DC | | | | D |
| | | **31(2)** | PW | Recom. | | | | D |
| | 2 | 39.7(0.6) | 05AIT/OGI | DC | | | | |
| | 3 | 44.3(1.3) | 05AIT/OGI | DC | | | | |
| | 4 | 45.1(0.3) | 05AIT/OGI | DC | | | | |
| | 5 | 53.6(0.9) | 05AIT/OGI | DC | | | | |
| 8 | 0 | 135.2(9.1) | 92TSU/SHI | DC | 125.3 | 06FAN/WÜN | AA | B |
| | 1 | 177.2(11.4) | 92TSU/SHI | DC | $\approx$155 | 92TSU/SHI | NA | B |

| $H^1\Sigma_g^+$,$v$,$N$ rovibronic levels | | | | | | | |
|---|---|---|---|---|---|---|---|
| $v$ | $N$ | Experimental studies | | | Non-empirical calculation | | | O-C |
| | | $\tau$, ns | Ref. | Method | $\tau$, ns | Ref. | Model | |
| 0 | 1 | 143.1(6.0) | 92TSU/ISH | DC | 140.3 | 06ROS/YOS | NA | A |
| | 2 | 121.9(5.1) | 92TSU/ISH | DC | 140.1 | 06ROS/YOS | NA | D |
| | 3 | 101.9(7.9) | 92TSU/ISH | DC | 139.8 | 06ROS/YOS | NA | D |
| | 4 | 108.7(8.5) | 92TSU/ISH | DC | 139.6 | 06ROS/YOS | NA | D |
| 2 | 0 | 104.8(3.5) | 05AIT/OGI | DC | 92.1 | 06ROS/YOS | NA | D |
| | 1 | 95.2(3.3) | 05AIT/OGI | DC | 91.3 | 06ROS/YOS | NA | B |
| | 2 | 88.6(2.1) | 05AIT/OGI | DC | 90.3 | 06ROS/YOS | NA | A |
| | 3 | 82.9(2.1) | 05AIT/OGI | DC | 90.3 | 06ROS/YOS | NA | D |
| | 5 | 72.6(0.5) | 05AIT/OGI | DC | 100.8 | 06ROS/YOS | NA | D |

| $\bar{H}\,^1\Sigma_g^+$,$v$,$N$ rovibronic levels | | | | | | | |
|---|---|---|---|---|---|---|---|
| $v$ | $N$ | Experimental studies | | | Non-empirical calculation | | | O-C |
| | | $\tau$, ns | Ref. | Method | $\tau$, ns | Ref. | Model | |
| 6 | 1 | *28(3)* | 00REI/HOG | TF | 37.0 | 06ROS/YOS | NA | D |
| | | *30.7(0.9)* | 06ROS/YOS | DC | | | | D |
| | | **29(2)** | PW | Recom. | | | | |
| | 2 | 32.8(0.8) | 06ROS/YOS | DC | 44.6 | 06ROS/YOS | NA | D |
| | 3 | *30(3)* | 00REI/HOG | TF | 61.6 | 06ROS/YOS | NA | D |
| | | *33.1(0.7)* | 06ROS/YOS | DC | | | | D |
| | | **32(2)** | PW | Recom. | | | | |
| 7 | 1 | *19(2)* | 00REI/HOG | TF | 43.2 | 06ROS/YOS | NA | D |
| | | *22.2(0.4)* | 06ROS/YOS | DC | | | | D |
| | | **21(2)** | PW | Recom. | | | | |
| | 2 | 28.0(0.7) | 06ROS/YOS | DC | 56.5 | 06ROS/YOS | NA | D |
| | 3 | *27(3)* | 00REI/HOG | TF | 80.0 | 06ROS/YOS | NA | D |
| | | *33.7(0.5)* | 06ROS/YOS | DC | | | | D |
| | | **31(3)** | PW | Recom. | | | | |
| 8 | 0 | 22(2) | 00REI/HOG | TF | 57.5 | 06ROS/YOS | NA | D |
| | 1 | *19(2)* | 00REI/HOG | TF | 66.8 | 06ROS/YOS | NA | D |
| | | *24.0(0.3)* | 06ROS/YOS | DC | | | | D |
| | | **22(2)** | PW | Recom. | | | | |
| | 2 | *28(3)* | 00REI/HOG | TF | 81.5 | 06ROS/YOS | NA | D |
| | | *30.3(0.3)* | 06ROS/YOS | DC | | | | D |
| | | **29(2)** | PW | Recom. | | | | |
| | 3 | *32(3)* | 00REI/HOG | TF | 73.4 | 06ROS/YOS | NA | D |
| | | *37.6(1.1)* | 06ROS/YOS | DC | | | | D |
| | | **35(2)** | PW | Recom. | | | | |

Table 6 – Continued.

| v | N | Experimental studies | | | Non-empirical calculation | | | O-C |
|---|---|---|---|---|---|---|---|---|
| | | τ, ns | Ref. | Method | τ, ns | Ref. | Model | |
| 9 | 0 | 28(3) | 00REI/HOG | TF | 74.8 | 06ROS/YOS | NA | D |
| | 1 | *32(3)* | 00REI/HOG | TF | 69.1 | 06ROS/YOS | NA | D |
| | | *34.4(1.0)* | 06ROS/YOS | DC | | | | D |
| | | **33(2)** | PW | Recom. | | | | |
| | 2 | *30(3)* | 00REI/HOG | TF | 50.8 | 06ROS/YOS | NA | D |
| | | *31.7(1.0)* | 06ROS/YOS | DC | | | | D |
| | | **31(2)** | PW | Recom. | | | | |
| | 3 | *42(4)* | 00REI/HOG | TF | 29.5 | 06ROS/YOS | NA | D |
| | | *45.0(0.5)* | 06ROS/YOS | DC | | | | D |
| | | **44(2)** | PW | Recom. | | | | |
| 10 | 0 | 41(4) | 00REI/HOG | TF | 21.0 | 06ROS/YOS | NA | D |
| | 1 | *29(3)* | 00REI/HOG | TF | 5.4 | 06ROS/YOS | NA | D |
| | | *33.3(0.4)* | 06ROS/YOS | DC | | | | D |
| | | **31(2)** | PW | Recom. | | | | |
| | 2 | *33(3)* | 00REI/HOG | TF | 10.1 | 06ROS/YOS | NA | D |
| | | *38.1(0.8)* | 06ROS/YOS | DC | | | | D |
| | | *36(2)* | PW | Recom. | | | | |
| | 3 | *27(3)* | 00REI/HOG | TF | 10.6 | 06ROS/YOS | NA | D |
| | | *25.9(0.6)* | 06ROS/YOS | DC | | | | D |
| | | **26(2)** | PW | Recom. | | | | |
| 11 | 0 | 23(2) | 00REI/HOG | TF | 10.4 | 06ROS/YOS | NA | D |
| | 1 | *18(2)* | 00REI/HOG | TF | 9.2 | 06ROS/YOS | NA | D |
| | | *20.3(0.6)* | 06ROS/YOS | DC | | | | D |
| | | **19 (1)** | PW | Recom. | | | | |
| | 2 | *14(1)* | 00REI/HOG | TF | 7.5 | 06ROS/YOS | NA | D |
| | | *16.6(0.5)* | 06ROS/YOS | DC | | | | D |
| | | **15(2)** | PW | Recom. | | | | |
| | 3 | *16(2)* | 00REI/HOG | TF | 5.8 | 06ROS/YOS | NA | D |
| | | *18.7(0.4)* | 06ROS/YOS | DC | | | | D |
| | | **17(2)** | PW | Recom. | | | | |
| 12 | 0 | 22(2) | 00REI/HOG | TF | 4.0 | 06ROS/YOS | NA | D |
| | 1 | 8(1) | 00REI/HOG | TF | 3.9 | 06ROS/YOS | NA | D |
| | 2 | 13(1) | 00REI/HOG | TF | 3.8 | 06ROS/YOS | NA | D |
| | 3 | 10(1) | 00REI/HOG | TF | 3.5 | 06ROS/YOS | NA | D |
| 14 | 0 | 0.021(4) | 00REI/HOG | TF | 0.0158 | 06ROS/YOS | NA | B |
| | 1 | 0.040(8) | 00REI/HOG | TF | 0.0098 | 06ROS/YOS | NA | D |
| | 2 | 0.034(7) | 00REI/HOG | TF | 0.0016 | 06ROS/YOS | NA | C |
| | 3 | 0.027(5) | 00REI/HOG | TF | 0.0037 | 06ROS/YOS | NA | B |
| | 4 | 0.048(10) | 00REI/HOG | TF | 0.0400 | 06ROS/YOS | NA | A |
| | 5 | 0.021(4) | 00REI/HOG | TF | 0.1305 | 06ROS/YOS | NA | C |
| 15 | 0 | 0.0079(16) | 00REI/HOG | TF | 0.0129 | 06ROS/YOS | NA | D |
| | 1 | 0.0037(7) | 00REI/HOG | TF | 0.0125 | 06ROS/YOS | NA | B |
| | 2 | 0.0055(11) | 00REI/HOG | TF | 0.0110 | 06ROS/YOS | NA | D |
| | 3 | 0.0056(11) | 00REI/HOG | TF | 0.0076 | 06ROS/YOS | NA | B |

$\overline{H}\ ^1\Sigma_g^+$,v,N rovibronic levels

| | | Experimental studies | | | Non-empirical calculation | | | O-C |
|---|---|---|---|---|---|---|---|---|
| $v$ | $N$ | $\tau$, ns | Ref. | Method | $\tau$, ns | Ref. | Model | |
| 0 | 1 | **20(2)** | 76GOM/CAM | DC | 16.0 | 92TSU/ISH | NA | B |
| | | ≈9 | 78CHI/DAL | Hanle | | | | |
| | | **21.7(1.7)** | 79DAY/AND | DC | | | | D |
| | | 36(6) | 80BRY/KOT | Hanle | | | | D |
| | | **23.3(1.6)** | 88SAN/CAM | DC | | | | D |
| | | **20.1(2.0)** | 92TSU/ISH | DC | | | | C |
| | | **21(2)** | PW | Recom. | | | | |
| | 2 | **20(2)** | 76GOM/CAM | DC | 16.4 | 92TSU/ISH | NA | B |
| | | 99(6) | 78CHI/DAL | Hanle | | | | D |
| | | **21.7(1.7)** | 79DAY/AND | DC | | | | D |
| | | **23.3(1.6)** | 88SAN/CAM | DC | | | | D |
| | | **20.6(1.6)** | 92TSU/ISH | DC | | | | C |
| | | **21(2)** | PW | Recom. | | | | |
| | 3 | **20(2)** | 76GOM/CAM | DC | 17.9 | 92TSU/ISH | NA | B |
| | | 85(10) | 80BRY/KOT | Hanle | | | | D |
| | | **25.3(1.6)** | 88SAN/CAM | DC | | | | D |
| | | **19.1(1.6)** | 92TSU/ISH | DC | | | | A |
| | | **21(2)** | PW | Recom. | | | | |
| | 4 | 21.8(1.9) | 92TSU/ISH | DC | 19.0 | 92TSU/ISH | NA | B |
| | 5 | ≈120 | 80BRY/KOT | Hanle | | | | |
| 1 | 1 | **20.0(1.5)** | 76GOM/CAM | DC | 18.0 | 90aQUA/DRE | NA | B |
| | | **20.9(1.4)** | 88SAN/CAM | DC | | | | C |
| | | **20.5(1.5)** | PW | Recom. | | | | |
| | 4 | **20.0(1.5)** | 76GOM/CAM | DC | | | | |
| | | **20.9(1.4)** | 88SAN/CAM | DC | | | | |
| | | **20.5(1.5)** | PW | Recom. | | | | |
| 3 | 1 | 33.0(1.7) | 88SAN/CAM | DC | 37.4 | 90aQUA/DRE | NA | C |
| | | 40.9(0.7) | 05AIT/OGI | DC | | | | D |
| | 2 | 35.7(0.9) | 05AIT/OGI | DC | | | | |
| | 3 | 54.4(3.3)[f] | 78CHI/DAL | Hanle | | | | |
| | | 42.1(0.8) | 05AIT/OGI | DC | | | | |
| | 4 | 43.1(0.7) | 05AIT/OGI | DC | | | | |
| 4[g] | 1 | 1.56(16)×10³ | 10ROS/AND | DC | 1.84×10³ | 10ROS/AND | AA | B |
| | 2 | 1.77(19)×10³ | 10ROS/AND | DC | 1.90×10³ | 10ROS/AND | AA | A |
| | 3 | 1.75(16)×10³ | 10ROS/AND | DC | 1.99×10³ | 10ROS/AND | AA | B |

| | | $I^1\Pi_g^-$, v,N rovibronic levels | | | | | | | | |
|---|---|---|---|---|---|---|---|---|---|---|
| v | N | Experimental studies | | | Semi-empirical determination | | | Non-empirical calculation | | | O-C |
| | | τ, ns | Ref. | Method | τ, ns | Ref. | Model | τ, ns | Ref. | Method | |
| 0 | 1 | *21.3(1.7)*[e] | 79DAY/AND | DC | 19(2) | 97AST/KOK | NA | 14.77 | 00bAST/LAV | NA | D |
| | | *18.6(1.9)* | 92TSU/ISH | DC | | | | | | | C |
| | | **20(2)** | PW | Recom. | | | | | | | |
| | 2 | 38.5(2.3) | 78CHI/DAL | Hanle | 20(2) | 97AST/KOK | NA | 15.15 | 00bAST/LAV | NA | D |
| | | *21.3(1.7)*[e] | 79DAY/AND | DC | | | | | | | D |
| | | 34(6) | 80BRY/KOT | Hanle | | | | | | | D |
| | | *18.7(1.2)* | 92TSU/ISH | DC | | | | | | | C |
| | | **20(2)** | PW | Recom. | | | | | | | |
| | 3 | 20.4(1.6) | 92TSU/ISH | DC | 20(2) | 97AST/KOK | NA | 15.45 | 00bAST/LAV | NA | D |
| | 4 | 74(28) | 80BRY/KOT | Hanle | 21(2) | 97AST/KOK | NA | 15.71 | 00bAST/LAV | NA | C |
| | 5 | | | | 21(2) | 97AST/KOK | NA | 15.93 | 00bAST/LAV | NA | |
| | 6 | 100(30) | 80BRY/KOT | Hanle | 22(2) | 97AST/KOK | NA | 16.18 | 00bAST/LAV | NA | C |
| 1 | 1 | 20.0(1.5) | 76GOM/CAM | DC | 20(2) | 97AST/KOK | NA | 15.04 | 00bAST/LAV | NA | D |
| | 2 | 40.2(2.4) | 78CHI/DAL | Hanle | 20(2) | 97AST/KOK | NA | 15.13 | 00bAST/LAV | NA | D |
| | 3 | 29.4(1.8) | 78CHI/DAL | Hanle | 20(2) | 97AST/KOK | NA | 15.34 | 00bAST/LAV | NA | D |
| | 4 | *20.0(1.5)* | 76GOM/CAM | DC | 21(2) | 97AST/KOK | NA | 15.54 | 00bAST/LAV | NA | C |
| | | *21.5(1.5)* | 88SAN/CAM | DC | | | | | | | D |
| | | **20.8(1.5)** | PW | Recom. | | | | | | | |
| | 5 | | | | 21(2) | 97AST/KOK | NA | 15.72 | 00bAST/LAV | NA | |
| | 6 | | | | 21(2) | 97AST/KOK | NA | 15.92 | 00bAST/LAV | NA | |
| 2 | 1 | | | | 21(2) | 97AST/KOK | NA | 15.44 | 00bAST/LAV | NA | |
| | 2 | | | | 21(2) | 97AST/KOK | NA | 15.54 | 00bAST/LAV | NA | |
| | 3 | | | | 21(2) | 97AST/KOK | NA | 15.67 | 00bAST/LAV | NA | |
| | 4 | 21.0(1.5) | 76GOM/CAM | DC | 22(2) | 97AST/KOK | NA | 15.81 | 00bAST/LAV | NA | D |
| 3 | 1 | 21.3(0.2) | 05AIT/OGI | DC | 24(2) | 97AST/KOK | NA | 17.40 | 00bAST/LAV | NA | D |
| | 2 | 21.5(0.1) | 05AIT/OGI | DC | 24(2) | 97AST/KOK | NA | 17.47 | 00bAST/LAV | NA | D |
| | 3 | 21.9(0.3) | 05AIT/OGI | DC | 25(2) | 97AST/KOK | NA | 17.61 | 00bAST/LAV | NA | D |
| | 4 | 22.0(0.3) | 05AIT/OGI | DC | | | | | | | |
| 4[g] | 1 | $1.95(6){\times}10^3$ | 10ROS/AND | DC | | | | $1.84{\times}10^3$ | 10ROS/AND | AA | B |
| | 2 | $1.93(12){\times}10^3$ | 10ROS/AND | DC | | | | $1.90{\times}10^3$ | 10ROS/AND | AA | A |

| | | $J^1\Delta_g^+,v,N$ rovibronic levels | | | | | | |
|---|---|---|---|---|---|---|---|---|
| $v$ | $N$ | Experimental studies | | | Non-empirical calculation | | | O-C |
| | | $\tau$, ns | Ref. | Method | $\tau$, ns | Ref. | Model | |
| 0 | 2 | 23.2(1.9) | 92TSU/ISH | DC | 22.0 | 92TSU/ISH | NA | A |
| | 3 | 24.3(2.0) | 92TSU/ISH | DC | 21.0 | 92TSU/ISH | NA | B |
| | 4 | 24.0(3.5) | 92TSU/ISH | DC | 18.9 | 92TSU/ISH | NA | B |
| 2 | 5 | 32.0(0.4) | 05AIT/OGI | DC | | | | |

| | | $J^1\Delta_g^-,v,N$ rovibronic levels | | | | | | | | | |
|---|---|---|---|---|---|---|---|---|---|---|---|
| $v$ | $N$ | Experimental studies | | | Semi-empirical determination | | | Non-empirical calculation | | | O-C |
| | | $\tau$, ns | Ref. | Method | O-C | Ref. | Model | $\tau$, ns | Ref. | Method | |
| 0 | 2 | 34.3(2.4) | 88SAN/CAM | DC | 27(2) | 97AST/KOK | NA | 17.97 | 00bAST/LAV | NA | D |
| | | **25.9(2.1)** | 92TSU/ISH | DC | | | | | | | D |
| | 3 | 26.2(1.8) | 92TSU/ISH | DC | 25(2) | 97AST/KOK | NA | 17.48 | 00bAST/LAV | NA | D |
| | 4 | | | | 25(2) | 97AST/KOK | NA | 17.07 | 00bAST/LAV | NA | |
| | 5 | | | | 24(2) | 97AST/KOK | NA | 16.71 | 00bAST/LAV | NA | |
| | 6 | | | | 23(2) | 97AST/KOK | NA | 16.31 | 00bAST/LAV | NA | |
| 1 | 2 | | | | 28(2) | 97AST/KOK | NA | 18.62 | 00bAST/LAV | NA | |
| | 3 | | | | 27(2) | 97AST/KOK | NA | 18.08 | 00bAST/LAV | NA | |
| | 4 | | | | 27(2) | 97AST/KOK | NA | 17.72 | 00bAST/LAV | NA | |
| | 5 | | | | 26(2) | 97AST/KOK | NA | 17.38 | 00bAST/LAV | NA | |
| | 6 | | | | 25(2) | 97AST/KOK | NA | 17.06 | 00bAST/LAV | NA | |
| 2 | 2 | | | | 30(3) | 97AST/KOK | NA | 18.87 | 00bAST/LAV | NA | |
| | 3 | | | | 29(3) | 97AST/KOK | NA | 18.64 | 00bAST/LAV | NA | |
| | 4 | | | | 29(2) | 97AST/KOK | NA | 18.45 | 00bAST/LAV | NA | |
| 3 | 2 | | | | 31(3) | 97AST/KOK | NA | 19.68 | 00bAST/LAV | NA | |
| | 3 | | | | 31(3) | 97AST/KOK | NA | 19.62 | 00bAST/LAV | NA | |

$O^1\Sigma_g^+,v,N$ rovibronic levels

| $v$ | $N$ | Experimental studies | | | Non-empirical calculation | | | O-C |
|---|---|---|---|---|---|---|---|---|
| | | $\tau$, ns | Ref. | Method | $\tau$, ns | Ref. | Model | |
| 0 | 0 | 172.8(7.4) | 05AIT/OGI | DC | 404.0 | 06FAN/WÜN | AA | D |
| | 1 | 193.7(13.2) | 05AIT/OGI | DC | | | | |
| | 3 | 68.8(0.9) | 05AIT/OGI | DC | | | | |

$P^1\Sigma_g^+,v,N$ rovibronic levels

| $v$ | $N$ | Experimental studies | | | Non-empirical calculation | | | O-C |
|---|---|---|---|---|---|---|---|---|
| | | $\tau$, ns | Ref. | Method | $\tau$, ns | Ref. | Model | |
| 0 | 1,5 | 51.3(3.6) | 88SAN/CAM | DC | 47.3[a] | 06FAN/WÜN | AA | B |
| 1 | 0 | 43.9(1.5) | 05AIT/OGI | DC | 58.6 | 06FAN/WÜN | AA | D |
| | 1 | 52.6(1.9) | 05AIT/OGI | DC | | | | |
| | 2 | 57.8(1.3) | 05AIT/OGI | DC | | | | |
| | 3 | 47.4(1.6) | 05AIT/OGI | DC | | | | |
| | 4 | 45.4(0.9) | 05AIT/OGI | DC | | | | |
| | 5 | 44.0(1.6) | 05AIT/OGI | DC | | | | |

$R^1\Pi_g^+,v,N$ rovibronic levels

| $v$ | $N$ | Experimental studies | | | Non-empirical calculation | | | O-C |
|---|---|---|---|---|---|---|---|---|
| | | $\tau$, ns | Ref. | Method | $\tau$, ns | Ref. | Model | |
| 0 | 1 | 49.2(1.4) | 05AIT/OGI | DC | 110.8[a] | 06FAN/WÜN | AA | D |
| | 2 | 48.0(0.4) | 05AIT/OGI | DC | | | | |
| | 3 | 84.0(1.5) | 05AIT/OGI | DC | | | | |

$R^1\Pi_g^-,v,N$ rovibronic level

| $v$ | $N$ | Experimental studies | | | Non-empirical calculation | | | O-C |
|---|---|---|---|---|---|---|---|---|
| | | $\tau$, ns | Ref. | Method | $\tau$, ns | Ref. | Model | |
| 0 | 1 | 43.7(0.6) | 05AIT/OGI | DC | 110.8[a] | 06FAN/WÜN | AA | D |

$S^1\Delta_g^+,v,N$ rovibronic level

| $v$ | $N$ | Experimental studies | | | Non-empirical calculation | | | O-C |
|---|---|---|---|---|---|---|---|---|
| | | $\tau$, ns | Ref. | Method | $\tau$, ns | Ref. | Model | |
| 0 | 4 | 52.2(0.2) | 05AIT/OGI | DC | 48.3[a] | 06FAN/WÜN | AA | D |

$S^1\Delta_g^-,v,N$ rovibronic levels

| $v$ | $N$ | Experimental studies | | | Non-empirical calculation | | | O-C |
|---|---|---|---|---|---|---|---|---|
| | | $\tau$, ns | Ref. | Method | $\tau$, ns | Ref. | Model | |
| 0 | 3 | 47.1(0.4) | 05AIT/OGI | DC | 48.3[a] | 06FAN/WÜN | AA | C |
| | 4 | 46.9(1.2) | 05AIT/OGI | DC | | | | |

| $a^3\Sigma_g^+, v$ vibronic levels | | | | | | | |
|---|---|---|---|---|---|---|---|
| $v$ | | Experimental studies | | Non-empirical calculation | | | O-C |
| | $\tau$, ns | Ref. | Method | $\tau$, ns | Ref. | Model | |
| | 35(8)[h] | 66FOW/HOL | PLASMA | | | | |
| | 10.45(25)[h] | 75KIN/REA | DC | | | | |
| | 11.4(8)[h] | 81BRE/GOD | DC | | | | |
| 0 | *11.0(4)* | 71IMH/REA | DC | 11.63 | 06FAN/WÜN | AA | B |
| | 26(2) | 72THO/FOW | DC | | | | D |
| | *11.9(1.2)* | 72SMI/CHE | PS | | | | A |
| | *9.94(39)* | 79MOH/KIN | DC | | | | D |
| | *11.1(3)* (N=1) | 88WED/PHE | DC | | | | B |
| | **11.0(6)** | PW | Recom. | | | | |
| 1 | *10.6(6)* | 71IMH/REA | DC | 10.21 | 06FAN/WÜN | AA | A |
| | *10.8(1.1)* | 72SMI/CHE | PS | | | | A |
| | *9.1(1.0)* | 79MOH/KIN | DC | | | | B |
| | *10.4(3)* (N=1) | 88WED/PHE | DC | | | | A |
| | **10.2(4)** | PW | Recom. | | | | |
| 2 | 10(2) | 72SMI/CHE | PS | 9.19 | 06FAN/WÜN | AA | A |
| 3 | 10(2) | 72SMI/CHE | PS | 8.44 | 06FAN/WÜN | AA | A |
| 0÷4 | 9.62(20)[i] | 79MOH/KIN | DC | | | | |

| $c^3\Pi_u^+, v, N$ rovibronic levels | | | | | | | | |
|---|---|---|---|---|---|---|---|---|
| $v$ | $N$ | Experimental studies | | | Non-empirical calculation | | | O-C |
| | | $\tau$, ns | Ref. | Method | $\tau$, ns | Ref. | Model | |
| 0 | | <10[a] | 77VOG/MEI | TF | | | | |
| | 1 | 6.2(5) | 84BRU/NEU | TF | 6.38 | 85COM/BRU | NA | A |
| | 2 | 2.1(2) | 84BRU/NEU | TF | 2.25 | 85COM/BRU | NA | A |
| 1 | 1 | 2.7(8) | 84BRU/NEU | TF | 1.44 | 85COM/BRU | NA | D |
| | 2 | 0.9(3) | 84BRU/NEU | TF | 0.509 | 85COM/BRU | NA | D |
| 2 | 1 | 2.1(8) | 84BRU/NEU | TF | 0.569 | 85COM/BRU | NA | D |
| | 2 | 0.7(3) | 84BRU/NEU | TF | 0.200 | 85COM/BRU | NA | D |
| 3 | 1 | 1.8(8) | 84BRU/NEU | TF | 0.295 | 85COM/BRU | NA | D |
| | 2 | 0.6(3) | 84BRU/NEU | TF | 0.103 | 85COM/BRU | NA | D |
| 4 | 1 | <1.5 | 84BRU/NEU | TF | 0.182 | 85COM/BRU | NA | |
| | 2 | <0.5 | 84BRU/NEU | TF | 0.0637 | 85COM/BRU | NA | |
| 5 | 1 | <1.5 | 84BRU/NEU | TF | 0.126 | 85COM/BRU | NA | |
| | 2 | <0.5 | 84BRU/NEU | TF | 0.0441 | 85COM/BRU | NA | |
| 6 | 1 | <1.5 | 84BRU/NEU | TF | 0.0954 | 85COM/BRU | NA | |
| | 2 | <0.5 | 84BRU/NEU | TF | 0.0333 | 85COM/BRU | NA | |
| 10 | 1 | <1.5 | 84BRU/NEU | TF | 0.0541 | 85COM/BRU | NA | |
| | 2 | <0.5 | 84BRU/NEU | TF | 0.0190 | 85COM/BRU | NA | |
| 15 | 1 | <1.5 | 84BRU/NEU | TF | 0.0831 | 85COM/BRU | NA | |
| | 2 | <0.5 | 84BRU/NEU | TF | 0.0292 | 85COM/BRU | NA | |

| v | N | Experimental studies | | | Non-empirical calculation | | | O-C |
|---|---|---|---|---|---|---|---|---|
| | | τ, ms | Ref. | Method | τ, ms | Ref. | Model | |
| 0 | 1 | 0.1÷0.5 | 62LIC | TF | 1.321 | 88CHO/FLE | AA | |
| | | 1.02(5) | 72JOH | TF | | | | D |
| | 1(J=1) | 0.145(50) | 94BER/OTT[j] | TF | | | | |
| | 1 (J=0,2) | >0.330 | 94BER/OTT[j] | TF | | | | |
| | | | | | 1.76 (J=0) | 79CHO/BHA | NA | |
| | | | | | 0.12 (J=1) | 79CHO/BHA | NA | |
| | | | | | 1.3 (J=2) | 79CHO/BHA | NA | |
| | 2 | 0.1÷0.5 | 62LIC | TF | | | | |
| | | 1.02(5) | 72JOH | TF | | | | |
| | 2 (J=2) | 0.203(34) | 94BER/OTT[j] | TF | 0.11 (J=2) | 79CHO/BHA | NA | C |
| | 2 (J=1,3) | 1.02(5) | 94BER/OTT[j] | TF | 1.32 (J=1) | 79CHO/BHA | NA | D |
| | | | | | 1.31 (J=3) | 79CHO/BHA | NA | D |
| | 3 (J=3) | 0.113(36) | 94BER/OTT[j] | TF | | | | |
| | 3 (J=2,4) | >0.250 | 94BER/OTT[j] | TF | | | | |
| | 4 (J=4) | >0.130 | 94BER/OTT[j] | TF | | | | |
| | 4 (J=3,5) | >0.170 | 94BER/OTT[j] | TF | | | | |
| 1 | 1 (J=1) | 0.0420(47) | 94BER/OTT[j] | TF | 0.166[a] | 92GUB/DAL | AA | D |
| | 1(J=0,2) | >0.530 | 94BER/OTT[j] | TF | | | | |
| | 2 (J=2) | 0.0340(35) | 94BER/OTT[j] | TF | | | | |
| | 2 (J=1,3) | 0.129(14) | 94BER/OTT[j] | TF | | | | |
| | 3 (J=3) | 0.0380(79) | 94BER/OTT[j] | TF | | | | |
| | 3 (J=2,4) | 0.129(28) | 94BER/OTT[j] | TF | | | | |
| | 4 (J=4) | 0.041(10) | 94BER/OTT[j] | TF | | | | |
| | 4 (J=3,5) | 0.099(14) | 94BER/OTT[j] | TF | | | | |
| 2 | 1 (J=1) | 0.0140(37) | 94BER/OTT[j] | TF | 0.100[a] | 92GUB/DAL | AA | D |
| | 1(J=0,2) | >0.170 | 94BER/OTT[j] | TF | | | | |
| | 2 (J=2) | 0.0160(24) | 94BER/OTT[j] | TF | | | | |
| | 2 (J=1,3) | 0.109(13) | 94BER/OTT[j] | TF | | | | |
| | 3 (J=2,4) | >0.110 | 94BER/OTT[j] | TF | | | | |
| | 4 (J=4) | 0.0170(47) | 94BER/OTT[j] | TF | | | | |
| | 4 (J=3,5) | 0.088(14) | 94BER/OTT[j] | TF | | | | |
| 3 | 1(J=0,2) | >0.300 | 94BER/OTT[j] | TF | 0.072[a] | 92GUB/DAL | AA | |
| | 2 (J=2) | 0.0100(35) | 94BER/OTT[j] | TF | | | | |
| | 2 (J=1,3) | 0.0680(74) | 94BER/OTT[j] | TF | | | | |
| | 3 (J=2,4) | >90 | 94BER/OTT[j] | TF | | | | |

| | | | | | | | | | | |
|---|---|---|---|---|---|---|---|---|---|---|
| | | $d^3\Pi_u^+$,$v$,$N$ rovibronic levels | | | | | | | | |
| $v$ | $N$ | Experimental studies | | | Semi-empirical determination | | | Non-empirical calculation | | O-C |
| | | $\tau$, ns | Ref. | Method | $\tau$, ns | Ref. | Model | $\tau$, ns | Ref. | Method |
| 0 | 1 | 31.0(3.0) | 72MAR/JOS | Hanle | 38.8(2.7) | 90POZ | NA | 38.8[a] | 06FAN/WÜN | AA | C |
| | | 29.4(3.2) | 73bFRE/MIL | MOMRIE | | | | | | C |
| | | 38.6(2.0) | 93KIY/SAT | DC | | | | | | A |
| | 2 | *39.2(1.9)* | 87KIY | DC | 38.9(2.7) | 90POZ | NA | | | |
| | | *36.0(2.0)* | 93KIY/SAT | DC | | | | | | |
| | | **38(2)** | PW | Recom. | | | | | | |
| | 3 | 34.0(2.5) | 93KIY/SAT | DC | 39.0(2.7) | 90POZ | NA | | | |
| | 4 | | | | 39.2(2.7) | 90POZ | NA | | | |
| | 5 | | | | 39.5(2.8) | 90POZ | NA | | | |
| | 6 | | | | 39.8(2.8) | 90POZ | NA | | | |
| 1 | 1 | 31.0(3.0) | 72MAR/JOS | Hanle | 39.5(2.8) | 90POZ | NA | 39.1[a] | 06FAN/WÜN | AA | C |
| | | 29.4(3.2) | 73bFRE/MIL | MOMRIE | | | | | | D |
| | | *34.6(1.9)* | 93KIY/SAT | DC | | | | | | C |
| | 2 | *28.7(1.7)* | 87KIY | DC | 39.2(2.7) | 90POZ | NA | | | |
| | | *27.7(1.5)* | 93KIY/SAT | DC | | | | | | |
| | | **28(2)** | PW | Recom. | | | | | | |
| | 3 | 16.9(2.0) | 93KIY/SAT | DC | 35.2(2.5) | 90POZ | NA | | | |
| | 4 | | | | 30.4(2.1) | 90POZ | NA | | | |
| | 5 | | | | 33.1(2.3) | 90POZ | NA | | | |
| | 6 | | | | 38.1(2.7) | 90POZ | NA | | | |
| 2 | 1 | 31.0(3.0) | 72MAR/JOS | Hanle | 40.5(2.8) | 90POZ | NA | 39.1[a] | 06FAN/WÜN | AA | C |
| | | 29.4(3.2) | 73bFRE/MIL | MOMRIE | | | | | | D |
| | | 38.2(1.6) | 93KIY/SAT | DC | | | | | | A |
| | 2 | *33.7(2.0)* | 87KIY | DC | 40.5(2.8) | 90POZ | NA | | | |
| | | *31.7(1.4)* | 93KIY/SAT | DC | | | | | | |
| | | **33(2)** | PW | Recom. | | | | | | |
| | 3 | 36.4(1.7) | 93KIY/SAT | DC | 40.6(2.8) | 90POZ | NA | | | |
| | 4 | 35.1(1.6) | 93KIY/SAT | DC | 40.8(2.8) | 90POZ | NA | | | |
| | 5 | | | | 41.0(2.9) | 90POZ | NA | | | |
| | 6 | | | | 41.3(2.9) | 90POZ | NA | | | |
| 3 | 1 | 31.0(3.0) | 72MAR/JOS | Hanle | 41.6(2.9) | 90POZ | NA | 39.7[a] | 06FAN/WÜN | AA | C |
| | | 29.4(3.2) | 73bFRE/MIL | MOMRIE | | | | | | D |
| | | 36.6(1.8) | 93KIY/SAT | DC | | | | | | C |
| | 2 | *38.8(2.2)* | 87KIY | DC | 41.6(2.9) | 90POZ | NA | | | |
| | | *39.5(1.5)* | 93KIY/SAT | DC | | | | | | |
| | | **39(2)** | PW | Recom. | | | | | | |
| | 3 | | | | 41.7(2.9) | 90POZ | NA | | | |
| | 4 | | | | 41.9(2.9) | 90POZ | NA | | | |
| | 5 | | | | 42.1(2.9) | 90POZ | NA | | | |
| | 6 | | | | 42.4(3.0) | 90POZ | NA | | | |

| | | $d^3\Pi_u^-$,v,N rovibronic levels | | | | | | | | |
|---|---|---|---|---|---|---|---|---|---|---|
| $v$ | $N$ | Experimental studies | | | Semi-empirical determination | | | Non-empirical calculation | | | O-C |
| | | $\tau$, ns | Ref. | Method | $\tau$, ns | Ref. | Model | $\tau$, ns | Ref. | Method | |
| 0 | 1÷3 | 68(5) | 69CAH | DC | | | | 37.8 | 99KIY/SAT | NA | D |
| | 2 | 31.5(3.0) | 72MAR/JOS | Hanle | | | | | | | C |
| | 1 | 32.0(5.0) | 73aFRE/MIL | MOMRIE | | | | | | | B |
| | 1÷3 | 42.2(2.6) | 78DAY/AND | DC | | | | | | | B |
| | 1 | *39.4(2.0)* | 78DAY/AND | DC | 38.7(0.5) | 87DRA/LAV | AA | | | | A |
| | 2 | 44.2(2.7)[i] | 78DAY/AND | DC | 40.7(1.4) | 87DRA/LAV | AA | | | | C |
| | 3 | 43.3(3.0) | 78DAY/AND | DC | 40.7(1.4) | 87DRA/LAV | AA | | | | B |
| | 1,3 | 35(5) | 81BOG/EFR | DC | | | | | | | A |
| | 1 | *40.8(2.0)* | 87KIY | DC | | | | | | | B |
| | 2 | 41.8(3.0) | 87KIY | DC | | | | | | | B |
| | 1 | 30.0(2.1) | 88SAN/CAM | DC | | | | | | | D |
| | 1,3 | *39.5(1.2)* | 90BUR/LAV | DC | | | | | | | B |
| | 1 | *39.1(1.4)* | 93KIY/SAT | DC | | | | | | | A |
| | 1,3 | *39.7(1.7)* | PW | Recom. | | | | | | | |
| 1 | 1÷3 | 58(5) | 69CAH | DC | | | | 38.2 | 99KIY/SAT | NA | D |
| | 2 | 31.5(3.0) | 72MAR/JOS | Hanle | | | | | | | C |
| | 1 | 32.0(5.0) | 73aFRE/MIL | MOMRIE | | | | | | | B |
| | 1÷3 | 39.0(2.3) | 78DAY/AND | DC | | | | | | | A |
| | 1 | *36.5(1.9)* | 78DAY/AND | DC | 38.9(1.3) | 87DRA/LAV | AA | | | | A |
| | 2 | 42.0(2.5)[k] | 78DAY/AND | DC | 38.9(1.3) | 87DRA/LAV | AA | | | | B |
| | 3 | *38.8(2.6)* | 78DAY/AND | DC | 39.5(1.9) | 87DRA/LAV | AA | | | | A |
| | 2 | *40.7(1.9)* | 87KIY | DC | | | | | | | B |
| | 1 | 26.8(1.9) | 88SAN/CAM | DC | | | | | | | D |
| | 1,3 | *38.2(0.8)* | 90BUR/LAV | DC | | | | | | | A |
| | 1 | *37.7(1.3)* | 93KIY/SAT | DC | | | | | | | A |
| | 1,3 | *38.4(1.7)* | PW | Recom. | | | | | | | |
| 2 | 1÷3 | 62(5) | 69CAH | DC | | | | 38.8 | 99KIY/SAT | NA | D |
| | 2 | 31.5(3.0) | 72MAR/JOS | Hanle | | | | | | | C |
| | 1 | 32.0(5.0) | 73aFRE/MIL | MOMRIE | | | | | | | B |
| | 1÷3 | 39.0(2.3) | 78DAY/AND | DC | | | | | | | A |
| | 1 | *36.2(2.0)* | 78DAY/AND | DC | 39.5(1.9) | 87DRA/LAV | AA | | | | B |
| | 2 | 40.5(2.5)[i] | 78DAY/AND | DC | 39.5(1.9) | 87DRA/LAV | AA | | | | A |
| | 3 | *40.6(2.5)* | 78DAY/AND | DC | 39.5(1.9) | 87DRA/LAV | AA | | | | A |
| | 2 | *39.0(2.9)* | 87KIY | DC | | | | | | | A |
| | 1 | *22.3(1.6)* | 88SAN/CAM | DC | | | | | | | D |
| | 1,3 | *42.2(0.8)* | 90BUR/LAV | DC | | | | | | | D |
| | 1 | *37.8(1.2)* | 93KIY/SAT | DC | | | | | | | A |
| | 1,3 | *39.2(1.9)* | PW | Recom. | | | | | | | |

Table 6 – Continued.

| | | \multicolumn{9}{c}{$d^3\Pi_u^-,v,N$ rovibronic levels} | | | | | | | | |
|---|---|---|---|---|---|---|---|---|---|---|---|
| $v$ | $N$ | Experimental studies | | | Semi-empirical determination | | | Non-empirical calculation | | | O-C |
| | | $\tau$, ns | Ref. | Method | $\tau$, ns | Ref. | Model | $\tau$, ns | Ref. | Method | |
| 0 | 1 | | | | 38.5(2.7) | 90POZ | NA | | | | |
| | 2 | 33.9(3.8) | 99KIY/SAT | DC | 38.8(2.7) | 90POZ | NA | 30.7 | 99KIY/SAT | NA | A |
| | 3 | | | | 39.1(2.7) | 90POZ | NA | | | | |
| | 4 | | | | 39.6(2.8) | 90POZ | NA | | | | |
| | 5 | | | | 40.2(2.8) | 90POZ | NA | | | | |
| | 6 | | | | 40.9(2.8) | 90POZ | NA | | | | |
| 1 | 1 | | | | 44.0(3.1) | 90POZ | NA | | | | |
| | 2 | 30.3(1.1) | 99KIY/SAT | DC | 44.3(3.1) | 90POZ | NA | 24.6 | 99KIY/SAT | NA | D |
| | 3 | | | | 44.6(3.1) | 90POZ | NA | | | | |
| | 4 | | | | 45.2(3.2) | 90POZ | NA | | | | |
| | 5 | | | | 45.9(3.2) | 90POZ | NA | | | | |
| | 6 | | | | 46.7(3.3) | 90POZ | NA | | | | |
| 2 | 0 | 16.1(1.6) | 99KIY/SAT | DC | | | | 13.7 | 99KIY/SAT | NA | B |
| | 1 | | | | | | | | | | |
| | 2 | 11.9(1.4) | 99KIY/SAT | DC | | | | 14.4 | 99KIY/SAT | NA | B |
| | 3 | | | | | | | | | | |
| | 4 | | | | | | | | | | |
| | 5 | | | | | | | | | | |
| | 6 | | | | | | | | | | |
| 3 | 0 | 7.6(0.5) | 99KIY/SAT | DC | | | | 7.4 | 99KIY/SAT | NA | A |
| | 1 | | | | | | | | | | |
| | 2 | 10.4(0.7) | 99KIY/SAT | DC | | | | 7.8 | 99KIY/SAT | NA | D |
| | 3 | | | | | | | | | | |
| | 4 | | | | | | | | | | |
| | 5 | | | | | | | | | | |
| | 6 | | | | | | | | | | |
| 4 | 1 | 4.7(1.2) | 99KIY/SAT | DC | | | | 4.7 | 99KIY/SAT | NA | A |
| | 2 | 5.4(0.7) | 99KIY/SAT | DC | | | | 4.9 | 99KIY/SAT | NA | A |
| | 3 | 11.5(0.6) | 99KIY/SAT | DC | | | | 5.1 | 99KIY/SAT | NA | D |
| 5 | 2 | 4.5(1.5) | 99KIY/SAT | DC | | | | 3.3 | 99KIY/SAT | NA | A |

Table 6 – Continued.

| | | $e^3\Sigma_u^+, v, N$ rovibronic levels | | | | | | | | | |
|---|---|---|---|---|---|---|---|---|---|---|---|
| $v$ | $N$ | Experimental studies | | | Semi-empirical determination | | | Non-empirical calculation | | | O-C |
| | | τ, ns | Ref. | Method | τ, ns | Ref. | Model | τ, ns | Ref. | Method | |
| 0 | 1 | | | | 38.5(2.7) | 90POZ | NA | | | | |
| | 2 | 33.9(3.8) | 99KIY/SAT | DC | 38.8(2.7) | 90POZ | NA | 30.7 | 99KIY/SAT | NA | A |
| | 3 | | | | 39.1(2.7) | 90POZ | NA | | | | |
| | 4 | | | | 39.6(2.8) | 90POZ | NA | | | | |
| | 5 | | | | 40.2(2.8) | 90POZ | NA | | | | |
| | 6 | | | | 40.9(2.8) | 90POZ | NA | | | | |
| 1 | 1 | | | | 44.0(3.1) | 90POZ | NA | | | | |
| | 2 | 30.3(1.1) | 99KIY/SAT | DC | 44.3(3.1) | 90POZ | NA | 24.6 | 99KIY/SAT | NA | D |
| | 3 | | | | 44.6(3.1) | 90POZ | NA | | | | |
| | 4 | | | | 45.2(3.2) | 90POZ | NA | | | | |
| | 5 | | | | 45.9(3.2) | 90POZ | NA | | | | |
| | 6 | | | | 46.7(3.3) | 90POZ | NA | | | | |
| 2 | 0 | 16.1(1.6) | 99KIY/SAT | DC | | | | 13.7 | 99KIY/SAT | NA | B |
| | 1 | | | | | | | | | | |
| | 2 | 11.9(1.4) | 99KIY/SAT | DC | | | | 14.4 | 99KIY/SAT | NA | B |
| | 3 | | | | | | | | | | |
| | 4 | | | | | | | | | | |
| | 5 | | | | | | | | | | |
| | 6 | | | | | | | | | | |
| 3 | 0 | 7.6(0.5) | 99KIY/SAT | DC | | | | 7.4 | 99KIY/SAT | NA | A |
| | 1 | | | | | | | | | | |
| | 2 | 10.4(0.7) | 99KIY/SAT | DC | | | | 7.8 | 99KIY/SAT | NA | D |
| | 3 | | | | | | | | | | |
| | 4 | | | | | | | | | | |
| | 5 | | | | | | | | | | |
| | 6 | | | | | | | | | | |
| 4 | 1 | 4.7(1.2) | 99KIY/SAT | DC | | | | 4.7 | 99KIY/SAT | NA | A |
| | 2 | 5.4(0.7) | 99KIY/SAT | DC | | | | 4.9 | 99KIY/SAT | NA | A |
| | 3 | 11.5(0.6) | 99KIY/SAT | DC | | | | 5.1 | 99KIY/SAT | NA | D |
| 5 | 2 | 4.5(1.5) | 99KIY/SAT | DC | | | | 3.3 | 99KIY/SAT | NA | A |

| | | $h^3\Sigma_g^+$,$v$,$N$ rovibronic levels | | | | | | | | | |
|---|---|---|---|---|---|---|---|---|---|---|---|
| $v$ | $N$ | Experimental studies | | | Semi-empirical determination | | | Non-empirical calculation | | | O-C |
| | | $\tau$, ns | Ref. | Method | $\tau$, ns | Ref. | Model | $\tau$, ns | Ref. | Method | |
| 0 | 1 | 35.1(1.8) | 02PAZ/PUP | DC | 25(3) | PW | NA | 38.0 | 02PAZ/PUP | NA | B |
| | 2 | 48.8(3.0) | 81EYL/PIP | DC | 51(5) | PW | NA | 33.9 | 02PAZ/PUP | NA | D |
| | | 34.2(4.6) | 02PAZ/PUP | DC | | | | | | | A |
| | 3 | | | | 80(8) | PW | NA | | | | |
| | 4 | | | | 98(10) | PW | NA | | | | |
| 1 | 1 | 15.7(2.4) | 02PAZ/PUP | DC | 21(2) | PW | NA | 11.2 | 02PAZ/PUP | NA | B |
| | 2 | 17.0(1.6) | 81EYL/PIP | DC | 30(3) | PW | NA | 13.7 | 02PAZ/PUP | NA | C |
| | 3 | | | | 36(4) | PW | NA | | | | |
| | 4 | | | | 38(4) | PW | NA | | | | |
| | 5 | | | | 37(4) | PW | NA | | | | |
| | 6 | | | | 35(4) | PW | NA | | | | |
| 2 | 1 | | | | 27(3) | PW | NA | | | | |
| | 2 | | | | 20(2) | PW | NA | | | | |

| | | $g^3\Sigma_g^+$,$v$,$N$ rovibronic levels | | | | | | | | | |
|---|---|---|---|---|---|---|---|---|---|---|---|
| $v$ | $N$ | Experimental studies | | | Semi-empirical determination | | | Non-empirical calculation | | | O-C |
| | | $\tau$, ns | Ref. | Method | $\tau$, ns | Ref. | Model | $\tau$, ns | Ref. | Method | |
| 0 | 1 | 6.3(6) | 02PAZ/PUP | DC | 14(1) | PW | NA | 6.3 | 02PAZ/PUP | NA | A |
| | 2 | 11.1(5) | 81EYL/PIP | DC | 12(1) | PW | NA | 6.6 | 02PAZ/PUP | NA | D |
| | 3 | | | | 12(1) | PW | NA | | | | |
| | 4 | | | | 12(1) | PW | NA | | | | |
| | 5 | | | | 13(1) | PW | NA | | | | |
| 1 | 1 | 5.6(4) | 02PAZ/PUP | DC | 10(1) | PW | NA | 5.5 | 02PAZ/PUP | NA | A |
| | 2 | 11.5(4) | 81EYL/PIP | DC | 11(1) | PW | NA | 5.8 | 02PAZ/PUP | NA | D |
| | | 6.05(11) | 96RAY/LAF | DC | | | | | | | C |
| | | 5.6(1.0) | 02PAZ/PUP | DC | | | | | | | A |
| | 3 | | | | 11(1) | PW | NA | | | | |
| | 4 | | | | 11(1) | PW | NA | | | | |
| | 5 | | | | 11(1) | PW | NA | | | | |
| | 6 | | | | 11(1) | PW | NA | | | | |
| 2 | 1 | | | | 10(1) | PW | NA | | | | |
| | 2 | | | | 10(1) | PW | NA | | | | |
| | 3 | | | | 10(1) | PW | NA | | | | |
| | 4 | | | | 10(1) | PW | NA | | | | |
| | 5 | | | | 10(1) | PW | NA | | | | |
| 3 | 1 | | | | 9(1) | PW | NA | | | | |
| | 2 | | | | 9(1) | PW | NA | | | | |
| | 3 | | | | 9(1) | PW | NA | | | | |
| | 4 | | | | 9(1) | PW | NA | | | | |

Table 6 − Continued.

| | | | $i^3\Pi_g^+,v,N$ rovibronic levels | | | | | | | | |
|---|---|---|---|---|---|---|---|---|---|---|---|
| | | Experimental studies | | | Semi-empirical determination | | | Non-empirical calculation | | | O-C |
| $v$ | $N$ | τ, ns | Ref. | Method | τ, ns | Ref. | Model | τ, ns | Ref. | Method | |
| 0 | 1 | 11.5(1.9) | 02PAZ/PUP | DC | 13(1) | PW | NA | 8.8 | 02PAZ/PUP | NA | B |
| | 2 | 13.5(4) | 81EYL/PIP | DC | 14(1) | PW | NA | 9.5 | 02PAZ/PUP | NA | D |
| | 3 | | | | 14(1) | PW | NA | | | | |
| | 4 | | | | 14(1) | PW | NA | | | | |
| | 5 | | | | 13(1) | PW | NA | | | | |
| 1 | 1 | 20(2) | 89KOO/ZAN | TF | 24(2) | PW | NA | 15.1 | 02PAZ/PUP | NA | C |
| | | 18.1(7) | 02PAZ/PUP | DC | | | | | | | D |
| | 2 | 15.2(1.3) | 81EYL/PIP | DC | 17(2) | PW | NA | 11.5 | 02PAZ/PUP | NA | C |
| | 3 | | | | 16(2) | PW | NA | | | | |
| | 4 | | | | 16(2) | PW | NA | | | | |
| | 5 | | | | 16(2) | PW | NA | | | | |
| 2 | 1 | | | | 10(1) | PW | NA | | | | |
| | 2 | | | | 13(1) | PW | NA | | | | |
| | 3 | | | | 17(2) | PW | NA | | | | |
| | 4 | | | | 22(2) | PW | NA | | | | |
| 3 | 1 | | | | 10(1) | PW | NA | | | | |
| | 2 | | | | 10(1) | PW | NA | | | | |
| | 3 | | | | 10(1) | PW | NA | | | | |
| | 4 | | | | 10(1) | PW | NA | | | | |
| | | \multicolumn{11}{c}{$i^3\Pi_g^-$,$v$,$N$ rovibronic levels} | | | | | | | | |
| $v$ | $N$ | Experimental studies | | | Semi-empirical determination | | | Non-empirical calculation | | | O-C |
| | | $\tau$, ns | Ref. | Method | $\tau$, ns | Ref. | Model | $\tau$, ns | Ref. | Method | |
| 0 | 1 | 13.5(1.0) | 81EYL/PIP | DC | 12.4(8) | 96AST/KÄN | NA | 8.67 | 00ADA/PAZ | NA | D |
| | | 8.66(16) | 96RAY/LAF | DC | | | | | | | A |
| | 2 | | | | 12.5(7) | 96AST/KÄN | NA | 9.03 | 00ADA/PAZ | NA | |
| | 3 ($J$=3) | 11.6(1.0)[l] | 81EYL/PIP | DC | 12.6(7) | 96AST/KÄN | NA | 9.29 | 00ADA/PAZ | NA | C |
| | 3 ($J$=4) | 12.9(0.3)[m] | 81EYL/PIP | DC | | | | | | | D |
| | 4 | | | | 12.7(7) | 96AST/KÄN | NA | 9.48 | 00ADA/PAZ | NA | |
| | 5 | | | | 12.8(7) | 96AST/KÄN | NA | 9.57 | 00ADA/PAZ | NA | |
| | 6 | | | | 12.9(7) | 96AST/KÄN | NA | 9.63 | 00ADA/PAZ | NA | |
| 1 | 1 | 13.1(1.1) | 81EYL/PIP | DC | 11.2(8) | 96AST/KÄN | NA | 7.88 | 00ADA/PAZ | NA | D |
| | 2 | | | | 11.4(8) | 96AST/KÄN | NA | 8.10 | 00ADA/PAZ | NA | |
| | 3 | 12.0(0.7) | 81EYL/PIP | DC | 11.5(7) | 96AST/KÄN | NA | 8.29 | 00ADA/PAZ | NA | D |
| | 4 | | | | 11.7(7) | 96AST/KÄN | NA | 8.46 | 00ADA/PAZ | NA | |
| | 5 | | | | 11.8(7) | 96AST/KÄN | NA | 8.58 | 00ADA/PAZ | NA | |
| | 6 | | | | 12.0(7) | 96AST/KÄN | NA | 8.65 | 00ADA/PAZ | NA | |
| 2 | 1 | | | | 10.6(8) | 96AST/KÄN | NA | 6.94 | 00ADA/PAZ | NA | |
| | 2 | | | | 10.8(8) | 96AST/KÄN | NA | 7.05 | 00ADA/PAZ | NA | |
| | 3 | | | | 11.0(8) | 96AST/KÄN | NA | 7.18 | 00ADA/PAZ | NA | |
| | 4 | | | | 11.2(8) | 96AST/KÄN | NA | 7.30 | 00ADA/PAZ | NA | |
| | 5 | | | | 11.4(8) | 96AST/KÄN | NA | 7.40 | 00ADA/PAZ | NA | |
| | 6 | | | | 11.6(8) | 96AST/KÄN | NA | 7.51 | 00ADA/PAZ | NA | |
| 3 | 1 | | | | 9.7(8) | 96AST/KÄN | NA | 5.81 | 00ADA/PAZ | NA | |
| | 2 | | | | 9.8(8) | 96AST/KÄN | NA | 5.86 | 00ADA/PAZ | NA | |
| | 3 | | | | 10.0(8) | 96AST/KÄN | NA | 5.92 | 00ADA/PAZ | NA | |
| | 4 | | | | 10.2(8) | 96AST/KÄN | NA | 5.98 | 00ADA/PAZ | NA | |

| | | $j^3\Delta_g^+,v,N$ rovibronic levels | | | | | | | | |
|---|---|---|---|---|---|---|---|---|---|---|
| | | Experimental studies | | | Semi-empirical determination | | | Non-empirical calculation | | | O-C |
| $v$ | $N$ | $\tau$, ns | Ref. | Method | $\tau$, ns | Ref. | Model | $\tau$, ns | Ref. | Method | |
| 0 | 2 | 15.4(8) | 81EYL/PIP | DC | 13(1) | PW | NA | 12.3 | 02PAZ/PUP | NA | D |
| | | 12.4(1.9) | 02PAZ/PUP | DC | 13(1) | PW | NA | | | | A |
| | 3 | 14(1) | 89KOO/ZAN | TF | 13(1) | PW | NA | 10.8 | 02PAZ/PUP | NA | D |
| | 4 | | | | 13(1) | PW | NA | | | | |
| | 5 | | | | 12(1) | PW | NA | | | | |
| 1 | 2 | 15.6(9) | 81EYL/PIP | DC | 13(1) | PW | NA | 13.1 | 02PAZ/PUP | NA | D |
| | 3 | | | | 13(1) | PW | NA | 12.1 | 02PAZ/PUP | NA | |
| | 4 | | | | 13(1) | PW | NA | | | | |
| | 5 | | | | 13(1) | PW | NA | | | | |
| | 6 | | | | 13(1) | PW | NA | | | | |
| 2 | 2 | 12.5(6) | 81MON | Hanle | 14(1) | PW | NA | 13.6 | 02PAZ/PUP | NA | B |
| | 3 | 11.5(2) | 81MON | Hanle | 13(1) | PW | NA | 12.5 | 02PAZ/PUP | NA | D |
| | 4 | | | | 13(1) | PW | NA | | | | |
| | 5 | | | | 13(1) | PW | NA | | | | |
| 3 | 2 | 12.7(8) | 81MON | Hanle | 15(1) | PW | NA | 13.8 | 02PAZ/PUP | NA | B |
| | 3 | | | | 14(1) | PW | NA | | | | |
| | 4 | | | | 14(1) | PW | NA | | | | |

| | | $j^3\Delta_g^-,v,N$ rovibronic levels | | | | | | | | |
|---|---|---|---|---|---|---|---|---|---|---|
| | | Experimental studies | | | Semi-empirical determination | | | Non-empirical calculation | | | O-C |
| $v$ | $N$ | $\tau$, ns | Ref. | Method | $\tau$, ns | Ref. | Model | $\tau$, ns | Ref. | Method | |
| 0 | 2 | *15.2(1.2)* | 81EYL/PIP | DC | 13.2(4) | 96AST/KÄN | NA | 13.3 | 00ADA/PAZ | NA | B |
| | | *15.3(1.1)* | 88SAN/CAM | DC | | | | | | | B |
| | | **15(1)** | PW | Recom. | | | | | | | B |
| | 3 | 14.8(0.8) | 81EYL/PIP | DC | 13.2(4) | 96AST/KÄN | NA | 12.7 | 00ADA/PAZ | NA | C |
| | 4 | | | | 13.1(4) | 96AST/KÄN | NA | 12.3 | 00ADA/PAZ | NA | |
| | 5 | | | | 13.0(4) | 96AST/KÄN | NA | 12.0 | 00ADA/PAZ | NA | |
| | 6 | | | | 13.0(5) | 96AST/KÄN | NA | 11.8 | 00ADA/PAZ | NA | |
| 1 | 2 | *14.9(0.8)* | 81MON | Hanle | 13.4(5) | 96AST/KÄN | NA | 13.6 | 00ADA/PAZ | NA | B |
| | | *14(1)* | 89KOO/ZAN | TF | | | | | | | A |
| | | *14.5(1.4)* | PW | Recom. | | | | | | | B |
| | 3 | *12.7(2.0)* | 81EYL/PIP | DC | 13.3(4) | 96AST/KÄN | NA | 12.9 | 00ADA/PAZ | NA | A |
| | | *12.77(0.30)* | 96RAY/LAF | DC | | | | | | | A |
| | | **12.7(1)** | PW | Recom. | | | | | | | B |
| | 4 | | | | 13.2(4) | 96AST/KÄN | NA | 12.4 | 00ADA/PAZ | NA | |
| | 5 | | | | 13.1(4) | 96AST/KÄN | NA | 12.0 | 00ADA/PAZ | NA | |
| | 6 | | | | 13.0(4) | 96AST/KÄN | NA | 11.6 | 00ADA/PAZ | NA | |
| 2 | 2 | *14.7(1.0)* | 81MON | Hanle | 13.9(4) | 96AST/KÄN | NA | 13.8 | 00ADA/PAZ | NA | A |
| | | *14(1)* | 89KOO/ZAN | TF | | | | | | | A |
| | | **14.4(1.0)** | PW | Recom. | | | | | | | B |
| | 3 | 14.7(1.0) | 81MON | Hanle | 13.6(4) | 96AST/KÄN | NA | 13.1 | 00ADA/PAZ | NA | B |
| | 4 | | | | 13.5(4) | 96AST/KÄN | NA | 12.5 | 00ADA/PAZ | NA | |
| | 5 | | | | 13.3(4) | 96AST/KÄN | NA | 11.9 | 00ADA/PAZ | NA | |
| | 6 | | | | 13.2(4) | 96AST/KÄN | NA | 11.3 | 00ADA/PAZ | NA | |
| 3 | 2 | 15.1(1.0) | 81MON | Hanle | 14.4(4) | 96AST/KÄN | NA | 13.9 | 00ADA/PAZ | NA | B |
| | 3 | 12.5(1.4) | 81MON | Hanle | 14.1(4) | 96AST/KÄN | NA | 13.1 | 00ADA/PAZ | NA | A |
| | 4 | | | | 13.9(4) | 96AST/KÄN | NA | 12.3 | 00ADA/PAZ | NA | |

| | | Experimental studies | | | Non-empirical calculation | | | |
|---|---|---|---|---|---|---|---|---|
| $v$ | $N$ | $k^3\Pi_u^-,v,N$ rovibronic levels | | | | | | O-C |
| | | $\tau$, ns | Ref. | Method | $\tau$, ns | Ref. | Model | |
| 0 | 1 | 50.0(3.5)[n] | 78DAY/AND | DC | 65.5 | 03KIY/SAT | NA | D |
| | | **66.3(1.1)** | 03KIY/SAT | DC | | | | A |
| | 2 | 54.9(4.2)[n] | 78DAY/AND | DC | 65.6 | 03KIY/SAT | NA | C |
| 1 | 1 | 32(5) | 74MIL/FRE | MOMRIE | 66.3 | 03KIY/SAT | NA | D |
| | | 47.6(3.6)[n] | 78DAY/AND | DC | | | | D |
| | | **64.9(0.6)** | 03KIY/SAT | DC | | | | C |
| | 2 | 49.8(3.7)[n] | 78DAY/AND | DC | 66.3 | 03KIY/SAT | NA | D |
| 2 | 1 | 32(5) | 74MIL/FRE | MOMRIE | 67.5 | 03KIY/SAT | NA | D |
| | | 52.1(4.1)[n] | 78DAY/AND | DC | | | | D |
| | | **66.1(0.8)** | 03KIY/SAT | DC | | | | B |
| | 2 | 49.3(3.6)[n] | 78DAY/AND | DC | 67.5 | 03KIY/SAT | NA | D |
| | | **66.7(0.8)** | 03KIY/SAT | DC | | | | A |
| | 5 | 65.6(1.9) | 03KIY/SAT | DC | 67.6 | | | B |
| 3 | 1 | 32(5) | 74MIL/FRE | MOMRIE | 68.4 | 03KIY/SAT | NA | D |
| | | 46.9(3.3)[n] | 78DAY/AND | DC | | | | D |
| | | **63.8(1.3)** | 03KIY/SAT | DC | | | | D |
| | 3 | 63.9(0.9) | 03KIY/SAT | DC | 68.5 | 03KIY/SAT | NA | D |
| 4 | 1 | 60.2(1.2) | 03KIY/SAT | DC | 49.9 | 03KIY/SAT | NA | D |
| | 2 | 57.8(1.4) | 03KIY/SAT | DC | 50.2 | 03KIY/SAT | NA | D |
| 5 | 1 | 33.6(0.5) | 03KIY/SAT | DC | 33.4 | 03KIY/SAT | NA | A |
| 6 | 1 | 31.1(0.4) | 03KIY/SAT | DC | 24.0 | 03KIY/SAT | NA | D |

[a] – this value is received without taking into account the rotation of the molecule;

[b] – reported in [01KIY/SAT] estimates based on experimental data in the framework of non-adiabatic model;

[c] – the value obtained for the temperature T=300° K;

[d] – the value obtained for the temperature T=80° K;

[e] – measurements obtained from observation of a rotational line blend;

[f] – the values obtained without corrections taking into account the fine structure of spectral lines;

[g] - the levels of outer potential well of the electronic state;

[h] – this value is received without taking into account an existence of vibrational and rotational structure of the levels;

[i] – the value is average over $v$=0÷4 levels;

[j] – the measured value of lifetime for $c^3\Pi_u^-$,0,$N$=2 ($J$=1,3) levels reported in [72JOH] was used for the calibration of experimental setup in [94BER/OTT];

[k] – spectral resolution of setup (4 Å) [78DAY/AND] is insufficiently to select measured Q2-line of ($v$−$v$) bands of $d^3\Pi_u^- \rightarrow a^3\Sigma_g^+$ transition;

[l] – the value obtained for rotational level with $J$=3;

[m] – the value obtained for rotational level with $J$=4;

[n] – the values have been obtained in present work from the data on the decay rates reported in [78DAY/AND].

Table 7. All experimental and most reliable semi-empiric and non-empirical lifetime values for vibro-rotational levels of various excited electronic states of the HD molecule. If several experimental lifetime values were reported for the same rovibronic level then the data that are more reliable are italicized. Recommended data are marked as PW and printed in the bold face. Letters A, B, C and D characterize the difference between observed and calculated values (O-C) when it is less or equal to 1σ, 2σ, 3σ and greater than 3σ correspondingly.

| $EF^1\Sigma_g^+,v,N$ rovibronic levels | | | | | | | |
|---|---|---|---|---|---|---|---|
| $v$ | $N$ | Experimental studies | | | Non-empirical calculation | | O-C |
| | | τ, ns | Ref. | Method | τ, ns | Ref. | Model |
| 0 | 0 | 201(4) | 86CHA/THO | TF | 220.7 | 06FAN/WÜN | AA | D |
| | 1 | 209(5) | 86CHA/THO | TF | | | | |
| 3 | 0 | 156(10) | 86CHA/THO | TF | 155.8 | 06FAN/WÜN | AA | A |
| | 1 | 158(5) | 86CHA/THO | TF | | | | |
| 6 | 0 | 106(6) | 86CHA/THO | TF | 112.9 | 06FAN/WÜN | AA | B |
| 31 | 3 | 103.4(1.4) | 04YOS/OGI | DC | | | | |
| 33 | 2 | 66.6(5.9) | 04YOS/OGI | DC | | | | |
| | 3 | 49.6(1.7) | 04YOS/OGI | DC | | | | |
| | 4 | 73.0(1.8) | 04YOS/OGI | DC | | | | |
| 35 | 1 | 121.7(6.5) | 04YOS/OGI | DC | | | | |
| | 2 | 113.9(3.3) | 04YOS/OGI | DC | | | | |
| | 3 | 98.9(0.4) | 04YOS/OGI | DC | | | | |
| 36 | 1 | 150.4(7.4) | 04YOS/OGI | DC | | | | |
| | 2 | 146.4(2.5) | 04YOS/OGI | DC | | | | |
| | 3 | 144.3(4.8) | 04YOS/OGI | DC | | | | |
| 37 | 1 | 79.6(13.6) | 04YOS/OGI | DC | | | | |

| $GK^1\Sigma_g^+,v,N$ rovibronic levels | | | | |
|---|---|---|---|---|
| $v$ | $N$ | Experimental studies | | |
| | | τ, ns | Ref. | Method |
| 6 | 1 | 31.8(1.1) | 04YOS/OGI | DC |
| | 2 | 15.6(0.1) | 04YOS/OGI | DC |
| | 3 | 17.5(0.1) | 04YOS/OGI | DC |
| | 4 | 15.0(0.3) | 04YOS/OGI | DC |
| 7 | 1 | 23.2(0.3) | 04YOS/OGI | DC |
| | 2 | 32.4(0.1) | 04YOS/OGI | DC |
| | 3 | 32.9(0.3) | 04YOS/OGI | DC |
| | 4 | 23.2(0.4) | 04YOS/OGI | DC |
| 8 | 1 | 49.2(0.9) | 04YOS/OGI | DC |
| | 2 | 29.8(1.4) | 04YOS/OGI | DC |
| | 3 | 30.6(0.8) | 04YOS/OGI | DC |
| | 4 | 61.6(1.6) | 04YOS/OGI | DC |
| 9 | 1 | 25.4(0.6) | 04YOS/OGI | DC |
| | 2 | 45.9(1.5) | 04YOS/OGI | DC |

Table 7. – Continued.

| | | $H^1\Sigma_g^+$, $v$, $N$ rovibronic levels | | |
|---|---|---|---|---|
| $v$ | $N$ | Experimental studies | | |
| | | $\tau$, ns | Ref. | Method |
| 0 | 1 | 108.2(4.8) | 04YOS/OGI | DC |
| | 2 | 95.9(0.8) | 04YOS/OGI | DC |
| 1 | 1 | 132.5(5.5) | 04YOS/OGI | DC |
| | 2 | 121.2(0.8) | 04YOS/OGI | DC |
| | 3 | 123.4(3.7) | 04YOS/OGI | DC |
| 2 | 1 | 84.3(1.7) | 04YOS/OGI | DC |
| | 2 | 88.9(1.2) | 04YOS/OGI | DC |
| 16 | 0 | 0.036(7) | 00REI/HOG | TF |
| | 1 | 0.034(7) | 00REI/HOG | TF |
| | 2 | 0.022(4) | 00REI/HOG | TF |
| | 3 | 0.032(6) | 00REI/HOG | TF |
| 17 | 0 | 0.0044(9) | 00REI/HOG | TF |
| | 1 | 0.0027(5) | 00REI/HOG | TF |
| | 2 | 0.0085(17) | 00REI/HOG | TF |
| | 3 | 0.011(2) | 00REI/HOG | TF |
| 18 | 0 | 0.020(4) | 00REI/HOG | TF |
| | 1 | 0.0082(16) | 00REI/HOG | TF |
| | 2 | 0.0052(10) | 00REI/HOG | TF |
| | 3 | 0.0010(2) | 00REI/HOG | TF |
| 19 | 0 | 0.0009(2) | 00REI/HOG | TF |
| | 1 | 0.0079(16) | 00REI/HOG | TF |
| | 2 | 0.0006(1) | 00REI/HOG | TF |
| | 3 | 0.0019(4) | 00REI/HOG | TF |

Table 7. – Continued.

| $v$ | $N$ | $\overline{B}\ ^1\Sigma_u^+,v,N$ rovibronic levels | | |
|---|---|---|---|---|
| | | Experimental studies | | |
| | | $\tau$, ns | Ref. | Method |
| 16 | 0 | 0.045(9) | 00REI/HOG | TF |
| | 1 | 0.047(9) | 00REI/HOG | TF |
| | 2 | 0.061(12) | 00REI/HOG | TF |
| | 3 | 0.036(7) | 00REI/HOG | TF |
| 17 | 0 | 0.020(4) | 00REI/HOG | TF |
| | 1 | 0.012(2) | 00REI/HOG | TF |
| | 2 | 0.024(5) | 00REI/HOG | TF |
| | 3 | 0.052(10) | 00REI/HOG | TF |
| 18 | 0 | 0.040(8) | 00REI/HOG | TF |
| | 1 | 0.030(6) | 00REI/HOG | TF |
| | 2 | 0.024(5) | 00REI/HOG | TF |
| | 3 | 0.016(3) | 00REI/HOG | TF |
| 19 | 0 | 0.0058(12) | 00REI/HOG | TF |
| | 1 | 0.059(12) | 00REI/HOG | TF |
| | 2 | 0.0043(9) | 00REI/HOG | TF |
| | 3 | 0.0015(3) | 00REI/HOG | TF |
| 20 | 0 | 0.004(2) | 00REI/HOG | TF |
| | 1 | 0.0024(5) | 00REI/HOG | TF |
| | 2 | 0.0034(7) | 00REI/HOG | TF |
| | 3 | 0.0025(5) | 00REI/HOG | TF |
| 21 | 0 | 0.0021(11) | 00REI/HOG | TF |
| | 1 | 0.0015(8) | 00REI/HOG | TF |
| | 2 | 0.0021(11) | 00REI/HOG | TF |
| | 3 | 0.0015(8) | 00REI/HOG | TF |
| 22 | 0 | 0.00030(15) | 00REI/HOG | TF |
| | 1 | 0.00030(15) | 00REI/HOG | TF |
| | 2 | 0.00030(15) | 00REI/HOG | TF |
| | 3 | 0.00030(15) | 00REI/HOG | TF |

$I^1\Pi_g^+, v, N$ rovibronic levels

| $v$ | $N$ | Experimental studies | | | Non-empirical calculation | | | O-C |
|---|---|---|---|---|---|---|---|---|
| | | $\tau$, ns | Ref. | Method | $\tau$, ns | Ref. | Model | |
| 3 | 1 | 29.6(2.5) | 04YOS/OGI | DC | 17.0[a] | 06FAN/WÜN | AA | D |
| | 2 | 31.7(0.7) | 04YOS/OGI | DC | | | | |
| 5 | 3 | 57.8(1.9) | 04YOS/OGI | DC | | | | |

$I^1\Pi_g^-, v, N$ rovibronic levels

| $v$ | $N$ | Experimental studies | | | Non-empirical calculation | | | O-C |
|---|---|---|---|---|---|---|---|---|
| | | $\tau$, ns | Ref. | Method | $\tau$, ns | Ref. | Model | |
| 3 | 1 | 20.7(0.1) | 04YOS/OGI | DC | 17.0[a] | 06FAN/WÜN | AA | D |
| | 2 | 20.0(0.1) | 04YOS/OGI | DC | | | | |
| | 3 | 19.6(0.6) | 04YOS/OGI | DC | | | | |
| 4 | 1 | 23.2(0.4) | 07ROS/FUJ | DC | 16.1 | 07ROS/FUJ | AA | D |
| | 2 | 21.5(0.4) | 07ROS/FUJ | DC | 15.5 | 07ROS/FUJ | AA | D |
| | 3 | 18.7(0.5) | 07ROS/FUJ | DC | 13.3 | 07ROS/FUJ | AA | D |
| | 4 | <15 | 07ROS/FUJ | DC | 8.8 | 07ROS/FUJ | AA | |

$J^1\Delta_g^+, v, N$ rovibronic levels

| $v$ | N | Experimental studies | | | Non-empirical calculation | | | O-C |
|---|---|---|---|---|---|---|---|---|
| | | $\tau$, ns | Ref. | Method | $\tau$, ns | Ref. | Model | |
| 3 | 2 | 23.6(0.1) | 04YOS/OGI | DC | 19.8[a] | 06FAN/WÜN | AA | D |
| | 3 | 25.4(0.5) | 04YOS/OGI | DC | | | | |
| | 4 | 23.8(1.1) | 04YOS/OGI | DC | | | | |

$J^1\Delta_g^-, v, N$ rovibronic level

| $v$ | $N$ | Experimental studies | | | Non-empirical calculation | | | O-C |
|---|---|---|---|---|---|---|---|---|
| | | $\tau$, ns | Ref. | Method | $\tau$, ns | Ref. | Model | |
| 3 | 2 | 25.2(0.2) | 04YOS/OGI | DC | 19.8[a] | 06FAN/WÜN | AA | D |

$c^3\Pi_u^-, v, N$ rovibronic levels

| $v$ | $N$ | Experimental studies | | | Non-empirical calculation | | | O-C |
|---|---|---|---|---|---|---|---|---|
| | | $\tau$, ms | Ref. | Method | $\tau$, ms | Ref. | Model | |
| 0 | 1 | 1.02(5) | 72JOH | TF | $\infty$ | 06FAN/WÜN | AA | D |
| | 2 | 1.02(5) | 72JOH | TF | | | | |

[a] – this value is calculated without taking into account the rotation of the molecule.

Table 8. All experimental and most reliable semi-empiric and non-empirical lifetime values for vibro-rotational levels of various excited electronic states of the $D_2$ molecule. If several experimental lifetime values were reported for the same rovibronic level then the data that are more reliable are italicized. Recommended data are marked as PW and printed in the bold face. Letters A, B, C and D characterize the difference between observed and calculated values (O-C) when it is less or equal to 1σ, 2σ, 3σ and greater than 3σ correspondingly.

| | | | | | | | | |
|---|---|---|---|---|---|---|---|---|
| | | | $EF\,{}^1\Sigma_g{}^+$,$v$,$N$ rovibronic levels | | | | | |
| $v$ | $N$ | Experimental studies | | | Non-empirical calculation | | | O-C |
| | | $\tau$, ns | Ref. | Method | $\tau$, ns | Ref. | Model | |
| 28 | 2 | 166.1(2.7) | 98SUZ/NAK | DC | | | | |
| 32 | 2 | 216.1(13.9) | 98SUZ/NAK | DC | | | | |
| 41 | 0 | 120.5(2.7) | 05AIT/YOS | DC | 197 | 90aQUA/DRE | NA | D |
| | 1 | 141.1(4.9) | 05AIT/YOS | DC | 136 | 90aQUA/DRE | NA | B |
| | 2 | 42.3(4.3) | 05AIT/YOS | DC | | | | |
| | 3 | 79.2(4.6) | 05AIT/YOS | DC | | | | |
| | 4 | 111.9(4.5) | 05AIT/YOS | DC | | | | |
| | 5 | 107.9(6.3) | 05AIT/YOS | DC | | | | |
| 44 | 0 | 472(5) | 11ROS/TSU | DC | 771 | 11ROS/TSU | NA[a] | D |
| | | | | | 771 | 11ROS/TSU | NA[b] | D |
| | 1 | 461(7) | 11ROS/TSU | DC | 772 | 11ROS/TSU | NA[a] | D |
| | | | | | 781 | 11ROS/TSU | NA[b] | D |
| | 2 | 404(13) | 11ROS/TSU | DC | 610 | 11ROS/TSU | NA[a] | D |
| | | | | | 638 | 11ROS/TSU | NA[b] | D |
| | 3 | 458(12) | 11ROS/TSU | DC | 757 | 11ROS/TSU | NA[a] | D |
| | | | | | 815 | 11ROS/TSU | NA[b] | D |
| | 4 | 457(19) | 11ROS/TSU | DC | 730 | 11ROS/TSU | NA[a] | D |
| | | | | | 827 | 11ROS/TSU | NA[b] | D |
| | 5 | 386(9) | 11ROS/TSU | DC | 662 | 11ROS/TSU | NA[a] | D |
| | | | | | 817 | 11ROS/TSU | NA[b] | D |
| 45 | 0 | 405(3) | 11ROS/TSU | DC | 644 | 11ROS/TSU | NA[a] | D |
| | | | | | 644 | 11ROS/TSU | NA[b] | D |
| | 1 | 442(9) | 11ROS/TSU | DC | 657 | 11ROS/TSU | NA[a] | D |
| | | | | | 684 | 11ROS/TSU | NA[b] | D |
| | 2 | 467(8) | 11ROS/TSU | DC | 679 | 11ROS/TSU | NA[a] | D |
| | | | | | 747 | 11ROS/TSU | NA[b] | D |
| | 3 | 430(8) | 11ROS/TSU | DC | 678 | 11ROS/TSU | NA[a] | D |
| | | | | | 790 | 11ROS/TSU | NA[b] | D |
| | 4 | 333(2) | 11ROS/TSU | DC | 362 | 11ROS/TSU | NA[a] | D |
| | | | | | 381 | 11ROS/TSU | NA[b] | D |
| | 5 | 114(4) | 11ROS/TSU | DC | 143 | 11ROS/TSU | NA[a] | D |
| | | | | | 194 | 11ROS/TSU | NA[b] | D |

| | | Experimental studies | | | Non-empirical calculation | | | O-C |
|---|---|---|---|---|---|---|---|---|
| $v$ | $N$ | $\tau$, ns | Ref. | Method | $\tau$, ns | Ref. | Model | |
| 46 | 0 | 315(4) | 11ROS/TSU | DC | 511 | 11ROS/TSU | NA[a] | D |
| | | | | | 511 | 11ROS/TSU | NA[b] | D |
| | 1 | 476(5) | 11ROS/TSU | DC | 661 | 11ROS/TSU | NA[a] | D |
| | | | | | 693 | 11ROS/TSU | NA[b] | D |
| | 2 | 641(15) | 11ROS/TSU | DC | 894 | 11ROS/TSU | NA[a] | D |
| | | | | | 1017 | 11ROS/TSU | NA[b] | D |
| | 3 | 883(14) | 11ROS/TSU | DC | 1077 | 11ROS/TSU | NA[a] | D |
| | | | | | 1333 | 11ROS/TSU | NA[b] | D |
| | 4 | 1236(72) | 11ROS/TSU | DC | 987 | 11ROS/TSU | NA[a] | D |
| | | | | | 1496 | 11ROS/TSU | NA[b] | D |
| | 5 | 366(3) | 11ROS/TSU | DC | 280 | 11ROS/TSU | NA[a] | D |
| | | | | | 660 | 11ROS/TSU | NA[b] | D |

The table header spans: $EF{}^1\Sigma_g{}^+,v,N$ rovibronic levels

Table 8 – Continued.

| $GK\ ^1\Sigma_g^+$,$v$,$N$ rovibronic levels | | | | | | | |
|---|---|---|---|---|---|---|---|
| $v$ | $N$ | Experimental studies | | | Non-empirical calculation | | O-C |
| | | $\tau$, ns | Ref. | Method | $\tau$, ns | Ref. | Model |
| 2 | 0 | 72.5(6.9) | 98SUZ/NAK | DC | 101 | 90aQUA/DRE | NA | D |
| | 2 | 73.6(6.6) | 98SUZ/NAK | DC | | | | |
| 8 | 0 | 50.2(2.3) | 05AIT/YOS | DC | 51.9 | 90aQUA/DRE | NA | A |
| | 1 | 54.0(1.2) | 05AIT/YOS | DC | 51.4 | 90aQUA/DRE | NA | C |
| | 2 | 52.6(1.4) | 05AIT/YOS | DC | | | | |
| | 3 | 50.8(2.3) | 05AIT/YOS | DC | | | | |
| | 4 | 52.5(1.9) | 05AIT/YOS | DC | | | | |
| | 5 | 56.9(1.9) | 05AIT/YOS | DC | | | | |
| 9 | 0 | 65.1(1.8) | 05AIT/YOS | DC | 48.2 | 90aQUA/DRE | NA | D |
| | 1 | 59.3(1.7) | 05AIT/YOS | DC | 43.2 | 90aQUA/DRE | NA | D |
| | 2 | 54.0(1.0) | 05AIT/YOS | DC | | | | |
| | 3 | 50.3(2.7) | 05AIT/YOS | DC | | | | |
| | 4 | 49.2(0.6) | 05AIT/YOS | DC | | | | |
| | 5 | 49.6(1.9) | 05AIT/YOS | DC | | | | |
| 10 | 0 | 60.6(0.3) | 05AIT/YOS | DC | 63.2 | 90aQUA/DRE | NA | D |
| | 1 | 62.7(1.0) | 05AIT/YOS | DC | 63.2 | 90aQUA/DRE | NA | A |
| | 2 | 62.9(1.0) | 05AIT/YOS | DC | | | | |
| | 3 | 63.3(0.8) | 05AIT/YOS | DC | | | | |
| | 4 | 68.8(1.8) | 05AIT/YOS | DC | | | | |
| | 5 | 112.9(2.2) | 05AIT/YOS | DC | | | | |
| 11 | 1 | 71.0(2.0) | 05AIT/YOS | DC | 71 | 11ROS/TSU | NA[a] | A |
| | | | | | 72 | 11ROS/TSU | NA[b] | A |
| | 2 | 70.8(1.4) | 05AIT/YOS | DC | 72 | 11ROS/TSU | NA[a] | A |
| | | | | | 75 | 11ROS/TSU | NA[b] | C |
| | 3 | 73.7(4.0) | 05AIT/YOS | DC | 73 | 11ROS/TSU | NA[a] | A |
| | | | | | 79 | 11ROS/TSU | NA[b] | B |
| | 4 | 72.3(0.9) | 05AIT/YOS | DC | 72 | 11ROS/TSU | NA[a] | A |
| | | | | | 85 | 11ROS/TSU | NA[b] | D |
| | 5 | 125(3) | 11ROS/TSU | DC | 131 | 11ROS/TSU | NA[a] | B |
| | | | | | 179 | 11ROS/TSU | NA[b] | D |

| $H\ ^1\Sigma_g^+$,$v$,$N$ rovibronic levels | | | | | | | |
|---|---|---|---|---|---|---|---|
| $v$ | $N$ | Experimental studies | | | Non-empirical calculation | | O-C |
| | | $\tau$, ns | Ref. | Method | $\tau$, ns | Ref. | Model |
| 0 | 0 | 170.8(4.5) | 98SUZ/NAK | DC | 142 | 90aQUA/DRE | NA | D |
| | 1 | 165.4(12.2) | 98SUZ/NAK | DC | 139 | 90aQUA/DRE | NA | C |
| | 2 | 156.5(3.3) | 98SUZ/NAK | DC | | | | |
| 1 | 2 | 115.9(1.5) | 98SUZ/NAK | DC | | | | |
| 3 | 0 | 65.4(0.7) | 05AIT/YOS | DC | 35.5 | 90aQUA/DRE | NA | D |
| | 1 | 34.4(1.0) | 05AIT/YOS | DC | 18.3 | 90aQUA/DRE | NA | D |
| | 2 | 43.1(1.4) | 05AIT/YOS | DC | | | | |
| | 3 | 36.1(1.0) | 05AIT/YOS | DC | | | | |
| | 4 | 34.3(0.2) | 05AIT/YOS | DC | | | | |

| | | $\bar{H}\ ^1\Sigma_g^+, v, N$ rovibronic levels | | | | | | |
|---|---|---|---|---|---|---|---|---|
| $v$ | $N$ | Experimental studies | | | Non-empirical calculation | | | O-C |
| | | $\tau$, ns | Ref. | Method | $\tau$, ns | Ref. | Model | |
| 9 | 0 | 78.9(1.1) | 06ROS/YOS | DC | 93.6 | 06ROS/YOS | NA | D |
| | 1 | 78.2(1.1) | 06ROS/YOS | DC | 93.6 | 06ROS/YOS | NA | D |
| | 2 | 77.4(1.6) | 06ROS/YOS | DC | 93.6 | 06ROS/YOS | NA | D |
| | 3 | 76.3(1.8) | 06ROS/YOS | DC | 93.7 | 06ROS/YOS | NA | D |
| | 4 | 77.3(3.1) | 06ROS/YOS | DC | 93.9 | 06ROS/YOS | NA | D |
| 10 | 0 | 73.9(2.2) | 06ROS/YOS | DC | 90.7 | 06ROS/YOS | NA | D |
| | 1 | 74.9(1.2) | 06ROS/YOS | DC | 90.7 | 06ROS/YOS | NA | D |
| | 2 | 74.0(0.6) | 06ROS/YOS | DC | 90.8 | 06ROS/YOS | NA | D |
| | 3 | 74.0(1.0) | 06ROS/YOS | DC | 90.9 | 06ROS/YOS | NA | D |
| | 4 | 73.7(1.5) | 06ROS/YOS | DC | 91.0 | 06ROS/YOS | NA | D |
| 11 | 0 | 73.3(1.9) | 06ROS/YOS | DC | 85.8 | 06ROS/YOS | NA | D |
| | 1 | 73.8(0.9) | 06ROS/YOS | DC | 85.9 | 06ROS/YOS | NA | D |
| | 2 | 72.0(1.9) | 06ROS/YOS | DC | 86.2 | 06ROS/YOS | NA | D |
| | 3 | 74.5(0.6) | 06ROS/YOS | DC | 86.6 | 06ROS/YOS | NA | D |
| | 4 | 71.4(2.3) | 06ROS/YOS | DC | 87.2 | 06ROS/YOS | NA | D |
| 12 | 0 | 70.3(1.4) | 06ROS/YOS | DC | 81.4 | 06ROS/YOS | NA | D |
| | 1 | 71.0(0.6) | 06ROS/YOS | DC | 81.3 | 06ROS/YOS | NA | D |
| | 2 | 69.8(0.5) | 06ROS/YOS | DC | 81.8 | 06ROS/YOS | NA | D |
| | 3 | 70.7(1.6) | 06ROS/YOS | DC | 82.3 | 06ROS/YOS | NA | D |
| | 4 | 69.3(1.2) | 06ROS/YOS | DC | 82.9 | 06ROS/YOS | NA | D |
| 13 | 0 | 69.1(2.2) | 06ROS/YOS | DC | 77.7 | 06ROS/YOS | NA | D |
| | 1 | 67.8(0.7) | 06ROS/YOS | DC | 77.9 | 06ROS/YOS | NA | D |
| | 2 | 66.9(1.3) | 06ROS/YOS | DC | 78.3 | 06ROS/YOS | NA | D |
| | 3 | 66.6[c] | 06ROS/YOS | DC | 78.7 | 06ROS/YOS | NA | |
| | 4 | 68.1[c] | 06ROS/YOS | DC | 79.2 | 06ROS/YOS | NA | |
| 14 | 0 | 63.3(1.6) | 06ROS/YOS | DC | 74.5 | 06ROS/YOS | NA | D |
| | 1 | 63.3(1.1) | 06ROS/YOS | DC | 74.7 | 06ROS/YOS | NA | D |
| | 2 | 66.0(2.1) | 06ROS/YOS | DC | 75.0 | 06ROS/YOS | NA | D |
| | 3 | 64.9(0.3) | 06ROS/YOS | DC | 75.4 | 06ROS/YOS | NA | D |
| | 4 | 65.9(2.1) | 06ROS/YOS | DC | 75.9 | 06ROS/YOS | NA | D |
| 15 | 0 | *56(6)* | 00REI/HOG | TF | 71.9 | 06ROS/YOS | NA | C |
| | | *62.1(1.7)* | 06ROS/YOS | DC | | | | D |
| | | *59(4)* | PW | Recom. | | | | D |
| | 1 | *58(6)* | 00REI/HOG | TF | 72.0 | | | C |
| | | *62.8(0.4)* | 06ROS/YOS | DC | | | | D |
| | | *60(3)* | PW | Recom. | | | | D |
| | 2 | *58(6)* | 00REI/HOG | TF | 72.1 | 06ROS/YOS | NA | C |
| | | *62.6[c]* | 06ROS/YOS | DC | | | | |
| | | *60(6)* | PW | Recom. | | | | D |
| | 3 | *55(6)* | 00REI/HOG | TF | 72.1 | 06ROS/YOS | NA | C |
| | | *61.7[c]* | 06ROS/YOS | DC | | | | |
| | | *58(6)* | PW | Recom. | | | | D |
| | 4 | 65.5[c] | 06ROS/YOS | DC | 71.9 | 06ROS/YOS | NA | |

| | | $\bar{H}$ $^1\Sigma_g^+$,$v$,$N$ rovibronic levels | | | | | |
|---|---|---|---|---|---|---|---|
| $v$ | $N$ | Experimental studies | | | Non-empirical calculation | | O-C |
| | | $\tau$, ns | Ref. | Method | $\tau$, ns | Ref. | Model |
| 16 | 0 | *56(6)* | 00REI/HOG | TF | 62.2 | 06ROS/YOS | NA | B |
| | | *57.1(1.9)* | 06ROS/YOS | DC | | | | C |
| | | **57(4)** | PW | Recom. | | | | D |
| | 1 | *48(5)* | 00REI/HOG | TF | 55.7 | 06ROS/YOS | NA | B |
| | | *59.5(0.2)* | 06ROS/YOS | DC | | | | D |
| | | **55(5)** | PW | Recom. | | | | D |
| | 2 | *50(5)* | 00REI/HOG | TF | 1.5[d] | 06ROS/YOS | NA | D |
| | | *60.2(2.3)* | 06ROS/YOS | DC | | | | D |
| | | *55(4)* | PW | Recom. | | | | D |
| | 3 | *56(6)* | 00REI/HOG | TF | 62.2 | 06ROS/YOS | NA | B |
| | | *59.9*[c] | 06ROS/YOS | DC | | | | |
| | | **58(6)** | PW | Recom. | | | | D |
| | 4 | 58.8[c] | 06ROS/YOS | DC | 62.7 | 06ROS/YOS | NA | |
| 17 | 0 | *46(5)* | 00REI/HOG | TF | 60.5 | 06ROS/YOS | NA | C |
| | | *54.3(1.9)* | 06ROS/YOS | DC | | | | D |
| | | **50(4)** | PW | Recom. | | | | |
| | 1 | *50(5)* | 00REI/HOG | TF | 60.2 | 06ROS/YOS | NA | C |
| | | *57.2(1.3)* | 06ROS/YOS | DC | | | | C |
| | | **54(3)** | PW | Recom. | | | | |
| | 2 | *51(5)* | 00REI/HOG | TF | 59.4 | 06ROS/YOS | NA | B |
| | | *55.0(1.2)* | 06ROS/YOS | DC | | | | D |
| | | **53(3)** | PW | Recom. | | | | |
| | 3 | *57(6)* | 00REI/HOG | TF | 57.7 | 06ROS/YOS | NA | A |
| | | *55.6(0.5)* | 06ROS/YOS | DC | | | | D |
| | | **56(3)** | PW | Recom. | | | | |
| | 4 | 55.6(0.8) | 06ROS/YOS | DC | 52.3 | 06ROS/YOS | NA | D |
| 18 | 0 | *35(4)* | 00REI/HOG | TF | 40.8 | 06ROS/YOS | NA | B |
| | | *40.5(3.6)* | 06ROS/YOS | DC | | | | A |
| | | **48(4)** | PW | Recom. | | | | D |
| | 1 | *10(1)* | 00REI/HOG | TF | 41.6 | 06ROS/YOS | NA | D |
| | | *11.7(0.1)* | 06ROS/YOS | DC | | | | D |
| | | **10.9(7)** | PW | Recom. | | | | |
| | 2 | <9.4 | 06ROS/YOS | DC | 42.7 | 06ROS/YOS | NA | |
| | 3 | *28(3)* | 00REI/HOG | TF | 43.6 | 06ROS/YOS | NA | D |
| | | *30.0(0.8)* | 06ROS/YOS | DC | | | | D |
| | | **29(2)** | PW | Recom. | | | | |
| | 4 | 51.1(0.7) | 06ROS/YOS | DC | 43.5 | 06ROS/YOS | NA | D |

Table 8. – Continued.

| | | $\overline{H}$ $^1\Sigma_g^+$,$v$,$N$ rovibronic levels | | | | | | |
|---|---|---|---|---|---|---|---|---|
| $v$ | $N$ | Experimental studies | | | Non-empirical calculation | | | O-C |
| | | $\tau$, ns | Ref. | Method | $\tau$, ns | Ref. | Model | |
| 19 | 0 | 10(1) | 00REI/HOG | TF | 2.2 | 06ROS/YOS | NA | D |
| | 1 | *16(2)* | 00REI/HOG | TF | 2.9 | 06ROS/YOS | NA | D |
| | | *16.9(0.4)* | 06ROS/YOS | DC | | | | D |
| | | **16.5(1.2)** | PW | Recom. | | | | |
| | 2 | *23(2)* | 00REI/HOG | TF | 4.5 | 06ROS/YOS | NA | D |
| | | *25.8(0.7)* | 06ROS/YOS | DC | | | | D |
| | | **24.4(1.4)** | PW | Recom. | | | | |
| | 3 | <5 | 00REI/HOG | TF | 7.2 | 06ROS/YOS | NA | |
| | | <9.8 | 06ROS/YOS | DC | | | | |
| | 4 | <9.8 | 06ROS/YOS | DC | 11.0 | 06ROS/YOS | NA | |
| 21 | 0 | 0.061(12) | 00REI/HOG | TF | 0.1459 | 06ROS/YOS | NA | D |
| | 1 | >0.100 | 00REI/HOG | TF | 0.1452 | 06ROS/YOS | NA | |
| | 2 | 0.047(9) | 00REI/HOG | TF | 0.1400 | 06ROS/YOS | NA | D |
| | 3 | 0.028(6) | 00REI/HOG | TF | 0.1230 | 06ROS/YOS | NA | D |
| | 4 | 0.027(5) | 00REI/HOG | TF | 0.0866 | 06ROS/YOS | NA | D |
| 22 | 0 | 0.0010(2) | 00REI/HOG | TF | 0.0116 | 06ROS/YOS | NA | D |
| | 1 | 0.0068(14) | 00REI/HOG | TF | 0.0132 | 06ROS/YOS | NA | D |
| | 2 | 0.0028(6) | 00REI/HOG | TF | 0.0166 | 06ROS/YOS | NA | D |
| | 3 | 0.0070(14) | 00REI/HOG | TF | 0.0214 | 06ROS/YOS | NA | D |
| | 4 | 0.024(5) | 00REI/HOG | TF | 0.0265 | 06ROS/YOS | NA | A |
| | 5 | 0.0077(15) | 00REI/HOG | TF | 0.0286 | 06ROS/YOS | NA | |
| 23 | 0 | 0.00054(11) | 00REI/HOG | TF | 0.0001 | 06ROS/YOS | NA | D |
| | 2 | 0.00054(11) | 00REI/HOG | TF | 0.0002 | 06ROS/YOS | NA | D |

| $I^1\Pi_g^+,v,N$ rovibronic levels | | | | | | | |
|---|---|---|---|---|---|---|---|
| $v$ | $N$ | Experimental studies | | | Non-empirical calculation | | O-C |
| | | $\tau$, ns | Ref. | Method | $\tau$, ns | Ref. | Model |
| 4 | 2 | 31.5(0.5) | 05AIT/YOS | DC | 31.6 | 90aQUA/DRE | NA | A |
| | 3 | 44.4(1.1) | 05AIT/YOS | DC | | | | |
| | 4 | 38.6(1.2) | 05AIT/YOS | DC | | | | |
| | 5 | 39.2(0.6) | 05AIT/YOS | DC | | | | |
| $5^e$ | 1 | $1.53(12)\times10^3$ | 10ROS/AND | DC | $1.64\times10^3$ | 10ROS/AND | AA | A |
| | 2 | $1.43(20)\times10^3$ | 10ROS/AND | DC | $1.67\times10^3$ | 10ROS/AND | AA | B |
| | 3 | $1.45(12)\times10^3$ | 10ROS/AND | DC | $1.70\times10^3$ | 10ROS/AND | AA | C |
| $6^e$ | 1 | $1.97(16)\times10^3$ | 10ROS/AND | DC | $2.81\times10^3$ | 10ROS/AND | AA | D |
| | 2 | $2.00(24)\times10^3$ | 10ROS/AND | DC | $2.87\times10^3$ | 10ROS/AND | AA | D |
| | 3 | $2.01(18)\times10^3$ | 10ROS/AND | DC | $2.96\times10^3$ | 10ROS/AND | AA | D |
| | 4 | $2.22(14)\times10^3$ | 10ROS/AND | DC | $3.08\times10^3$ | 10ROS/AND | AA | D |

| $I^1\Pi_g^-,v,N$ rovibronic levels | | | | | | | |
|---|---|---|---|---|---|---|---|
| $v$ | $N$ | Experimental studies | | | Non-empirical calculation | | O-C |
| | | $\tau$, ns | Ref. | Method | $\tau$, ns | Ref. | Model |
| 4 | 1 | 21.3(0.4) | 05AIT/YOS | DC | $17.2^f$ | 06FAN/WÜN | AA | D |
| | 2 | 20.3(1.1) | 05AIT/YOS | DC | | | | |
| | 3 | 21.1(0.2) | 05AIT/YOS | DC | | | | |
| | 4 | 20.4(0.9) | 05AIT/YOS | DC | | | | |
| 5 | 1 | 23.1(0.5) | 05AIT/YOS | DC | 16.9 | 07ROS/FUJ | AA | D |
| | 2 | 22.8(0.6) | 05AIT/YOS | DC | 16.9 | 07ROS/FUJ | AA | D |
| | 3 | 22.4(0.2) | 05AIT/YOS | DC | 16.9 | 07ROS/FUJ | AA | D |
| | 4 | 20.7(1.1) | 05AIT/YOS | DC | 16.9 | 07ROS/FUJ | AA | D |
| $5^e$ | 1 | $1.57(18)\times10^3$ | 10ROS/AND | DC | $1.64\times10^3$ | 10ROS/AND | AA | A |
| | 2 | $1.43(14)\times10^3$ | 10ROS/AND | DC | $1.67\times10^3$ | 10ROS/AND | AA | B |
| | 3 | $1.36(16)\times10^3$ | 10ROS/AND | DC | $1.70\times10^3$ | 10ROS/AND | AA | C |
| $6^e$ | 1 | $2.15(10)\times10^3$ | 10ROS/AND | DC | $2.81\times10^3$ | 10ROS/AND | AA | D |
| | 2 | $1.99(14)\times10^3$ | 10ROS/AND | DC | $2.87\times10^3$ | 10ROS/AND | AA | D |
| | 3 | $1.74(18)\times10^3$ | 10ROS/AND | DC | $2.96\times10^3$ | 10ROS/AND | AA | D |

| $P^1\Sigma_g^+,v,N$ rovibronic levels | | | | | | | |
|---|---|---|---|---|---|---|---|
| $v$ | $N$ | Experimental studies | | | Non-empirical calculation | | O-C |
| | | $\tau$, ns | Ref. | Method | $\tau$, ns | Ref. | Model |
| 0 | 1 | 47.7(2.9) | 05AIT/YOS | DC | $45.3^f$ | 06FAN/WÜN | AA | A |
| | 2 | 44.9(0.4) | 05AIT/YOS | DC | | | | |
| | 3 | 44.2(0.9) | 05AIT/YOS | DC | | | | |
| | 4 | 42.2(1.2) | 05AIT/YOS | DC | | | | |
| | 5 | 43.2(0.9) | 05AIT/YOS | DC | | | | |

| $R\,^1\Pi_g^+$,$v$,$N$ rovibronic levels | | | | | | | |
|---|---|---|---|---|---|---|---|
| $v$ | $N$ | Experimental studies | | | Non-empirical calculation | | | O-C |
| | | $\tau$, ns | Ref. | Method | $\tau$, ns | Ref. | Model | |
| 0 | 1 | 42.0(2.0) | 05AIT/YOS | DC | 109.8[f] | 06FAN/WÜN | AA | D |
| | 2 | 38.5(5.5) | 05AIT/YOS | DC | | | | |

| $R\,^1\Pi_g^-$,$v$,$N$ rovibronic levels | | | | | | | |
|---|---|---|---|---|---|---|---|
| $v$ | $N$ | Experimental studies | | | Non-empirical calculation | | | O-C |
| | | $\tau$, ns | Ref. | Method | $\tau$, ns | Ref. | Model | |
| 0 | 2 | 52.5(1.4) | 05AIT/YOS | DC | 109.8[f] | 06FAN/WÜN | AA | D |
| 1 | 1 | 48.3(0.6) | 05AIT/YOS | DC | 114.9[f] | 06FAN/WÜN | AA | D |
| | 2 | 48.1(1.4) | 05AIT/YOS | DC | | | | |
| | 3 | 48.6(0.6) | 05AIT/YOS | DC | | | | |
| | 4 | 51.6(0.1) | 05AIT/YOS | DC | | | | |
| 4 | 1 | 38.1(1.0) | 05AIT/YOS | DC | 140.6[f] | 06FAN/WÜN | AA | D |
| | 2 | 42.2(0.5) | 05AIT/YOS | DC | | | | |
| | 3 | 43.2(2.8) | 05AIT/YOS | DC | | | | |
| | 4 | 49.7(1.3) | 05AIT/YOS | DC | | | | |
| 5 | 1 | 44.4(0.3) | 05AIT/YOS | DC | 154.5[f] | 06FAN/WÜN | AA | D |
| | 2 | 33.6(1.6) | 05AIT/YOS | DC | | | | |
| | 3 | 18.0(0.6) | 05AIT/YOS | DC | | | | |
| | 4 | <10 | 05AIT/YOS | DC | | | | |
| 6 | 1 | 38.2(0.7) | 05AIT/YOS | DC | 167.6[f] | 06FAN/WÜN | AA | D |
| | 2 | 41.6(0.2) | 05AIT/YOS | DC | | | | |
| | 3 | 52.0(0.1) | 05AIT/YOS | DC | | | | |
| 7 | 1 | <10 | 05AIT/YOS | DC | 164.4[f] | 06FAN/WÜN | AA | |
| | 3 | <10 | 05AIT/YOS | DC | | | | |

| $S\,^1\Delta_g^+$,$v$,$N$ rovibronic levels | | | | | | | |
|---|---|---|---|---|---|---|---|
| $v$ | $N$ | Experimental studies | | | Non-empirical calculation | | | O-C |
| | | $\tau$, ns | Ref. | Method | $\tau$, ns | Ref. | Model | |
| 0 | 2 | 37.0(1.1) | 05AIT/YOS | DC | 48.2[f] | 06FAN/WÜN | AA | D |
| | 3 | 50.3(0.7) | 05AIT/YOS | DC | | | | |
| | 4 | 44.2(1.1) | 05AIT/YOS | DC | | | | |
| | 5 | 47.3(1.5) | 05AIT/YOS | DC | | | | |

| $S\,^1\Delta_g^-$,$v$,$N$ rovibronic levels | | | | | | | |
|---|---|---|---|---|---|---|---|
| $v$ | $N$ | Experimental studies | | | Non-empirical calculation | | | O-C |
| | | $\tau$, ns | Ref. | Method | $\tau$, ns | Ref. | Model | |
| 0 | 3 | 44.8(0.2) | 05AIT/YOS | DC | 48.2[f] | 06FAN/WÜN | AA | D |

**Table 8. – Continued.**

| | | | | | | | |
|---|---|---|---|---|---|---|---|
|  | $a^3\Sigma_g^+,v$ vibronic levels | | | | | | |
| $v$ | Experimental studies | | | Non-empirical calculation | | | O-C |
| | $\tau$, ns | Ref. | Method | $\tau$, ns | Ref. | Model | |
| 0 | 35(8) | 66FOW/HOL | PLASMA | 11.92 | 06FAN/WÜN | AA | D |
| | **12.9(1.6)** | 72SMI/CHE | PS | | | | A |
| | $c^3\Pi_u^-,v,N$ rovibronic levels | | | | | | |
| $v$ | $N$ | Experimental studies | | | Non-empirical calculation | | O-C |
| | | $\tau$, ms | Ref. | Method | $\tau$, ms | Ref. | Model |
| 0 | 1 | 1.02(5) | 72JOH | TF | $\infty$ | | AA |
| | 2 | 1.02(5) | 72JOH | TF | | | |

| | | | | | | | | | | |
|---|---|---|---|---|---|---|---|---|---|---|
| | | $d^3\Pi_u^+,v,N$ rovibronic levels | | | | | | | | |
| $v$ | $N$ | Experimental studies | | | Semi-empirical determination | | | Non-empirical calculation | | O-C |
| | | $\tau$, ns | Ref. | Method | $\tau$, ns | Ref. | Model | $\tau$, ns | Ref. | Method |
| 0 | 1 | | | | 38.5(0.5) | 90POZ | AA | 38.8[f] | 06FAN/WÜN | AA |
| 1 | 1 | 38.4(1.5) | 93KIY/SAT | DC | 39.0(0.5) | 90POZ | AA | 39.0[f] | 06FAN/WÜN | AA | A |
| | 2 | 39.2(1.4) | 93KIY/SAT | DC | | | | | | |
| | 3 | 39.7(1.8) | 93KIY/SAT | DC | | | | | | |
| 2 | 1 | | | | 39.6(0.5) | 90POZ | AA | 39.2[f] | 06FAN/WÜN | AA |
| | 2 | 40.1(1.5) | 93KIY/SAT | DC | | | | | | | A |
| 3 | 1 | | | | 40.3(0.5) | 90POZ | AA | 39.6[f] | 06FAN/WÜN | AA |
| 4 | 1 | 37(1)[g] | 99KIY/SAT | DC | 40.9(0.5) | 90POZ | AA | 34.7[g] | 99KIY/SAT | NA | C |
| 5 | 1 | 36(3)[g] | 99KIY/SAT | DC | 41.7(0.6) | 90POZ | AA | 34.8[g] | 99KIY/SAT | NA | A |
| 6 | 1 | 37(3)[g] | 99KIY/SAT | DC | 42.4(0.6) | 90POZ | AA | 35.0[g] | 99KIY/SAT | NA | A |

**Table 8. – Continued.**

| | | $d^3\Pi_u^-,v,N$ rovibronic levels | | | | | | | | |
|---|---|---|---|---|---|---|---|---|---|---|
| | | Experimental studies | | | Semi-empirical determination | | | Non-empirical calculation | | O-C |
| $v$ | $N$ | τ, ns | Ref. | Method | τ, ns | Ref. | Model | τ, ns | Ref. | Method | |
| 0 | 1 | | | | 38.5(0.5) | 90POZ | AA | 38.8[f] | 06FAN/WÜN | AA | |
| 1 | 1 | 38.4(1.5) | 93KIY/SAT | DC | 39.0(0.5) | 90POZ | AA | 39.0[f] | 06FAN/WÜN | AA | A |
| | 2 | 39.2(1.4) | 93KIY/SAT | DC | | | | | | | |
| | 3 | 39.7(1.8) | 93KIY/SAT | DC | | | | | | | |
| 2 | 1 | | | | 39.6(0.5) | 90POZ | AA | 39.2[f] | 06FAN/WÜN | AA | |
| | 2 | 40.1(1.5) | 93KIY/SAT | DC | | | | | | | A |
| 3 | 1 | | | | 40.3(0.5) | 90POZ | AA | 39.6[f] | 06FAN/WÜN | AA | |
| 4 | 1 | 37(1)[g] | 99KIY/SAT | DC | 40.9(0.5) | 90POZ | AA | 34.7[g] | 99KIY/SAT | NA | C |
| 5 | 1 | 36(3)[g] | 99KIY/SAT | DC | 41.7(0.6) | 90POZ | AA | 34.8[g] | 99KIY/SAT | NA | A |
| 6 | 1 | 37(3)[g] | 99KIY/SAT | DC | 42.4(0.6) | 90POZ | AA | 35.0[g] | 99KIY/SAT | NA | A |

| | | $e^3\Sigma_u^+,v,N$ rovibronic levels | | | | | | | | |
|---|---|---|---|---|---|---|---|---|---|---|
| | | Experimental studies | | | Semi-empirical determination | | | Non-empirical calculation | | O-C |
| $v$ | $N$ | τ, ns | Ref. | Method | τ, ns | Ref. | Model | τ, ns | Ref. | Method | |
| 0 | 1 | | | | 37.3(3.0) | 90POZ | NA | 31.9[f] | 06FAN/WÜN | AA | |
| | 2 | | | | 37.5(3.0) | 90POZ | NA | | | | |
| | 3 | | | | 37.6(3.0) | 90POZ | NA | | | | |
| | 4 | | | | 37.9(3.0) | 90POZ | NA | | | | |
| | 5 | | | | 38.2(3.0) | 90POZ | NA | | | | |
| | 6 | | | | 38.5(3.0) | 90POZ | NA | | | | |
| 1 | 1 | | | | 41.1(3.0) | 90POZ | NA | 33.9[f] | 06FAN/WÜN | AA | |
| | 2 | | | | 41.3(3.0) | 90POZ | NA | | | | |
| | 3 | | | | 41.5(3.0) | 90POZ | NA | | | | |
| | 4 | | | | 41.7(3.0) | 90POZ | NA | | | | |
| | 5 | | | | 42.1(3.0) | 90POZ | NA | | | | |
| | 6 | | | | 42.5(3.0) | 90POZ | NA | | | | |
| 2 | 1 | | | | 45.5(3.0) | 90POZ | NA | 36.2[f] | 06FAN/WÜN | AA | |
| | 2 | | | | 45.7(3.0) | 90POZ | NA | | | | |
| | 3 | | | | 45.9(3.0) | 90POZ | NA | | | | |
| | 4 | | | | 46.2(3.0) | 90POZ | NA | | | | |
| | 5 | | | | 46.6(3.0) | 90POZ | NA | | | | |
| | 6 | | | | 47.0(3.0) | 90POZ | NA | | | | |
| 3 | 0 | 45(1)[h] | 99KIY/SAT | DC | | | | 36[h] | 99KIY/SAT | NA | D |
| | 1 | | | | 50.6(4.0) | 90POZ | NA | | | | |
| | 2 | | | | 50.8(4.0) | 90POZ | NA | | | | |
| | 3 | | | | 51.0(4.0) | 90POZ | NA | | | | |
| | 4 | | | | 51.3(4.0) | 90POZ | NA | | | | |
| | 5 | | | | 51.8(4.0) | 90POZ | NA | | | | |
| | 6 | | | | 52.3(4.0) | 90POZ | NA | | | | |
| 4 | 0 | 41(1)[h] | 99KIY/SAT | DC | | | | 39[h] | 99KIY/SAT | NA | B |
| 5 | 0 | 42(6)[h] | 99KIY/SAT | DC | | | | 42[h] | 99KIY/SAT | NA | A |

Table 8. – Continued.

| $v$ | $N$ | $k^3\Pi_u^-,v,N$ rovibronic levels | | | | | | |
|-----|-----|------------------------------------|---|---|---|---|---|---|
| | | Experimental studies | | | Non-empirical calculation | | | O-C |
| | | $\tau$, ns | Ref. | Method | $\tau$, ns | Ref. | Model | |
| 0 | 2 | 68.0(2.5) | 03KIY/SAT | DC | 65.4 | 03KIY/SAT | NA | B |
| 1 | 2 | 67.2(2.3) | 03KIY/SAT | DC | 65.3 | 03KIY/SAT | NA | A |
| 2 | 2 | 67.4(2.0) | 03KIY/SAT | DC | 65.7 | 03KIY/SAT | NA | A |
| 3 | 2 | 65.6(2.5) | 03KIY/SAT | DC | 67.1 | 03KIY/SAT | NA | A |
| 4 | 2 | 65.2(1.9) | 03KIY/SAT | DC | 68.0 | 03KIY/SAT | NA | B |
| 5 | 2 | 64.6(1.9) | 03KIY/SAT | DC | 67.7 | 03KIY/SAT | NA | B |
| 6 | 2 | 64.5(1.5) | 03KIY/SAT | DC | 69.5 | 03KIY/SAT | NA | D |

[a] – the non-adiabatic calculation taking into account the interaction with levels of inner well of $I^1\Pi_g$ state [11ROS/TSU];

[b] – the non-adiabatic calculation taking into account the interaction with levels of outer well of $I^1\Pi_g$ state [11ROS/TSU];

[c] – results of singular records, author's error estimates are not reported in [06ROS/YOS];

[d] – this level is subject to strongly enhanced tunneling to the inner well, according to non-adiabatic calculations [06ROS/YOS];

[e] – the levels of outer potential well of the electronic state;

[f] – this value is calculated without taking into account the rotation of the molecule;

[g] – the data are taken from fig.5b from [99KIY/SAT];

[h] – the data are taken from fig. 5a from [99KIY/SAT].